\documentclass[reprint,aps,prd,amssymb, amsmath, nofootinbib, onecolumn, superscriptaddress]{revtex4-2}

\usepackage{graphicx}
\usepackage{subcaption}

\usepackage{mathtools}
\usepackage[svgnames]{xcolor} 

\usepackage[colorlinks=True, linkcolor=SlateBlue,
            citecolor=SteelBlue,urlcolor=BlueViolet]{hyperref}
\usepackage[capitalise]{cleveref}
\usepackage{orcidlink}
\usepackage{extpfeil}

\usepackage{aas_macros}

\newcommand{\vect}[1]{\boldsymbol{#1}}
\newcommand{\dvect}[1]{\dot{\boldsymbol{#1}}}
\newcommand{\uvect}[1]{\vect{\hat{#1}}}

\newcommand{\vxi}{\boldsymbol{\xi}}
\newcommand{\Ma}{\mathcal{M}_\ast}
\newcommand{\If}{\mathcal{I}_f}

\newcommand{\II}{\mathcal{I}}
\newcommand{\JJ}{\mathcal{J}}

\begin{document}

\author{Hang Yu\,\orcidlink{0000-0002-6011-6190}}
\email{hang.yu2@montana.edu}

\affiliation{eXtreme Gravity Institute, Department of Physics, Montana State University,
Bozeman, Montana 59717, USA}

\author{Giorgio Nicolini\, \orcidlink{0009-0007-7862-1789}}
\affiliation{eXtreme Gravity Institute, Department of Physics, Montana State University,
Bozeman, Montana 59717, USA}

\author{Shu Yan Lau\,\orcidlink{0000-0002-8239-0174}}
\affiliation{eXtreme Gravity Institute, Department of Physics, Montana State University,
Bozeman, Montana 59717, USA}

\author{K.J. Kwon\,\orcidlink{0000-0001-9802-362X}}
\affiliation{Department of Physics, University of California, Santa Barbara, CA 93106, USA}

\author{Tejaswi Venumadhav\,\orcidlink{0000-0002-1661-2138}}
\affiliation{Department of Physics, University of California, Santa Barbara, CA 93106, USA}
\affiliation{International Centre for Theoretical Sciences, Tata Institute of Fundamental Research, Bangalore 560089, India}

\author{Nils Andersson\,\orcidlink{0000-0001-8550-3843}}
\affiliation{Mathematical Sciences and STAG Research Centre,
University of Southampton, Southampton SO17 1BJ, United Kingdom}

\author{Pantelis Pnigouras\,\orcidlink{0000-0003-1895-9431}}
\affiliation{Departamento de F\'isica, Universidad de Alicante, Campus de San Vicente del Raspeig, E-03690 Alicante, Spain}

\author{Fabian Gittins\,\orcidlink{0000-0002-9439-7701}}
\affiliation{Institute for Gravitational and Subatomic Physics (GRASP),
Utrecht University, Princetonplein 1, 3584 CC Utrecht, Netherlands}
\affiliation{Nikhef, Science Park 105, 1098 XG Amsterdam, Netherlands}

\author{Amlan Nanda}
\affiliation{Department of Physics and Astronomy, Washington State University, Pullman, Washington 99164–2814, USA}


\title{Nonlinear hydrodynamics in spinning neutron stars: \\
Theoretical universal relations and equilibrium solutions}

\begin{abstract} 
We study tides during the inspiral of a binary neutron star system, including nonlinear hydrodynamical interactions.  
Using an affine approximation that treats the perturbed neutron star as an ellipsoid, we analytically derive coupling coefficients among the quadrupolar f-modes and the radial mode to the four-wave order (next-to-next-to-leading order) in the Hamiltonian, allowing for arbitrary (aligned or anti-aligned) spin of the background star. 
Our model reveals a series of universal relations from first-principles arguments.
We show that the three-wave (next-to-leading-order) interaction coefficients in a non-spinning star are fully determined by the properties of the linear tide. Therefore, they do not probe new physics of the neutron star. 
Nonetheless, three-wave nonlinear tides are significant corrections to the gravitational waveform. 
We support this via a hybrid approach that simultaneously captures mode resonances expected in Newtonian hydrodynamics and is consistent with relativistic calculations in the low-frequency expansion. 
The nonlinear tide in a single neutron star can cause a phase shift of around 1.8 radians accumulated up to merger compared to the linear tide model; for a binary of similar masses, the phase shift is approximately doubled.
In a low-frequency expansion, the nonlinear tide is degenerate with the finite-frequency correction of the linear tide, introducing systematic bias when ignored. 
Our calculation extends to four-wave interactions, which, for a slowly spinning neutron star, provide only small corrections. 
For a rapidly rotating neutron star, the nonlinear centrifugal drive of the f-mode provides a window to study the adiabatic exponent related to internal buoyancy that cannot be probed by the linear and three-wave tides in slowly spinning systems.
The four-wave anharmonicity cannot lead to resonance locking of the f-mode. 
\end{abstract}

\maketitle

\section{Introduction}

Tidal interactions in a binary neutron star (BNS) or a neutron star-black hole (NSBH) system inspiralling due to gravitational waves (GWs) offer a valuable window to simultaneously study the neutron star (NS) equation of state at extreme densities \cite{Lattimer:12} and the dynamics at extreme spacetime curvatures \cite{Silva:21}. 

So far, theoretical studies of BNS tidal interactions have been treating the tide as a small perturbation to the background star (e.g., \cite{Flanagan:08, Damour:09, Binnington:09, Hinderer:10, Bini:12, Damour:12, Bini:14, Bernuzzi:15,  Hinderer:16, Steinhoff:16, Nagar:18, Andersson:20, Ma:20, Gamba:21b, Gamba:23b, Mandal:23, Mandal:24, Pitre:24, Pnigouras:24, Yu:24a, AbhishekHegade:24, Pnigouras:25, Yu:25a, Haberland:25, Miao:25, Hegade:26}). In this case, the NS's response has been calculated only to the linear order in the perturbed displacement over the background radius in the fluid's equation of motion, or quadratic order in the potential energies (or the Lagrangian or Hamiltonian). This linear theory works well in the early inspiral when the binary components are far apart.  
However, as the orbit decays to smaller separations, the tidal bulge increases to a significant fraction of the unperturbed radius. Moreover, the tidal forcing frequency approaches the frequency of an NS eigenmode in the late inspiral, significantly amplifying the tidal bulge while creating a lag in it via mode resonance. Nonlinear hydrodynamical effects are thus expected to become significant. 

The authors of \cite{Yu:23a} studied large-scale nonlinear tides (i.e., nonlinear interactions among fundamental f-modes) in BNSs within Newtonian hydrodynamics. They showed that the nonlinear tide can cause a GW phase shift of $\mathcal{O}(1)\,{\rm rad}$. 
With conservative assumptions, \cite{Bretz:26} showed that ignoring the nonlinear tide can cause a significant bias in the data analysis of BNSs. The bias mainly enters through the inferred f-mode eigenfrequency, which, in most BNS waveforms, is not a free parameter. Instead, it is treated as a function of the tidal deformability via their linear universal relation \cite{Chan:14}. This treatment transfers the systematic error from mode frequency into the inferred tidal deformability, which was estimated to be at least 8\% higher than the actual value.    

Recently, \cite{Pitre:25} extended the calculation of \cite{Yu:23a} to a relativistic NS, and \cite{Pani:25} provided the associated GW waveform, confirming the significance of nonlinear tides in general relativity (GR). However, both studies were based on low-frequency or even adiabatic approximations where the tidal forcing frequency is assumed to be much lower than the f-mode frequency (or the dynamical frequency of the NS), with an expansion in their ratio made. 
However, \cite{Yu:23a} showed that the dominant nonlinear correction is, in fact, a decrease in the f-mode frequency, which becomes most significant during the late inspiral, when the orbit approaches resonance with the f-mode, and the tidal response is sensitive to small perturbations to the frequency. Therefore, a low-frequency expansion likely underestimates the impact of the nonlinear tide by an appreciable amount.

In this work, we will apply a hybrid approach where we evolve the hydrodynamical equations of \cite{Yu:23a} that capture all dynamical effects expected in Newtonian theory (e.g., mode resonance), while applying a calibration procedure to the coefficients such that under a low-frequency expansion, our result is identical to the GR calculation of \cite{Pitre:25}. 
This approach is consistent with the resummation attempted by \cite{Pitre:24} to resum the expansion into a Lorentzian form before mode resonance and extends further to the post-resonance region, where a Lorentzian is no longer sufficient. 
While not exactly rigorous in full GR  (but at least the functional forms are exact in the Newtonian limit), it may provide a theoretically motivated estimate of the impact of the nonlinear tide beyond the low-frequency limit. 

To compute the nonlinear coupling coefficients analytically and allow an arbitrary value of the spin of the background star, we will adopt the affine model proposed by \cite{Carter:83, Carter:85}. This model was used to study the inspiral of BNSs and NSBHs in the Newtonian limit by \cite{Lai:93, Lai:94a, Lai:94b} and in post-Newtonian (PN) theory by \cite{Ferrari:09, Ferrari:12}; these studies used fluid variables (e.g., the major axes of the ellipsoids) that did not make the hydrodynamical nature explicit. \cite{Yu:26} recently revisited the affine model and instead linked the fluid variables to amplitudes of the NS monopolar (with a polar degree $l_a=0$) radial mode and quadrupolar ($l_a=2$) f-modes. This representation made the appearance of the f-mode resonance transparent, while simultaneously offering an intuitive interpretation of the dominant nonlinear hydrodynamical effects as corrections to the natural frequencies of the modes \cite{Yu:23a}. Nonetheless, in the low-frequency limit, corrections to the coupling to the external field enter at the same order as the frequency shift. Thus, when resumming the low-frequency expansion into the Lorentzian form expected for a forced harmonic oscillator, as attempted by \cite{Pitre:25}, these two effects must be separated. 
The approach of \cite{Yu:26} further provides a direct way to extract various nonlinear coupling coefficients \cite{VanHoolst:93, VanHoolst:94, Schenk:02, Weinberg:12, Venumadhav:14, Weinberg:16} among the f- and radial modes. In this work, we will follow \cite{Yu:26} to study the nonlinear tides in an NS during its inspiral with another compact object companion.

The affine approximation of \cite{Carter:85, Yu:26} contains only a finite number of internal NS parameters (two in a Newtonian model, the polytropic and adiabatic exponents, $\Gamma$ and $\Gamma_{\rm ad}$, respectively; an additional compactness in a relativistic one). 
Consequently, various linear and nonlinear coefficients derived from it will naturally be related, revealing a series of universal relations from first-principles calculations (similar to \cite{Chan:14, Sham:15, Yagi:14}). In addition to the well-known relations, such as the tidal deformability--moment of inertia \cite{Yagi:13} and the tidal deformability--f-mode frequency relations \cite{Chan:14} (see \cite{Yagi:17} for a review), the affine model further connects three-wave couplings (i.e., the next-to-leading-order, or NLO tidal effect) to the properties of the linear tide, which are ultimately determined by only the polytropic exponent and compactness of the NS when it has no background spin. 
This connection has important implications for GW data analysis. Specifically, it means that the parameters required to describe the NLO nonlinear tides in a slowly spinning system do not represent fundamentally new properties of the NS, and therefore can be expressed in terms of the linear tidal deformability, thereby reducing the dimensionality of the parameter space explored during inference. On the other hand, ignoring the nonlinear tide (as nearly all current BNS and NSBH waveforms do) will lead to biases in the inferred deformability. Such biases will be significant in the era of third-generation GW detectors (including the Einstein Telescope \cite{Punturo:10, ET:25} and the Cosmic Explorer \cite{Evans:17, Evans:19, Evans:21}), where both a large number of total detections and loud individual events are expected, as recently investigated by \cite{Bretz:26}.

Note that the affine model considered in this work is a one-mode approximation that considers only the tidal coupling to the f-modes.
It cannot capture smaller-scale perturbations caused by, e.g., the gravity modes (g-modes), whose nonlinear effects may also be significant. For example, a low-order (long-wavelength) g-mode could experience resonance locking due to the conservative anharmonicity \cite{Kwon:24, Kwon:25}, whereas a high-order g-mode (short-wavelength) might lead to significant fluid dissipation via the pg-instability \cite{Weinberg:13, Venumadhav:14, Weinberg:16, Essick:16, GW170817pg}. We also ignore inertial \cite{Flanagan:07, Ma:21, Gupta:21, Gupta:23} and interface modes \cite{Tsang:12, Pan:20, Passamonti:21} in this study. 

Throughout this analysis, we adopt geometric units, setting $G=c=1$. Einstein summation is assumed, with Latin indices $i, j,...,$ running from 1 to 3 denoting the three spatial components expressed in a Cartesian coordinate system and used interchangeably in superscripts and subscripts.  
Summation over NS eigenmodes will always be made explicit. The configuration-space eigenmodes will be labeled by subscripts $a, b, \ldots$. 
For a particular mode, its label $a$ will be replaced either by its azimuthal quantum number $m_a$ (when referring to the quadrupolar f-modes) or a label $r$ (when referring to the monopolar, or radial mode). 
Each configuration-space mode can be further split into two phase-space modes labeled by both $a$ and $s_a$, with $s_a=\pm$ the sign of the mode's eigenfrequency. When the $s_a$ subscript is omitted, it will always refer to the positive-frequency mode. 
We use subscripts $A$ and $B$ to label the tidally perturbed star and the companion, respectively. 
By default, we will use $M_A=M_B=1.35\,M_\odot$ and $R_A=11.7\,{\rm km}$, with a compactness $M_A/R_A=0.17$. The mass-radius relation \cite{Read:09a} of NS $A$ is consistent with an NS described by the SLy equation of state \cite{Douchin:01}, though we will also approximate it as a $\Gamma=2$ polytrope. We introduce $E_A\equiv M_A^2/R_A$ and $\omega_A^2=M_A/R_A^3$ as natural units for energies and frequencies, respectively. Tides on the companion will be ignored throughout this work, so the tidal phase shifts quoted here should be multiplied by $\sim 2$ for a typical BNS system with a mass ratio close to unity. 

The rest of the paper is organized as follows. 
We first provide a road map to the main results of the study in Sec. \ref{sec:exe_sum}.
In Sec. \ref{sec:affine}, we present the Hamiltonian derived from the affine model that is used to generate the equations of motion. The theoretical origin of various universal relations is described in Sec. \ref{sec:UR_nl_tide_N} for a Newtonian model and in Sec. \ref{sec:UR_in_GR} for the (empirical) extension to GR. A calibration procedure is introduced in Sec. \ref{sec:calibrations}, which ensures the consistency of our calculations with \cite{Pitre:24, Pitre:25} in the low-frequency limit.
We first solve the internal problem of the NS tidal response in terms of mode amplitudes in Sec. \ref{sec:eq_sol_hydro}. Analytical solutions including four-wave interactions in the Hamiltonian (the next-to-next-to-leading-order, or NNLO tide) are presented in Sec. \ref{sec:nl_mode_amp}. We then use a combination of theoretical and numerical tools to examine whether an f-mode can experience resonance locking in a rapidly spinning NS in Sec. \ref{sec:res_locking}. The possibility of probing new NS physics from the nonlinear tide is explored in Sec. \ref{sec:Gamma_ad}. 
We then solve the external problem of the tidal back-reaction on the orbit in Sec. \ref{sec:sol_orb}. Solutions based on the commonly adopted effective Love number are presented in Sec. \ref{sec:effective_Love}, with caveats regarding its usage in a Hamiltonian system discussed in Sec. \ref{sec:issue_w_effective_Love}. We then compare the low-frequency limit of the NLO tidal phase shift in GW waveforms with numerical results in Sec. \ref{sec:GW_phase}.
Finally, we discuss our results in Sec. \ref{sec:discussion}.

\subsection{Road map to the main results}
\label{sec:exe_sum}
As this paper contains an in-depth discussion of a number of issues relevant to realistic tidal dynamics, it may be challenging to digest through a single ``linear" reading. Moreover, different readers may want to focus on specific aspects. In order to help the reader navigate the material and identify the main results of the paper, we provide the following road map. 

\emph{1) System dynamics.} 
The non-point particle (non-PP) part of the Hamiltonian is given by Eqs. (\ref{eq:H_npp})-(\ref{eq:H_rot}), describing the internal hydrodynamics and the fluid's tidal coupling to the orbit to fourth order in the fluid displacement (i.e., NNLO, also referred to as four-wave interactions in the paper). 
The non-PP Hamiltonian is combined with the point-particle (PP) orbital Hamiltonian (Eq. \ref{eq:full_Ham}) to complete the conservative dynamics. 
We couple the conservative dynamics with an energy loss from the system, given by Eq. (\ref{eq:dot_E_sys}), which is implemented as Burke-Thorne accelerations following \cite{Yu:23a}. 
Together, the results provide a set of differential equations that can be numerically solved, providing the most accurate predictions of this study. 

\emph{2) Analytical nonlinear coupling coefficients.} 
The nonlinear coupling coefficients are derived from the affine model (defined in Eq. \ref{eq:qij_def}). Of particular importance are $\kappa_2$ and $\JJ_2$, given by Eqs. (\ref{eq:kap2_vs_I_w}) and (\ref{eq:J2_vs_I_w}). These parameters determine the NLO tidal dynamics in a slowly spinning NS \cite{Yu:23a}. We also highlight the mapping between coefficients used in this study, $(\If, \omega_{f0}, \kappa_2, \JJ_2)$, for the linear tidal overlap, the non-spinning NS f-mode frequency, the three-wave coupling coefficient between f-modes, and the nonlinear tidal overlap, with those used in \cite{Pitre:25}, $(k_{2A}, \ddot{k}_{2A}, p_{2A})$, for the Love number, the coefficient describing the leading-order finite-frequency correction, and the coefficient for the NLO tide. In the Newtonian limit, the different quantities are related by
\begin{align}
    &k_{2A} = \frac{4\pi}{5} \If^2, \\
    &\ddot{k}_{2A} = \frac{4\pi}{5} \If^2 \frac{\omega_A^2}{\omega_{f0}^2}, \\
    &p_{2A} = -\frac{8\pi}{5}\sqrt{\frac{\pi}{5}} \If^2 (3\JJ_2 + 4\kappa_2 \If),
\end{align}
see Eqs. (\ref{eq:lamA_vs_k2A_vs_If}), (\ref{eq:ddk2_vs_k2_wf}), and (\ref{eq:p2_vs_J2_kap2}). We use these relations to calibrate our coupling coefficients to account for relativistic effects of the NS in Eqs. (\ref{eq:kap2_cal_GR}) and (\ref{eq:J2_cal_GR}) so that our calculation is fully consistent with \cite{Pitre:25} in the low-frequency expansion.

\emph{3) New theoretical universal relations.} 
Since our hydrodynamics are derived from a low-dimensional affine model, various coupling coefficients are naturally connected, revealing the theoretical origin of NS universal relations. Specifically, we highlight a new relation between $p_{2A}$ and $\ddot{k}_{2A}$ in Eq. (\ref{eq:p2_vs_ddk2_N}). We further empirically extend the Newtonian result to (also Eq. \ref{eq:p2_vs_ddk2_GR}), 
\begin{equation}
    p_{2A} = \frac{20}{7}\ddot{k}_{2A} \exp[-4(M_A/R_A)],
\end{equation}
which holds for NSs with arbitrary compactness in GR.
Besides the relation above, we also present a practical relation between $\bar{p}_{2A} = p_{2A}/(M_A/R_A)^8$, a specific combination that shows up in the GW phase (Eq. \ref{eq:dPsi_LF_barp2A}), and the dimensionless tidal deformability $\bar{\lambda}_A=\lambda_A/M_A^5$ in Eq. (\ref{eq:p2bar_vs_lambar}). 
This result suggests that the NLO tide in a slowly spinning NS is determined by the properties of the linear tide and does not probe new physics of the NS, allowing for a reduction in the parameter space dimensionality in BNS parameter estimations.

\emph{4) Analytical solutions of internal hydrodynamics to NNLO.} 
The tidal response of an NS is derived analytically in terms of the amplitudes of its eigenmodes (denoted by $q_a$ for a configuration space expansion shown in Eq. \ref{eq:mode_amp_def}; equivalently, $q_a$ and $\dot{q}_a$ can also be described by two phase-space modes as in Eqs. \ref{eq:q_vs_c} and \ref{eq:dq_vs_c}). Each mode can be viewed as a harmonic oscillator driven by the tidal field at a frequency $m_a\omega$, with $m_a$ the azimuthal quantum number and $\omega\equiv\dot{\phi}$ the orbital frequency ($\phi$ is the orbital phase). The particular solution of the oscillator, with an inertial-frame phase evolution $\exp(-im_a\phi)$ and referred to as the equilibrium tide in this work following \cite{Yu:24a}, is characterized by a Lorentzian form, 
\begin{equation}
    q_a = \frac{2\omega_{a0}^2 V_a }{\omega_{a0}^2 - m_a^2 [\omega^2 - 2\omega(\Omega_A + C_a \Omega_A)]}e^{-im_a\phi},
\end{equation}
with $2V_a$ denoting the amplitude of the drive (the factor of 2 is for our normalization; Eq. \ref{eq:normalization}). In the denominator, the $\Omega_A$ and $C_a\Omega_A$ terms are spin corrections due to the Doppler and (the leading-order) Coriolis effects, respectively. 
Compared to the baseline solution above, our main result, Eq. (\ref{eq:b2_p_bnc}), takes a different form, 
\begin{equation}
    q_a = \frac{2\omega_{a0}^2 (V_a + \Delta V_a) }{[\omega_{a0}^2(1+\Delta \omega_a^2/\omega_{a0}^2) - m_a^2 (\omega-\Omega_A)(\omega-\Omega_A-2C_a\Omega_A)] + i\omega_{a0}\gamma_{m_a d} }e^{-im_a\phi},
    \label{eq:q_a_eq_sol}
\end{equation}
with four key modifications. First, the $\mathcal{O}(\Omega_A^2)$ Coriolis effect is incorporated (Appendix \ref{appx:Coriolis}) so that the validity of our result can be pushed beyond the slow rotation limit. The rest three modifications are respectively denoted by $\Delta V_a$, $\Delta \omega_a^2$, and $\gamma_{m_a d}$. The $\Delta V_a$ term (lines \ref{eq:dV2_3m} and \ref{eq:dV2}) describes the nonlinear correction to the mode's coupling to the tidal field, while the $\Delta \omega_a^2$ term (lines \ref{eq:dw2_wf_3m}-\ref{eq:dw2_wf}) the nonlinear shift of the mode's natural frequency squared. Importantly, nonlinear hydrodynamics enter both the numerator and denominator of the Lorentzian. 
Because the orbit is evolving due to GW emission, it introduces a dynamical lag in the tidal bulge \cite{Lai:94c} that behaves like an effective damping \cite{Yu:24a}, which we characterize through the $\gamma_{m_a d}$ term (Eq. \ref{eq:gam_2d}). The effective damping depends on the evolution of the orbital separation ($\dot{r}$) and frequency ($\dot{\omega}$), and becomes increasingly important as the binary inspirals. It also diverges at a mode's resonance, leading to the equilibrium tide vanishing at that point. The physical solution, of course, is non-zero at resonance. The surviving part corresponds to the dynamical tide in \cite{Yu:24a} excited at resonance, whose analytical solution is not discussed in this work. 

We validate our analytical result, Eq. (\ref{eq:q_a_eq_sol}) or (\ref{eq:b2_p_bnc}), against numerical solutions to the differential equations in Fig. \ref{fig:b2_vs_f_G}. We note that it is sufficient to include the nonlinear effects to three-wave order (i.e., NLO) for a typical, slowly-spinning relativistic NS. The effective damping $\gamma_{m_a d}$ plays a more significant role than the NNLO nonlinearity. It is crucial that tidal back-reactions are accounted for in the evaluation of $\dot{r}$ and $\dot{\omega}$ in $\gamma_{m_a d}$. Using PP GW predictions significantly underestimates the damping. This makes the problem second-order in the tidal back-reaction (to be distinguished from the nonlinearity in the fluid displacement studied here), representing a major technical challenge to be addressed in the future.

\emph{5) Effective Love number and GW phase shift.} 
Once the solutions to the modes are obtained, the tidal back-reaction on the orbit follows directly. One popular quantity used by the community is the effective Love number, defined through Eq. (\ref{eq:klm_eff_def}). By plugging the solutions of mode amplitudes into the modes' nonlinear mapping to the NS's multipolar moments, Eq. (\ref{eq:Q_lm}), or more explicitly, (\ref{eq:Q22}), we obtain 
\begin{equation}
    \kappa_{22, {\rm eff}} \simeq \frac{(1 -4\sqrt{\frac{\pi}{5}}\JJ_2\frac{M_B}{M_t} \frac{\omega^2}{\omega_A^2} )}{1- 4\frac{\omega^2}{\omega_{f0}^2}+2\sqrt{\frac{\pi}{5}}(\JJ_2+4\kappa_2\If)\frac{M_B}{M_t}\frac{\omega^2}{\omega_A^2}   + i\frac{\gamma_{2d}}{\omega_{f0}}},
\end{equation}
as the non-spinning NLO effective Love number for the $l=m=2$ harmonic (note that the values are different for different harmonics once one relaxes the adiabatic assumption). Compared to the resummation proposed in \cite{Pitre:25} (see their eqs. 1.8, 1.10, and 8.9), where all the nonlinear corrections enter through the denominator of the Lorentzian, our result differs in that the nonlinear hydrodynamics corrects both the numerator and denominator of the expression, as in the mode amplitudes. 
This highlights the need to have both $\JJ_2$ and $\kappa_2$ when describing the NLO dynamics, instead of a single combination of the two given by $p_{2A}$ (Eq. \ref{eq:p2_vs_J2_kap2}). 
We also include the effective damping, $\gamma_{m_a d}$, which is crucial in the late inspiral. Our result is validated against numerical calculations in Fig. \ref{fig:eff_love}. Eq. (\ref{eq:k22_resum_3m_w_spin}) extends the result to include background spins (aligned or anti-aligned). 

A more direct observable is the phase of the GW, $\phi_{\rm gw}=2\phi$, which we estimate through an energy-balancing argument (Eq. \ref{eq:delta_t_vs_f}). The tidal correction to the equilibrium energy is given by Eq. (\ref{eq:Delta_E}), where one can plug in equilibrium solutions of mode amplitudes into the Hamiltonian in Eq. (\ref{eq:H_npp}) to obtain the tidal interaction energy and energies stored inside the NS. The change in the orbital energy is computed through the change in the circular orbit radius, Eq. (\ref{eq:delta_r_r}). Similarly, the perturbation of the energy loss is computed from Eq. (\ref{eq:dot_E_sys}) by plugging in the mode solutions into the NS's mass quadrupole. To provide analytical insights, we provide a low-frequency expansion of the frequency-domain phase shift in Eqs. (\ref{eq:phase_shift_LF}) and (\ref{eq:k_psi_eff}). 
In particular, the three-wave (NLO) nonlinear tide's contribution to the phase shift is (last term of Eq. \ref{eq:k_psi_eff})
\begin{equation}
    \Delta \Psi_{\rm NLO, LF} (\omega) = -\frac{15}{352}\frac{M_A^5(17 M_B + M_t)}{M_t^6} \frac{p_{2A}}{(M_A/R_A)^8} (M_t\omega)^{11/3},
\end{equation}
under the low-frequency expansion (indicated by the ``LF'' in the subscript).
In comparison, the leading-order, finite-frequency correction of the linear tide leads to a phase shift of (the second last term of Eq. \ref{eq:k_psi_eff})
\begin{equation}
    \Delta \Psi_{\rm lin,FF, LF} = -\frac{15}{704}\frac{M_A^5(147 M_B + 8 M_t)}{M_B M_t^5}\frac{\ddot{k}_{2A}}{(M_A/R_A)^8} (M_t\omega)^{11/3}. 
\end{equation}
This makes the degeneracy between the two effects transparent. Currently, most waveform models adopted by GW data analysis ignore the nonlinear tide and uses the linear universal relation between $\ddot{k}_{2A}$ (or $\omega_{f0}$; Eq. \ref{eq:ddk2_vs_k2_wf}) and the tidal deformability \cite{Chan:14, Saes:25} to estimate the tidal coefficient of the $(M_t\omega)^{11/3}$ term (which is formally at an 8-PN order despite its Newtonian origin). The inferred tidal deformability using such a model will inevitably be biased \cite{Bretz:26}. 

Note, however, that the low-frequency expansion should only be used to reveal the degeneracy. It underestimates the actual phase shift by about a factor of 2 (left panels of Figs. \ref{fig:phase_shift} and \ref{fig:phase_shift_SLy} for two equations of states, $\Gamma=2$ and SLy). In the time domain, the nonlinear tide introduces a GW phase correction of $\sim1.8\,{\rm rad}$ relative to the linear tide (measured at a fixed orbital separation of $r=2R_A$) for the SLy equation of state from a single, non-spinning NS, highlighting its significance. This result is illustrated in the right panel of Fig. \ref{fig:phase_shift_SLy}. 

\emph{6) No resonance locking of the f-mode and a new probe of $\Gamma_{\rm ad}$.} 
While our analytical solutions focus on the equilibrium tides inside slowly-spinning NSs, our formulation allows for an arbitrary spin rate. Two sets of numerical experiments of rapidly spinning NSs with f-mode resonantly excited during the inspiral are conducted in Secs. \ref{sec:res_locking} and \ref{sec:Gamma_ad}. 
In the first experiment, we demonstrate that the f-mode does not experience resonance locking. An approximately linear-in-frequency increase in the tidal spin observed by \cite{Kuan:24} after resonance can already be reproduced by linear theory due to the overshooting of a second-order system, with nonlinear tide introducing corrections only at a quantitative level (see the lower panel of Fig. \ref{fig:no_res_lock}).  
In the second experiment, we demonstrate the dependence on $\Gamma_{\rm ad}$ through the nonlinear correction of the f-mode frequency due mainly to the centrifugal potential. 
This parameter is not probed by the linear and NLO tides in a slowly spinning NS. 
When the f-mode is resonantly excited, $> 1\,{\rm rad}$ change in the GW phase is observed as $\Gamma_{\rm ad}$ changes (Fig. \ref{fig:dphi_vs_Gam_ad}). Effectively, this means that the nonlinear tide involves aspects associated with the composition of the high-density nuclear matter.

\section{Nonlinear hydrodynamics from the affine approximation}
\label{sec:affine}

Consider first the fluid motion in a frame corotating with the NS. We use $\bar{\vect{x}}$ and $\vect{x}$ to respectively denote the original and perturbed location of a mass element in the deformed NS with mass $M_A$ and radius $R_A$. The Lagrangian displacement $\vxi$ and its modal decomposition are defined through
\begin{equation}
    \vect{x}(t, \bar{\vect{x}})=\bar{\vect{x}} + \vxi(t, \bar{\vect{x}})=\bar{\vect{x}} + \sum_a q_a(t)\vxi_a(\bar{\vect{x}}),
    \label{eq:mode_amp_def}
\end{equation}
where in the second equality, a configuration-space decomposition of $\vxi$ into eigenmodes is assumed. 
We further assume that the perturbation satisfies an affine model \cite{Carter:83, Carter:85, Diener:95, Yu:26}, 
\begin{equation}
    x_i(t, \bar{\vect{x}}) = q_{ij}(t) \bar{x}_j,
    \label{eq:qij_def}
\end{equation}
which is exact when the fluid is incompressible \cite{Chandrasekhar:87}, and remains a good approximation for fluids inside an NS when their velocities are low compared to the speed of sound. The matrix components $q_{ij}$ are viewed as functions of mode amplitudes $q_a$.   
In particular, we consider four modes in our analysis, including three $l_a=2$ f-modes with $m_a=2,0, -2$, as well as a radial ($l_a=m_a=0$) mode. The $l_a=2$, $m_a=\pm 1$ f-modes do not couple to the tide and remain zero even with nonlinear couplings. 
The f-modes are assumed to have eigenfunctions $\vxi_{m_a} \propto \nabla[\bar{x}_r^2 Y_{2m_a}(\bar{x}_\theta, \bar{x}_\phi)]$ to satisfy the affine model, where $(\bar{x}_r, \bar{x}_\theta, \bar{x}_\phi)$ are spherical coordinates of $\bar{\vect{x}}$.  They directly couple to the orbit via the tide at the linear order and NLO \cite{Yu:23a}, and map to the three semi-major axes of an ellipsoid (see later in Eq. \ref{eq:semiaxes}). The radial mode with $\vxi_{r}\propto \bar{\vect{x}}$ describes an overall expansion or contraction of the star, which is needed to account for the centrifugal force caused by rotation, as well as tides at NNLO. 
Together, they capture the symmetric trace-free and the trace components of $q_{ij}$, respectively. 
In principle, the affine model also allows an overall rotation of the NS around its spin axis (the z-axis of our coordinate system), which is also parallel to the orbital angular momentum. It is related to the anti-symmetric part of $q_{ij}$, or an axial mode \cite{Pnigouras:24} described by the vector spherical harmonic $\vect{\bar{x}}\times \nabla Y_{10}$. Such an overall rotation does not affect the internal energy, and its coupling to the $|m_a|=2$ f-modes enters beyond NNLO, hence it is ignored in the analysis (see Appendix \ref{appx:Coriolis} for further discussions). 
We normalize the modes so that
\begin{equation}
    2\omega_{a0}^2 \int dM \xi_{ai}^{\ast} \xi_{ai}= 2\omega_{a0}^2 (q_{ij}^{(a)\ast}-\delta_{ij})(q_{ik}^{(a)}-\delta_{ik})\int \bar{x}_j \bar{x}_k dM 
    = 2 \omega_{a0}^2\Ma (q_{ij}^{(a)\ast}-\delta_{ij})(q_{ik}^{(a)}-\delta_{ik}) \delta_{jk}= \frac{M_A^2}{R_A} = E_A. 
    \label{eq:normalization}
\end{equation}
where $q_{ij}^{(a)} = q_{ij}|_{q_b=\delta_{ab}}$ (i.e., setting the amplitude $q_a$ to 1 for mode $a$ and 0 for all other modes in $q_{ij}$; see later in Eq. \ref{eq:q_ij_vs_q_a}), and
the $\Ma$ factor is introduced through $\int \bar{x}_i \bar{x}_j dM = \Ma M_A R_A^2 \delta_{ij}$, which is related to the (unperturbed) NS' moment of inertia $I_A$ via
\begin{equation}
    I_A=2\Ma M_AR_A^2.
    \label{eq:IA_vs_Ma}
\end{equation}
We use $\omega_{a0}$ ($>0$) to represent the eigenfrequency of a mode in the non-spinning limit. For the $l_a=2$ f-modes and the $l_a=0$ radial mode, we will further write their eigenfrequencies as $\omega_{f0}$ and $\omega_{r}$ (both positive), with \cite{Yu:26} 
\begin{align}
    \frac{\omega_{f0}^2}{\omega_A^2}= \frac{4 (\Gamma-1)}{5[5(\Gamma-1) - 1] \Ma}  
    \text{ and }
    \frac{\omega_r^2}{\omega_{A}^2}= \frac{5(3\Gamma_{\rm ad} - 4)}{4} \frac{\omega_{f0}^2}{\omega_A^2},
    \label{eq:omega_a}
\end{align}
where $\Gamma = d \log P/d\log \rho$ and $\Gamma_{\rm ad}=(\partial \log P/\partial \log \rho)_{\rm entropy,...}$ are respectively polytropic and adiabatic exponents relating pressure $P$ and density $\rho$. Both $\Gamma$ and $\Gamma_{\rm ad}$ are assumed to be constants throughout the star. We further introduce the polytropic index $n$ through $\Gamma=1+1/n$.  
The factor of 2 in the normalization, Eq. (\ref{eq:normalization}), is for future convenience, when we transition to a phase-space description in terms of Hamiltonian for future integrations to an effective-one-body (EOB) framework with relativistic orbital dynamics \cite{Buonanno:99, Steinhoff:16, Steinhoff:21, Haberland:25, Yu:25a}. In this convention, the coupling coefficients can be directly compared to those in \cite{Yu:23a}. 
Under this normalization, the non-zero elements of $q_{ij}$ are
\begin{subequations}
\begin{align}
    q_{11} &= 1 + \frac{\omega_A}{\omega_{f0}}\frac{q_2+q_{-2} -\sqrt{2/3} q_0}{2\sqrt{2\Ma} } + \frac{\omega_A}{\omega_{r}}\frac{q_r}{\sqrt{6\Ma}},\\
    q_{12} &= q_{21}=\frac{\omega_A}{ \omega_{f0} } \frac{i(q_2-q_{-2}) }{2\sqrt{2\Ma}} ,\\
    q_{22} &= 1 - \frac{\omega_A}{\omega_{f0}} \frac{q_2+q_{-2} + \sqrt{2/3}q_0}{2\sqrt{2\Ma} } + \frac{\omega_A}{\omega_r}\frac{q_r}{\sqrt{6\Ma} },\\
    q_{33} &= 1+\frac{\omega_A}{\omega_{f0}}\frac{q_0}{\sqrt{3\Ma}} + \frac{\omega_A}{\omega_r}\frac{q_r}{\sqrt{6\Ma}}.
\end{align}
\label{eq:q_ij_vs_q_a}
\end{subequations}

Our procedure to obtain the equations of motion for the fluid is similar to \cite{Yu:26} but improves in the treatment of the Coriolis force. The formalism in this work captures the full effects and is valid beyond the linear-in-$\Omega_A$ order (detailed in Appendix \ref{appx:Coriolis}).  We also include the centrifugal potential as an external perturbation (i.e., on the same footing as the tidal potential; note that we hold the baryonic mass constant, which is different from \cite{Hartle:67, Kruger:20, Kruger:23} where the central energy density is held constant).  
The affine approximation allows all the kinetic and potential terms (hence the Lagrangian and Hamiltonian) to be computed analytically.

The configuration-space description is then transformed to a phase-space description to assist comparisons with the standard treatment of rotation in \cite{Schenk:02} (which is general and does not require the affine approximation) and previous work on nonlinear tides (\cite{Yu:23a} in particular).
This is done by splitting each mode according to 
\begin{equation}
    q_a = c_{a+} + c_{a-}=c_{a+}+c_{-a+}^\ast,
    \label{eq:q_vs_c}
\end{equation} 
where in the subscripts, we still use $a$ to indicate $m_a$ for the f-modes, and let $a=r$ for the radial mode. The $\pm$  following $a$ stands for the signs of the modes' frequencies and will be denoted by $s_a$. The phase-space description doubles the number of modes (the sign of the eigenfrequency, $s_a$, is now a new quantum number), with each mode's evolution described by a first-order differential equation. The mode amplitudes satisfy $c_{a,s_a} = c_{-a,-s_a}^\ast$ (note that for the radial model, $a=-a=r$, as it has $m_r=0$). After transferring to an inertial frame by applying an $e^{im_a\Omega t}$ phase shift, the equation of motion of a mode reads
\begin{equation}
    \dot{c}_{a+} + i (\omega_{a} + m_a \Omega_A)c_{a+}  + i\sigma_a c_{a-} =  i \omega_{a0} K_a, 
    \label{eq:EOM_phase}
\end{equation}
where $K_a$ includes the tidal, centrifugal, and nonlinear drive of the oscillator, whose explicit form will be provided later. 
The Coriolis force has two effects (see Appendix \ref{appx:Coriolis}). First, it modifies the mode frequency in the corotating frame as 
\begin{equation}
    \omega_a = \omega_{a0} + m_aC_a \Omega_A +\frac{m_a^2 C_a^2}{2} \frac{\Omega_A^2}{\omega_{a0}},
    \label{eq:freq_shift_coriolis}
\end{equation}
The structural constant $C_a$ is defined through
\begin{equation}
    C_a=-i\frac{\int dM \vxi_a^\ast\cdot (\vect{\Omega}_A \times \vxi_a)}{m_a\Omega_A\int dM \vxi^\ast_a \cdot \vxi_a} 
    \label{eq:Ca}
\end{equation}
In the Newtonian affine approximation, $C_f=-1/2$ for the $l_a=2$ f-modes, a decent approximation for more realistic models with $n\leq 1.5$ \cite{Lai:21}. Second, it couples the $c_{a\pm}$ modes via the $\sigma_a$ term, given by
\begin{equation}
    \sigma_a = \frac{m_a^2 C_a^2}{2}\frac{\Omega_A^2}{\omega_{a0}}.
    \label{eq:sigma_a_coriolis}
\end{equation}
To complete the mapping between configuration space and phase space, we also have 
\begin{equation}
    \dot{q}_a^{(\rm in)} \equiv \frac{d}{dt}(q_a e^{-im_a\Omega_At})
    =- i [\omega_{a0}+m_a(1+C_a)\Omega_A] c_{a+} + i[\omega_{a0} -m_a(1+C_a)\Omega_A] c_{-a+}^\ast.
    \label{eq:dq_vs_c}
\end{equation}
We remind the readers that $q_a$ stands for the configuration-space mode amplitude in the frame corotating with the NS, while the corresponding inertial frame amplitude is $q_a^{(\rm in)}$. The phase-space mode amplitudes $c_{a, s_a}$ will be evaluated in the inertial frame. Later, we will also introduce $b_{a, s_a}=c_{a, s_a}e^{i m_a \phi}$ as the phase-space mode amplitude corotating with the orbit.

The inertial frame equation of motion of Eq. (\ref{eq:EOM_phase}) can be derived from a non-PP Hamiltonian $H_{\rm npp}$, 
\begin{equation}
    H_{\rm npp}= H_{\rm mode} + H_{\rm tide} + H_{\rm cen}, 
    \label{eq:H_npp}
\end{equation}
describing respectively the internal fluid motion, the fluid's coupling to the tidal field, and the coupling to the centrifugal potential. The terms can be schematically written as \cite{Yu:23a, Yu:25a}
\begin{align}
    \frac{H_{\rm mode}}{E_A} =& \underbrace{\frac{1}{2} \sum 
    c_{a, s_a}c_{a,s_a}^\ast}_{\text{linear oscillator}} 
    + \underbrace{\frac{\Omega_A}{2} \sum \frac{\mathcal{S}_{a,s_a}}{E_A}}_{\text{frame transformation}}
    + \underbrace{\frac{C_f\Omega_A}{2}\sum \frac{\mathcal{S}_{a,s_a}}{E_A} +\frac{C_f^2\Omega_A^2}{8}\sum m_a^2(c_{a,s_a}+c_{a, -s_a})(c_{-a,s_a}+c_{-a, -s_a})}_{\text{Coriolis}} \nonumber \\
    &\underbrace{- \frac{1}{3}\sum  \kappa_{abc} c_{a, s_a}^\ast c_{b, s_b}^\ast c_{c, s_c}^\ast}_{\text{three-wave corrections}} 
    \underbrace{- \frac{1}{4}\sum\zeta_{abcd} c_{a, s_a}^\ast c_{b,s_b}^\ast c_{c, s_c}^\ast c_{d,s_d}^\ast, }_{\text{four-wave corrections}}
    \label{eq:H_mode}
    \\
    \frac{H_{\rm tide}}{E_A}=& \underbrace{- \sum U_a c_{a,s_a}^\ast}_{\text{linear tide }} \underbrace{- \frac{1}{2}\sum U_{a,bc}^\ast c_{b,s_b}^\ast c_{c,s_c}^\ast}_{\text{three-wave corrections}}, 
    \label{eq:H_tide}
    \\
    \frac{H_{\rm cen}}{E_A}=&\underbrace{-\sum R_a c_{a, s_a}^\ast}_{\text{linear tide}} \underbrace{- \frac{1}{2}\sum R_{a,bc} c_{b, s_b}^\ast c_{c,s_c}^\ast}_{\text{three-wave corrections}}, 
    \label{eq:H_rot}
\end{align}
where the summation over the modes runs over all the modes (8 in total, including 4 different angular harmonics indicated by $a, b, c,...$ and 2 signs of eigenfrequencies indicated by $s_a, s_b, s_c, ...$). When referring to a driving potential (the $a$ index in $U_a$, $R_a$, $U_{a,bc}$, and $R_{a,bc}$), it runs over $\{2, 0, -2, r\}$ with $r$ standing for $l=m=0$. This is because the tidal potential can be decomposed into harmonics with angular quantum numbers $l=2$ and $m=(2, 0, -2)$, assuming the companion is in the equatorial plane, and the centrifugal potential has both $l=m=0$ and $l=2, m=0$. 
For the mode degrees of freedom, the canonical displacements are the positive-frequency mode amplitudes $c_{a+}$, while the canonical momenta are $i E_Ac_{-a-}/\omega_{a0} = i E_Ac_{a+}^\ast/\omega_{a0}$. In other words, we have the Poisson bracket $\{c_{a+}, i(E_{A}/\omega_{a0}) c_{b+}^\ast\}=\delta_{ab}$.
The Hamiltonian is truncated at the four-wave (NNLO) level to $\epsilon_A^4$, with each term's order counting indicated by the under braces, counting
\begin{equation}
    \epsilon_A=\left(\frac{M_B}{M_A}\right) \left(\frac{R_A}{r}\right)^3 \sim |c_{a, s_a}| \sim \omega^2 /\omega_A^2 \sim \Omega_A^2/\omega_A^2
    \label{eq:order_counting}
\end{equation} as the same order, with $r$ the orbital separation. We further denote the orbital phase as $\phi$, and $\omega=\dot{\phi}$ the orbital frequency. Since the quadrupolar tidal potential and the centrifugal potential both scale as $\bar{x}_r^2$, they have no contributions beyond NLO. 

More specifically, in $H_{\rm mode}$, the six terms correspond to the linear harmonic oscillator in a non-spinning background (what defines the modes), a ``frame-dragging'' term arising from transforming reference frames (from corotating with the NS to an inertial frame; see sec. II B of \cite{Yu:25a}), 
the linear and quadratic in $\Omega_A$ corrections from the Coriolis effect, 
and the three- and four-wave interactions whose coupling coefficients ($\kappa_{abc}$ and $\zeta_{abcd}$) will be described later. 
We introduce
\begin{equation}
    \mathcal{S}_{a,s_a}= \frac{m_a}{s_a\omega_{a0}} c_{a,s_a} c_{a,s_a}^\ast E_A
    \label{eq:tidal_spin}
\end{equation}
as the canonical spin carried by each mode \cite{Friedman:78}.
Note that in the order counting scheme defined by Eq. (\ref{eq:order_counting}), the frame-dragging and the Coriolis corrections are of higher order than $\epsilon_A^2$. Nonetheless, these terms are only quadratic in the mode amplitude, and we will still group them into the linear tide. Previous analyses \cite{Ma:20, Steinhoff:21, Yu:24a, Yu:25a} of linear tide have captured these effects (except for the $\Omega_A^2$ Coriolis term). 

For the tidal interaction Hamiltonian, the first term describes the coupling between the tidal field and the unperturbed background, while the second term accounts for corrections due to perturbations of the background. For the linear and nonlinear driving potentials, we write 
\begin{align}
    &U_a e^{i m_a \phi} = V_{m_a} =W_a \mathcal{I}_a \epsilon_A, \\
    &U_{a,bc}e^{im_a\phi} = V_{a,bc} = W_a \mathcal{J}_{a,bc} \epsilon_A,
\end{align}
where $W_a=4\pi Y_{l_a m_a}(\pi/2, 0)/(2l_a+1)$ is a constant related to the decomposition of the tidal field whose non-zero components are $W_{\pm2}=\sqrt{3\pi/10}$, $W_{0}=-\sqrt{\pi/5}$, and $W_r=2\sqrt{\pi}$, and $\mathcal{I}_a=\If$ for the $l_a=2$ f-modes the tidal overlap integral, defined as \cite{Weinberg:12}
\begin{equation}
    \If =\frac{\int dM_A \vxi_m \cdot \nabla(\bar{x}_r^l Y_{lm}^\ast)}{M_A R_A^l}\Bigg{|}_{l=2}
    = \sqrt{\frac{15 \Ma}{4\pi}}\frac{\omega_A}{\omega_{f0}}.
    \label{eq:I_f}
\end{equation}
Its value is independent of $m_a$. 
The second equality follows from the normalization condition, Eq. (\ref{eq:normalization}). The linear tidal overlap is also related to the tidal deformability of an NS $\lambda_A$ (or its Love number $k_{2A}$) through \cite{Yu:23a}
\begin{equation}
    \lambda_{A}=\frac{2}{3}k_{2A} R_A^5= \frac{8\pi}{15} \If^2 R_A^5
    \label{eq:lamA_vs_k2A_vs_If}
\end{equation}
The nonlinear tidal coupling coefficient $\mathcal{J}_{a, bc}$ is defined through \cite{Weinberg:12} (for $l_a>0$)
\begin{equation}
    \JJ_{a,bc} \equiv \frac{\int dM \xi_b^i\xi_c^j \partial_i\partial_j (\bar{x}_r^{l_a} Y_{l_a m_a}) }{M_A R_A^2}.
\end{equation}
The coupling with the centrifugal potential is nearly identical to the tidal coupling, with 
\begin{align}
    &R_{a} = \frac{2}{3}\frac{\Omega_A^2}{\omega_A^2}W_{a}\mathcal{I}_a, \text{ and }
    R_{a, bc} = \frac{2}{3}\frac{\Omega_A^2}{\omega_A^2} W_{a} \mathcal{J}_{a,bc}.
\end{align}
To match the energy, we define for the $l_a=0$ ($a=r$ in the subscript) case
\begin{equation}
    \mathcal{I}_r = \sqrt{\frac{3\Ma}{8\pi}} \frac{\omega_A}{\omega_r} = \sqrt{\frac{1}{10}} \frac{\omega_{f0}}{\omega_r} \If.
    \label{eq:I_r}
\end{equation}
The value of $J_{r,bc}$ will be introduced later in Eq. (\ref{eq:J_r2}).

The similarity between the two is expected, as they can be respectively written as $H_{\rm tide} = E_{ij} Q_{ij}/2$ and $H_{\rm cen} =  R_{ij}Q_{ij}/2$, where 
\begin{equation}
    E_{ij} = -\frac{M_B}{r^3}
    \begin{bmatrix}
    \frac{1 + 3\cos 2\phi}{2}, & 3 \sin\phi\cos\phi, & 0\\
    3\sin\phi \cos\phi, & \frac{1-3\cos 2\phi}{2}, & 0\\
    0, & 0& -1
    \end{bmatrix}, 
    \text{ and }
    R_{ij} = -\Omega_A^2\begin{bmatrix}
        1 & 0 & 0\\
        0 & 1 & 0\\
        0 & 0 &0
    \end{bmatrix}.
\end{equation}
We also introduce the mass quadrupole
\begin{equation}
    Q_{ij} = \int x_i x_j dM = q_{ik} q_{jk} \Ma M_A R_A^2,
\end{equation}
as well as its spherical harmonic counterparts (for $l>0$)
\begin{equation}
    Q_{lm} = \mathcal{Y}_{lm}^{ij}Q_{ij} = M_AR_A^2\left[\sum_a^{l_a=l,\ m_a=m}\left( \II_a c_a +\frac{1}{2}\sum_{bc}\mathcal{J}_{a, bc}c_b^\ast c_c^\ast\right) \right], 
    \label{eq:Q_lm}
\end{equation}
where $\mathcal{Y}_{lm}^{ij}$ are components of the symmetric trace-free tensor defined through \cite{Poisson:14}
\begin{equation}
    \mathcal{Y}_{lm}^{ij} = \frac{15}{8\pi}\int n^i n^j Y^\ast_{lm} (\theta, \phi) \sin\theta d\theta d\phi,
\end{equation}
with $\boldsymbol{n}=(\sin\theta \cos\phi, \sin\theta\sin\phi, \cos\theta)$ in a Cartesian basis. We similarly define $E_{lm} = \mathcal{Y}_{lm}^{ij} E_{ij}$.

The nonlinear coupling coefficients $\kappa_{abc}$ and $\zeta_{abcd}$ can be derived by matching the nonlinear potentials in $H_{\rm mode}$ to those derived in \cite{Yu:26}.  
In particular, the affine model describes a perturbed ellipsoid whose energy can be analytically computed \cite{Chandrasekhar:87, Carter:83, Carter:85, Yu:26}.
The kinetic energy is 
\begin{equation}
    \frac{E_{\rm kinetic}}{E_A} = \frac{1}{2} \Ma \dot{q}_{ij}\dot{q}_{ij}.
\end{equation}
The potential energy is the sum of internal and gravitational energies. 
The internal energy is given by 
\begin{equation}
    \frac{E_{\rm internal}}{E_A} = \frac{(\Gamma - 1)({\rm det}[q_{ij}])^{1-\Gamma_{\rm ad}}}{(\Gamma_{\rm ad} - 1)(5\Gamma-6)},
    \label{eq:E_internal_affine}
\end{equation}
and the gravitational binding energy is 
\begin{equation}
    \frac{E_{\rm gravitational}}{E_A}=-\frac{3}{2}\frac{\Gamma-1}{5\Gamma-6} \sum_{i=1}^3 A_{i} a_i^2,
    \label{eq:E_grav_affine}
\end{equation}
where we have made the summation over $i$ explicit. We denote the semi-axes of the ellipsoid by $a_i$, with 
\begin{subequations}
\begin{align}
    a_1 &= 1 + \frac{\omega_A}{\omega_{f0}}\frac{2|q_2| - \sqrt{2/3}q_0 }{2\sqrt{2\Ma} }  
    + \frac{\omega_A}{\omega_{r}}\frac{q_r}{\sqrt{6\Ma}}, \\
    a_2 & = 1 - \frac{\omega_A}{\omega_{f0}}\frac{2|q_2| + \sqrt{2/3}q_0 }{2\sqrt{2\Ma} }  
    + \frac{\omega_A}{\omega_{r}}\frac{q_r}{\sqrt{6\Ma}}, \\
    a_3 & = 1+\frac{\omega_A}{\omega_{f0}}\frac{q_0}{\sqrt{3\Ma} \omega_{f0}} + \frac{\omega_A}{\omega_r}\frac{q_r}{\sqrt{6\Ma}},
\end{align}
\label{eq:semiaxes}
\end{subequations}
and 
\begin{subequations}
\begin{align}
    &A_1 = \frac{2}{a_1^3 \sin^3\phi_e} \frac{1}{\sin^2\theta_e}(F_e-E_e), \\
    &A_2 = \frac{2}{a_1^3 \sin^3\phi_e} \frac{1}{\cos^2\theta_e}(\frac{a_3}{a_2}\sin\phi_e - E_e), \\
    &A_3=\frac{2}{a_1^3 \sin^3\phi_e} \frac{1}{\sin^2\theta_e\cos^2\theta_e}(E_e-F_e \cos^2\theta_e - \frac{a_2}{a_3}\sin^2\theta_e\sin\phi_e), 
\end{align}
\end{subequations}
with 
\begin{equation}
    \sin^2\theta_e=\frac{a_1^2-a_3^2}{a_1^2-a_2^2}, 
    \text{ and } \cos\phi_e = \frac{a_2}{a_1}. 
\end{equation}
The incomplete elliptic integrals of the first and second kinds are denoted by
\begin{align*}
    &F_e=F(\phi_e, m=\sin^2\theta_e) = \int^{\phi_e}_0 (1-m \sin^2 \varphi)^{-1/2} d\varphi, \\
    &E_e=E(\phi_e, m=\sin^2\theta_e) = \int^{\phi_e}_0 (1-m \sin^2 \varphi)^{+1/2} d\varphi.
\end{align*}
Expanding these potentials to third and fourth orders of $c_a$ then gives the coupling coefficients $\kappa_{abc}$ and $\zeta_{abcd}$ in Eq. (\ref{eq:H_mode}). 

\subsection{Theoretical universal relations}
\label{sec:UR_nl_tide_N}

Numerous universal relations between different NS properties have been observed (see, e.g., \cite{Yagi:17} for a review), and multiple authors have examined the theoretical origin behind those relations \cite{Chan:14, Yagi:14, Sham:15}, which mainly relies on the fact that fluid is nearly incompressible (hence the affine approximation works well). 
We present derivations of some well-known universal relations and provide an intuitive argument for their existence based on counting the number of free parameters of the affine model. Adding to the literature, our formulation reveals new universal relations connecting the nonlinear tidal properties to the linear ones. 

In the affine approximation of \cite{Carter:83, Carter:85}, various energies (hence the fluid dynamics) are fully specified by only three parameters, $(\Ma, \Gamma, \Gamma_{\rm ad})$. They can be viewed as representing three degrees of freedom in the problem: a theory of gravity governing the hydrostatic equilibrium, an equation of state governing the equilibrium pressure-density relation, and the $P-\rho$ relation in an ``adiabatic'' (which can also mean constant composition in an NS) process. 
For a fixed theory of gravity (e.g., Newtonian or GR), we further have $\Ma = \Ma(\Gamma)$ despite the lack of a closed-form expression in general. In other words, the presence of various universal relations in NSs is not a surprise; they are natural consequences of the fact that there are only two intrinsic parameters, ($\Gamma, \Gamma_{\rm ad}$), describing a large-scale perturbation. 
Practically, we will use the Love number $k_{2A}$ (or the deformability $\lambda_A$) instead of $\Gamma$ to represent the first parameter. All properties related to the linear tide (including the quadrupole-monopole interaction from spin-induced quadrupole), as well as three-wave (NLO) interactions in a slowly spinning NS, are determined by it. The other parameter, $\Gamma_{\rm ad}$ (or equivalently, $\omega_r$; Eq. \ref{eq:omega_a}), can only be probed via the nonlinear interaction between f- and radial modes at NLO  in a rapidly spinning NS, or the four-wave (NNLO) interactions when the background is non-spinning. 

For example, we first demonstrate the appearance of the $I_A-\lambda_A$ relation \cite{Yagi:13} by eliminating $\omega_{f0}$ from $\If$ in Eq. (\ref{eq:I_f}) using Eq. (\ref{eq:omega_a}), which leads to 
\begin{equation}
    \Ma^2= \frac{4(\Gamma-1)}{15(5\Gamma-6)}k_{2A}. 
    \label{eq:k2_vs_Ma}
\end{equation}
This further translates to a relation between the tidal deformability and moment of inertia
\begin{equation}
    \left(\frac{I_A}{M_A R_A^2}\right)^2=\frac{8(\Gamma-1)}{5(5\Gamma-6)}\left(\frac{\lambda_A}{R_A^5}\right).
    \label{eq:I_A_vs_lam_A}
\end{equation}
This already shows a quasi-universal relation $I_A\propto \lambda_A^{1/2}$, as when $\Gamma$ varies from 2 to 3 ($n$ from 1 to 0.5, which are typical values for NSs), the $8(\Gamma-1)/[5(5\Gamma - 6)]$ pre-factor changes only slightly from 0.4 to 0.36. Furthermore, $k_{2A}$ and $\Gamma$ are not independent from each other, but instead with 
\begin{equation}
    \Gamma \simeq 2.0\left(\frac{k_{2A}}{0.26}\right)^{0.73}
    \label{eq:Gamma_vs_k2}
\end{equation}
in Newtonian gravity around values typical for an NS \cite{Poisson:14}. We can use this relation to to expand Eq. (\ref{eq:k2_vs_Ma}) around $k_{2A}=0.26$ or $\Gamma=2$, as
\begin{align}
    \Ma^2 \simeq \left(0.067 - 0.038 k_{2A}^{0.73} + 0.010 k_{2A}^{-0.27}\right)k_{2A}.
\end{align}

Closely related to the above relation, a coefficient $\mathcal{Q}_A$ governing the spin-induced quadrupole moment will be universally related to $k_{2A}$.
We write the spin-induced quadrupole moment scalar, as defined in \cite{Poisson:98}, as
\begin{align}
    -\mathcal{Q}_A \frac{I_A\Omega_A^2}{M_A^3R_A} M_A R_A^2 = 2\sqrt{\frac{\pi}{5}}\int x_ix_i Y_{20} dM_A= 2\sqrt{\frac{\pi}{5}}Q_{20}.
\end{align}
Plugging in the leading-order expression of $Q_{20}$ from Eq. (\ref{eq:Q_lm}) and the leading order solution of $c_0$ driven by the centrifugal potential (see later in Eq. \ref{eq:c0_sol_lin}), we have 
\begin{equation}
    \mathcal{Q}_A=\left(\frac{\lambda_A}{R_A^5}\right)\left(\frac{I_A}{M_A R_A^2}\right)^{-2}\simeq 
    (0.40 - 0.27k_{2A}^{0.73} +  0.06k_{2A}^{-0.27}).
    \label{eq:Q_A_vs_Lam_A_I_A}
\end{equation}
The second equality follows directly from Eqs. (\ref{eq:I_A_vs_lam_A}) and (\ref{eq:Gamma_vs_k2}).  

Similarly, one may use Eqs. (\ref{eq:omega_a}) and (\ref{eq:I_f}) to eliminate $\Ma \propto I_A$, leading to 
\begin{equation}
    \left(\frac{\omega_{f0}}{\omega_A}\right)^{-4} = \frac{5(5\Gamma-6)}{12(\Gamma-1)} k_{2A} 
    = \frac{5(5\Gamma-6)}{8(\Gamma-1)} \left(\frac{\lambda_A}{R_A^5}\right)
    \label{eq:omega_f_vs_k2}
\end{equation}
This reveals the quasi-universal relation between $\lambda_A$ and $\omega_{f0}$. Again, we can expand the right-hand side around $k_{2A}=0.26$ to get
\begin{equation}
    \left(\frac{\omega_{f0}}{\omega_A}\right)^{-4} \simeq (1.7 + 0.76 k_{2A}^{0.73} - 0.20k_{2A}^{-0.27}) k_{2A}. 
    \label{eq:omega_f_vs_k2_apprx}
\end{equation}
Interestingly, one may form an especially clean $I_A-\lambda_A-\omega_{f0}$ relation that eliminates the explicit appearance of $\Gamma$, 
\begin{equation}
    \left(\frac{\lambda_A}{R_A^5}\right)=\left(\frac{I_A}{M_AR_A^2} \right)\frac{\omega_A^2}{\omega_{f0}^2}.
\end{equation}
A similar result has been obtained by \cite{Chan:14}.   

Besides the known relations, we can also derive new ones for the nonlinear tide. 
Notably, the three-wave coupling coefficients among the f-modes can be expressed as functions of $\If$ (or equivalently, $k_{2A}$ or $\lambda_A/R_A^5$) and $\omega_{f0}^2/\omega_A^2$ (which itself is a function of $k_{2A}$ by Eq. \ref{eq:omega_f_vs_k2_apprx}) only, 
\begin{align}
    \kappa_2 & \equiv\kappa_{2-20} 
    = - \frac{19}{112}\sqrt{\frac{5}{\pi}} \frac{\omega_A^2}{\omega_{f0}^2 \If}
    \simeq -0.4380,
    \label{eq:kap2_vs_I_w}
    \\
    \mathcal{J}_2&\equiv \mathcal{J}_{2,-20}
    =-\frac{1}{4}\sqrt{\frac{5}{\pi}}\frac{\omega_A^2}{\omega_{f0}^2} \simeq -0.2061,
    \label{eq:J2_vs_I_w}\\
    \kappa_0&\equiv\kappa_{000}=-\kappa_2, \text{ and }\JJ_0\equiv\JJ_{0, 00}=-\JJ_2. 
\end{align}
The numerical values are for a Newtonian $\Gamma=1+1/n=2$ polytrope, agreeing well with the numerical calculations found in \cite{Yu:23a}.
The coupling is invariant under permutations of the subscripts. 
Since $\JJ_2$ and $\kappa_2$ are what determine the three-wave tidal interactions in a non-spinning NS \cite{Yu:23a}, the NLO tide does not provide new information about NSs under the affine approximation, though not including them will lead to large errors in the waveform (see later in Sec. \ref{sec:GW_phase} and \cite{Bretz:26}). 

New physics ($\Gamma_{\rm ad}$, or more conveniently, $\omega_r^2$, the eigenfrequency squared of the radial mode; Eq. \eqref{eq:omega_a}) is probed starting at the four-wave (NNLO) level for a slowly spinning NS.  
We have
\begin{align}
    &\zeta_{22}\equiv\zeta_{22-2-2} = -\frac{5(67\omega_{f0}^2 + 14 \omega_r^2)\omega_A^4}{672 \pi \If^2 \omega_{f0}^6}
    = -\frac{(67\omega_{f0}^2 + 14 \omega_r^2)\omega_A^2}{504 \Ma \omega_{f0}^4},\\
    &\zeta_{20} \equiv\zeta_{2-200} = \frac{1}{2}\zeta_{22}, \\
    &\zeta_{00} \equiv \zeta_{0000} = \frac{3}{2}\zeta_{22}.
\end{align}
Again, $\zeta_{abcd}$ are invariant under permutations of subscripts. 

When spin is significant, there will be an expansion of the star (of fixed baryonic mass) in the equatorial plane due to the centrifugal force that can be decomposed into two harmonics, one with $l=m=0$ and the other with $(l,m)=(2,0)$. The $l=0$ component of the expansion is captured by the radial mode in our model, which also probes $\omega_r$ (Eq. \ref{eq:I_r}). Such a mode does not directly couple to the external tidal field, so its impact on the orbit relies on its coupling to the $l_a=2$ f-modes starting at NLO. The relevant coefficients
\begin{align}
    &\kappa_{2r}\equiv\kappa_{r2-2}=\kappa_{r00} = \sqrt{\frac{5}{128\pi}} \frac{1}{\If}\left(\frac{3\omega_A^2}{\omega_{f0}\omega_r} + \frac{\omega_A^2\omega_r}{\omega_{f0}^3}\right), \\
    &\kappa_{rr} \equiv \kappa_{rrr}=\frac{1}{8\sqrt{10\pi}}\frac{1}{\If} \left(25\frac{\omega_A^2}{\omega_{f0}\omega_r} + 4\frac{\omega_A^2\omega_r}{\omega_{f0}^3}\right), \\
    &\zeta_{20r}\equiv\zeta_{2-20r}=-\zeta_{000r}=\frac{5(38\omega_{f0}^2+7\omega_r^2)\omega_A^4}{336\sqrt{2}\pi\If^2\omega_{f0}^5\omega_r}, \\
    &\zeta_{2r}\equiv\zeta_{2-2rr}=-\frac{(60\omega_{f0}^4 + 35\omega_{f0}^2\omega_r^2 + 4\omega_r^4)\omega_A^4}{96\pi\If^2\omega_{f0}^6\omega_r^2},\\
    &\zeta_{0r}\equiv\zeta_{00rr} = \zeta_{2r}. 
\end{align}

While most of the coefficients we use are invariant under permutations of the indices, the only exception is when a radial perturbation is involved in $\mathcal{J}_{a, bc}$ (i.e., when the centrifugal potential is involved), in which case the result is invariant only for permutations of the last two indices after the comma. To match the form of $H_{\rm tide}$ and $H_{\rm cen}$, we have
\begin{align}
    &\JJ_{2r}\equiv\mathcal{J}_{2,r-2}=\mathcal{J}_{0,r0}=\sqrt{\frac{5}{8\pi}} \frac{\omega_A^2}{\omega_{f0}\omega_{r}},\\
    &\JJ_{r2}\equiv\mathcal{J}_{r, 2-2} = \mathcal{J}_{r, 00} = \frac{1}{4\sqrt{\pi}} \frac{\omega_A^2}{\omega_{f0}^2}, 
    \label{eq:J_r2}
    \\
    &\JJ_{rr}\equiv\mathcal{J}_{r,rr} = \frac{1}{4\sqrt{\pi}} \frac{\omega_A^2}{\omega_r^2}.
\end{align}

\subsection{Extension to relativistic NSs}
\label{sec:UR_in_GR}

As shown in \cite{Yu:23a}, the dominant nonlinear tide in a non-spinning NS is specified by $(\If, \omega_{f0}, \JJ_{2}, \kappa_2)$. We will show later in Sec. \ref{sec:effective_Love} and \ref{sec:vs_PP25} that the coefficients used in this study map to those used in \cite{Pitre:24, Pitre:25} via
\begin{align}
    \ddot{k}_{2A}&= k_{2A} \frac{\omega_A^2}{\omega_{f0}^2} = \frac{4\pi}{5}\If^2\frac{\omega_A^2}{\omega_{f0}^2} \simeq 0.167, 
    \label{eq:ddk2_vs_k2_wf}\\
    p_{2A} &=-\frac{8 \pi}{5} \sqrt{\frac{\pi}{5}} \If^2(3 \mathcal{J}_2 + 4 \kappa_2 \If)
    = \frac{20}{7} k_{2A} \frac{\omega_A^2}{\omega_f^2}\simeq0.478,
    \label{eq:p2_vs_J2_kap2}
\end{align}
where our $(k_{2A}, \ddot{k}_{2A}, p_{2A})$ have the same meaning as $(k_2, \ddot{k}_2, p_2)$ in \cite{Pitre:25}, and the numerical values are for a Newtonian $\Gamma=1+1/n=2$ polytrope under the affine approximation. They agree well with table 1 of \cite{Pitre:25} in the low-compactness limit. 

As argued in Sec. \ref{sec:UR_nl_tide_N}, $(k_{2A}, \ddot{k}_{2A}, p_{2A})$ are not independent coefficients. Instead, they are all governed by a single underlying NS parameter, $\Gamma$, in the Newtonian limit. Once $k_{2A}$ is given, the other two can be determined. 
First, consider $\ddot{k}_{2A}$. From Eq. (\ref{eq:omega_f_vs_k2}), we have
\begin{equation}
    \ddot{k}_{2A} = \sqrt{\frac{5(5\Gamma-6)}{12(\Gamma - 1)}} k_{2A}^{3/2}
    \simeq \sqrt{\left(1.7 +0.76 k_{2A}^{0.73} - 0.20 k_{2A}^{-0.27}\right) k_{2A}^3 },
    \label{eq:ddk2_vs_k2_N}
\end{equation}
where the first equality is exact under the affine approximation and the second one from expanding $\Gamma(k_{2A})$. 

For $p_{2A}$ that sets the three-wave nonlinear tide's strength in the low-frequency limit (see later in Sec. \ref{sec:effective_Love} and \ref{sec:GW_phase}), its value is directly set by $\ddot{k}_{2A}$ via a simple relation:
\begin{equation}
    p_{2A} =  \frac{20}{7}\ddot{k}_{2A}. 
    \label{eq:p2_vs_ddk2_N}
\end{equation}
One can further check that this relation is well satisfied by all polytropic models reported in \cite{Pitre:25} under the Newtonian limit, hence indicating a universal relation. 

\begin{figure}
    \centering
    \includegraphics[width=0.95\linewidth]{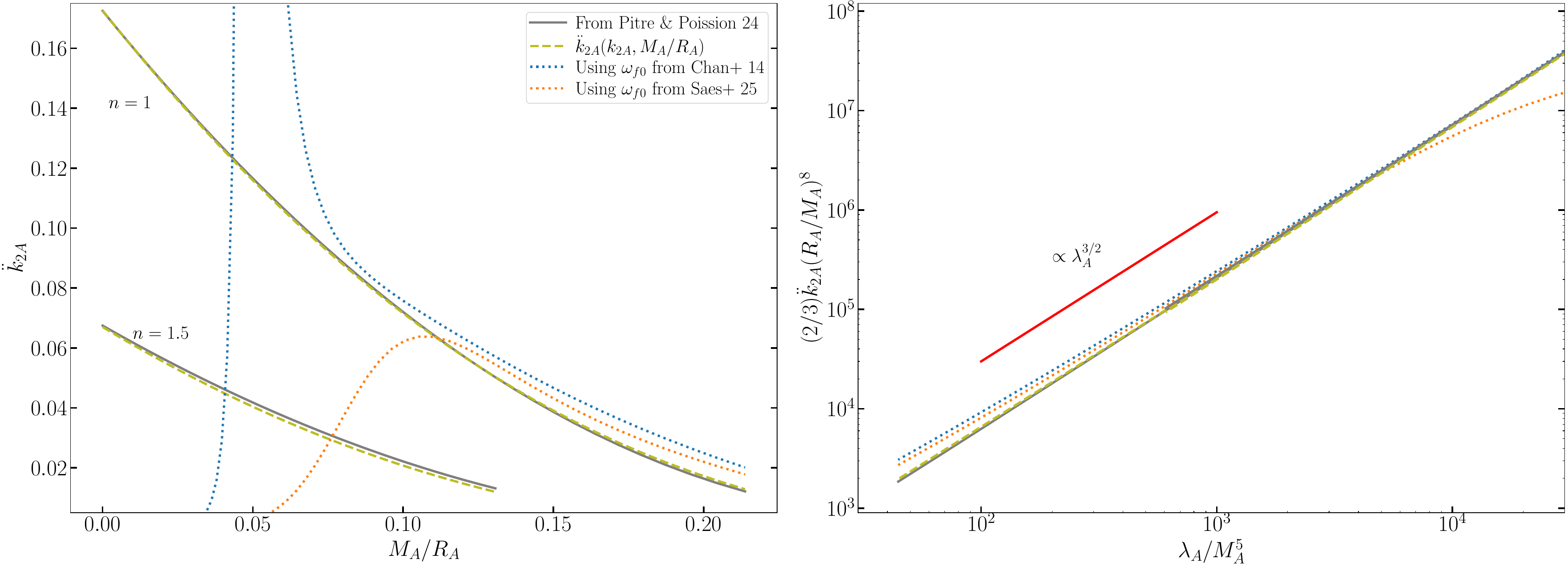}
    \caption{Comparison of the calculated $\ddot{k}_{2A}$ (shown in gray) with prediction from the universal relation (shown in yellow). 
    On the left, we show $\ddot{k}_{2A}$ as a function of the compactness $(M_A/R_A)$. 
    Unlike previous empirical relations (\cite{Saes:25} in orange-dotted, \cite{Chan:14} in blue-dotted) that have limited range of validity, our theoretically informed relation holds for the entire range of compactness. 
    There are theoretical uncertainties in the relativistic regime, leading to differences among predictions in the $M_A/R_A>0.1$ range. 
    On the right, we show $\bar{\lambda}_{A}^{(2)}$ as a function of $\bar{\lambda}_A$. 
    The approximate $\propto \lambda_A^{3/2}$ scaling can be analytically argued from Eqs. (\ref{eq:ddk2_vs_k2_wf}) and (\ref{eq:lambar2_vs_lambar}). 
    }
    \label{fig:ddk2_vs_k2}
\end{figure}

\begin{figure}
    \centering
    \includegraphics[width=0.95\linewidth]{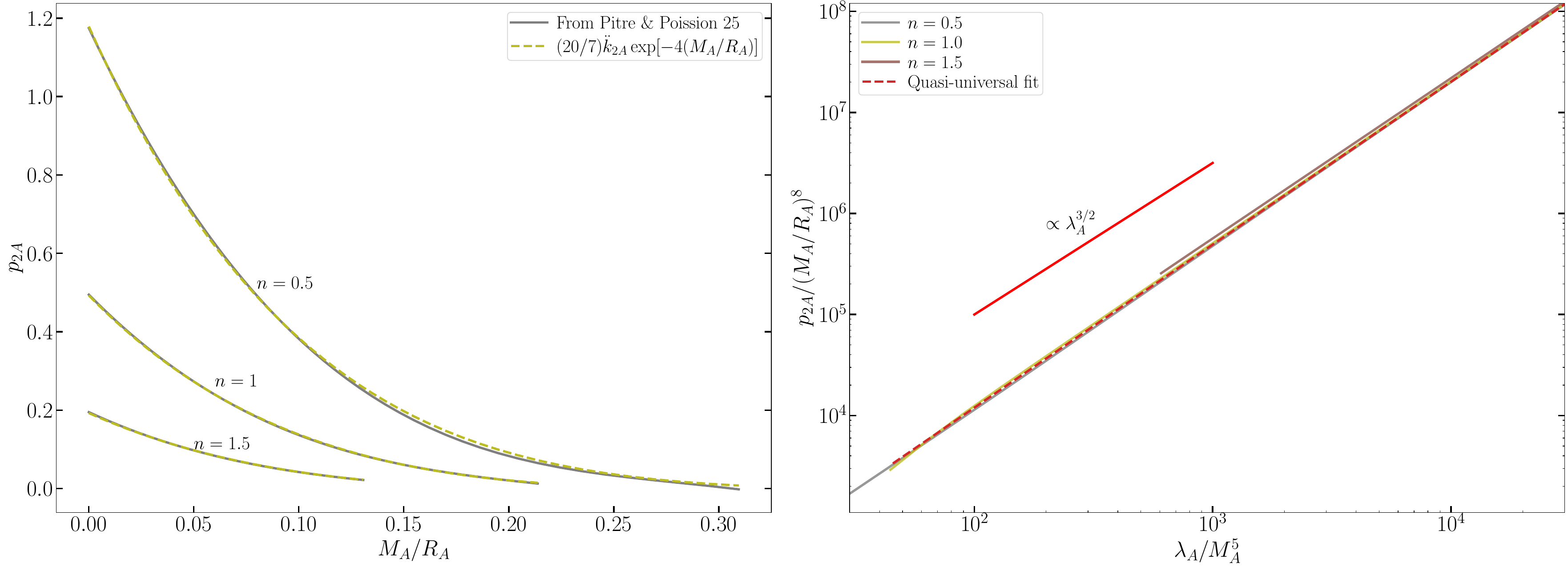}
    \caption{Left: Comparison of $p_{2A}$ obtained from \cite{Pitre:25} (solid-gray) and those predicted by the universal relations (Eq. \ref{eq:p2_vs_ddk2_GR}; dashed-yellow). 
    Right: Comparison of the $\bar{p}_2-\bar{\lambda}_A$ quasi-universal fit, Eq. (\ref{eq:p2bar_vs_lambar}) with tracks computed for different polytropes. 
    }
    \label{fig:p2_vs_ddk2}
\end{figure}

When GR is taken into account, the compactness of an NS, $M_A/R_A$, introduces an additional parameter that is not constrained by the Newtonian affine model in \ref{sec:UR_nl_tide_N}. Nonetheless, using the values provided in \cite{Pitre:24} for $\ddot{k}_{2A}$ and in \cite{Pitre:25} for $p_{2A}$, we empirically find that through introducing a simple scaling of the form $\exp(\alpha M_A/R_A)$, with $\alpha$ some numerical constant, Eqs. (\ref{eq:ddk2_vs_k2_N}) and (\ref{eq:p2_vs_ddk2_N}) still holds in a relativistic star; see Figs. \ref{fig:ddk2_vs_k2} and \ref{fig:p2_vs_ddk2}. In other words, we seek to express $\ddot{k}_{2A} = \ddot{k}_{2A} [k_{2A}(\Gamma, M_A/R_A), M_A/R_A]$ and similarly for $p_{2A}$ to highlight their dependence on the internal NS parameters. The compactness enters only via an exponential dependence, which represents a smooth deformation of the theoretical results derived in Newtonian gravity. 

First, we find that by replacing $k_{2A}$ by $\tilde{k}_{2A} = k_{2A} \exp[2.4(M_A/R_A)]$, we can still decently estimate the value of $\ddot{k}_{2A}$ using the same form (cf. Eq. \ref{eq:ddk2_vs_k2_N})
\begin{equation}
    \ddot{k}_{2A} 
    \simeq \sqrt{\left(1.7 +0.76 \tilde{k}_{2A}^{0.73} - 0.20 \tilde{k}_{2A}^{-0.27}\right) \tilde{k}_{2A}^3 }.
    \label{eq:ddk2_vs_k2_GR}
\end{equation}
The result is shown in the left panel of Fig. \ref{fig:ddk2_vs_k2} for $n=1$ ($\Gamma=2$) and $n=1.5$ ($\Gamma=5/3$) polytropes. The gray line is from \cite{Pitre:24} and the yellow-dashed line from Eq. (\ref{eq:ddk2_vs_k2_GR}), treating $k_{2A}$ as a function of $M_A/R_A$. Good agreement is seen across the entire range of compactness and for different values of polytropic indices. 
Also shown are the empirical quasi-universal fits from \cite{Chan:14} and \cite{Saes:25}, obtained by first using the $k_{2A}-(M_A/R_A)$ relation for an $n=1$ polytrope to compute $\bar{\lambda}_A\equiv \lambda_A/M_A^5=(2/3) k_{2A} (R_A/M_A)^5$, then using it to fit for $\omega_{f0}$, and lastly reconstructing the value of $\ddot{k}_{2A}$ from Eq. (\ref{eq:ddk2_vs_k2_wf}). Therefore, they should only be compared with the $n=1$ curve. 
Due to the empirical nature of the fits of \cite{Chan:14, Saes:25}, they lose validity when $M_A/R_A\lesssim 0.1$ (i.e., outside their range of calibrations).
When $M_A/R_A\gtrsim 0.1$, there are currently some theoretical uncertainties on how the third PN term of the adiabatic tide is handled \cite{AbhishekHegade:24}, causing a $\sim 10\%$  difference in $\ddot{k}_{2A}$ between \cite{Saes:25} and \cite{Pitre:24} when plotted as a function of $\bar{\lambda}_A$. At a given $M_A/R_A$, the difference can be more significant for high compactness. 
In this work, we choose to use values from \cite{Pitre:24}, which provides the smallest $\ddot{k}_{2A}$ and hence the \emph{most conservative} estimation of the finite-frequency corrections. 
Given the theoretical uncertainty, we do not fine-tune Eqs. (\ref{eq:omega_f_vs_k2_apprx}) and (\ref{eq:ddk2_vs_k2_GR}).

We can  rewrite Eq. (\ref{eq:ddk2_vs_k2_GR}) as 
\begin{align}
    \bar{\lambda}^{(2)}_A\equiv\frac{2}{3}\frac{\ddot{k}_{2A}}{(M_A/R_A)^8}\simeq \sqrt{\frac{3}{2}}\left\{\sqrt{\frac{5(5\Gamma-6)}{12(\Gamma-1)}} \frac{\exp[3.6(M_A/R_A) ]}{\sqrt{(M_A/R_A)}} \right\} \bar{\lambda}_A^{3/2}. 
    \label{eq:lambar2_vs_lambar}
\end{align}
The term inside the curly bracket (for a fixed $\Gamma$) varies by only $4\%$ as the compactness varies from 0.15 to 0.2, values typical for an NS. It also depends weakly on the value of $\Gamma$ (less than 5\% variation from $\Gamma=2$ to $\Gamma=3$ while holding the compactness constant). The variation can be further reduced by expanding $\Gamma(\bar{\lambda}_A, M_A/R_A)$ around values typical for an NS, $\bar{\lambda}_A\sim 300$, $M_A/R_A\sim 0.17$, similar to what we have done for Eq. (\ref{eq:omega_f_vs_k2_apprx}). Therefore, Eq. (\ref{eq:ddk2_vs_k2_GR}) implies that when $\bar{\lambda}^{(2)}_A$ is plotted as a function of $\bar{\lambda}_A$, it will follow a quasi-universal line that approximately scales with $\bar{\lambda}_A^{3/2}$, as observed in \cite{Saes:25} and shown in the right panel of Fig. \ref{fig:ddk2_vs_k2}.

For the nonlinear tide, we empirically find that Eq. (\ref{eq:p2_vs_ddk2_N}) is modified as 
\begin{equation}
    p_{2A} =  \frac{20}{7}\ddot{k}_{2A} \exp[-4(M_A/R_A)].
    \label{eq:p2_vs_ddk2_GR}
\end{equation}
The left panel of Fig. \ref{fig:p2_vs_ddk2} shows a comparison between $p_{2A}$ directly from \cite{Pitre:25} (shown in gray-solid lines) and that predicted by Eq. (\ref{eq:p2_vs_ddk2_GR}) shown in yellow-dashed lines (for $n=1$ and $n=1.5$). For $n=0.5$, direct computation of $\ddot{k}_{2A}$ is challenging due to numerical instability \cite{Pitre:24}, and therefore we first estimate its value using Eq. (\ref{eq:ddk2_vs_k2_GR}) and then get $p_{2A}$ from Eq. (\ref{eq:p2_vs_ddk2_GR}). Since Eq. (\ref{eq:ddk2_vs_k2_GR}) is a crude approximation, the $n=0.5$ yellow-dashed curve shows a larger deviation from the gray one than the other two. Again, the theoretical nature of  Eq. (\ref{eq:p2_vs_ddk2_N}) ensures its validity holds from Newtonian stars to highly compact NSs, representing one of the main results of this work. The reason why a simple $\exp[-4(M_A/R_A)]$ is sufficient to provide such a high level of agreement with values from \cite{Pitre:25}, and if it stays when different boundary conditions are used (see Fig. \ref{fig:ddk2_vs_k2}),
remain topics for future theoretical investigations. 

From Eqs. (\ref{eq:lambar2_vs_lambar}) and (\ref{eq:p2_vs_ddk2_GR}), it is naturally expected that $\bar{p}_{2A}\equiv p_{2A}/(M_A/R_A)^8$, a combination that shows up in the GW waveform (Sec. \ref{sec:GW_phase}), follows a quasi-universal relation with $\bar{\lambda}_A$. This is shown in the right panel of Fig. \ref{fig:p2_vs_ddk2}. 
Using an average between $n=1$ and $n=0.5$ polytropes, we find that
\begin{equation}
    \log \bar{p}_{2A} = 1.933 + 1.622\log\bar{\lambda}_A -6.676\times10^{-4} (\log\bar{\lambda}_A)^2. 
    \label{eq:p2bar_vs_lambar}
\end{equation}
The fitting is performed over $100 < \lambda_A< 10^4$.
Note that the term quadratic in $\log\bar{\lambda}_A$ has an especially small coefficient, while the linear term has a coefficient close to 1.5, both are consistent with the theoretical expectations (as $p_{2A}\propto \ddot{k}_{2A}\propto k_{2A}^{3/2}$). The linear coefficient is slightly greater than 1.5 to absorb the compactness dependence in Eq. (\ref{eq:p2_vs_ddk2_GR}). Since $\bar{\lambda}_A = \bar{\lambda}_A (\Gamma, M_A/R_A)$, inverting $M_A/R_A$ as a function of $\bar{\lambda}_A$ can only be done at a fixed value of $\Gamma$, which is the reason why the $n=1.5$ curve shows small deviations from the fit. For realistic NSs, however, the effective $\Gamma$ varies only slightly around 2 so $\bar{\lambda}_A\simeq \bar{\lambda}_A(M_A/R_A)$, and one would expect that Eq. (\ref{eq:p2bar_vs_lambar}) to be well satisfied.


\subsection{Calibrations of coupling coefficients}
\label{sec:calibrations}

To address the significance of the nonlinear tide at three- and four-wave orders (NLO and NNLO), we will proceed using a hybrid scheme. We will solve a set of coupled differential equations generated by the Hamiltonian 
\begin{equation}
    H=H_{\rm npp} + H_{\rm pp} = H_{\rm npp} + \frac{\left(p_r^2 + p_\phi^2/r^2\right)}{2\mu} - \frac{\mu M_t}{r}
    \label{eq:full_Ham}
\end{equation}
with $H_{\rm pp}$ the PP orbital Hamiltonian described by canonical displacements $(r, \phi)$ and canonical momenta $(p_r, p_\phi)$, and the $H_{\rm npp}$ term is given by Eq. (\ref{eq:H_npp}). 
We define $M_t=M_A+M_B$, $\mu = M_A M_B/M_t$, and $\eta=\mu/M_t$ as usual. 
The system evolves due to an energy loss given by the quadrupole formula (see later in Eq. \ref{eq:dot_E_sys}).  
This provides a first-principles description of hydrodynamics under Newtonian theory, capturing dynamical effects at arbitrary frequencies, even beyond mode resonance \cite{Yu:24a, Pnigouras:25} where no relativistic treatments are available.
For a typical NS with $M_A/R_A\sim0.15-0.2$, it may be a reasonable assumption that the form of hydrodynamical equations does not significantly deviate from its Newtonian form. 
Meanwhile, we apply a series of calibrations to various coefficients $(\If, \omega_{f0}, \JJ_{2}, \kappa_2, ...)$, such that our result, when expanded to the low-frequency limit, is identical to relativistic calculations obtained by \cite{Pitre:24, Pitre:25} (also \cite{Pani:25}). 
The calibration procedure is described in this subsection. 

We treat $(k_{2A}, \ddot{k}_{2A}, p_{2A})$ provided in \cite{Pitre:24, Pitre:25} as input parameters for the calibration, though fundamentally, $\ddot{k}_{2A}$ and $p_{2A}$ are functions of $k_{2A}$ and $M_A/R_A$ as described in Sec. \ref{sec:UR_in_GR}. By default, we will consider a polytropic model with $\Gamma=1+1/n=2$ that has $k_{2A}=0.056$, $\ddot{k}_{2A}=0.028$, and $p_{2A}=0.042$ at the specified compactness. 
The Love number and compactness further imply a dimensionless tidal deformability of $\bar{\lambda}_A=262$. 
Note that this corresponds to a very soft equation of state (hardly deformed by tide). In comparison, a soft equation of state like APR \cite{Akmal:98} has $\bar{\lambda}_A\simeq 340$ while a medium-soft one like SLy \cite{Douchin:01} (which has the same compactness as our fiducial model) has $\bar{\lambda}_A\simeq 390$ for NSs of similar masses. Together with the fact that we are using a conservative estimation of $\ddot{k}_{2A}$ (Fig. \ref{fig:ddk2_vs_k2}), our default model provides a highly conservative estimate of the nonlinear tide. 

In our formulation, the linear tide of a non-spinning NS is specified by $\If$ and $\omega_{f0}$, which we compute as 
\begin{align}
    &\If = \sqrt{\frac{5}{4\pi}k_{2A}}, \\
    &\frac{\omega_{f0}^2}{\omega_A^2}=\frac{k_{2A}}{\ddot{k}_{2A}}. 
\end{align}
This implicitly assumes an effective-one-mode approximation, which is appropriate for the affine model applied in this study. 
For the three-wave interactions among f-mode, we need to obtain separately $\kappa_2$ and $\JJ_2$, as they enter in different combinations in the frequency shift (line \ref{eq:dw2_wf_3m}) and the modifications to the overall driving magnitude (line \ref{eq:dV2_3m}; see later in Sec. \ref{sec:eq_sol_hydro}). In the low-frequency limit, however, only a specific combination, $p_{2A}$ (Eq. \ref{eq:p2_vs_J2_kap2}), is constrained by \cite{Pitre:25}. Nonetheless, from Eqs. (\ref{eq:kap2_vs_I_w}) and (\ref{eq:J2_vs_I_w}), we may expect that $\JJ_2$ and $\kappa_2\If$ scale the same way, we therefore write
\begin{align}
    &\kappa_2\simeq-\sqrt{\frac{5}{\pi}}\frac{5}{32\pi} \frac{p_{2A}}{\II_f^3}\frac{4\kappa_2^{\rm (N)}\If^{\rm (N)}}{3\JJ_2^{\rm (N)} + 4\kappa_2^{\rm (N)}\If^{\rm (N)}}=-\frac{19}{160}\frac{p_{2A}}{k_{2A}^{3/2}}, 
    \label{eq:kap2_cal_GR}
    \\
    &\JJ_2\simeq-\sqrt{\frac{5}{\pi}}\frac{5}{24\pi} \frac{p_{2A}}{\If^2}\frac{3\JJ_2^{\rm (N)}}{3\JJ_2^{\rm (N)} + 4\kappa_2^{(N)}\If^{\rm (N)}} = - \sqrt{\frac{5}{\pi}}\frac{7}{80}\frac{p_{2A}}{k_{2A}}.
    \label{eq:J2_cal_GR}
\end{align}
In this subsection, we use a superscript ``(N)'' to denote coupling coefficients derived in the Newtonian limit in Sec. \ref{sec:UR_nl_tide_N}. 
While there exists some uncertainty in the partition between $\JJ_2$ and $\kappa_2\If$ in $p_{2A}$, the above calibration at least ensures consistency with \cite{Pitre:25} in the low-frequency limit. With this construction, we have an empirical scaling that
\begin{align}
    \frac{\kappa_{2}}{\kappa_2^{(N)}} = \frac{\JJ_2/\If}{\JJ_2^{(\rm N)}/\II_f^{\rm (N)}} \simeq \exp[-(M_A/R_A)].
    \label{eq:kap2_vs_kap2_N}
\end{align}
The origin of this scaling remains to be explored. 

To account for rotation at the lowest order, we need to account for the modification of $C_f$ (Eq. \ref{eq:freq_shift_coriolis}). We assume
\begin{align}
    C_f\simeq C_f^{\rm (N)}\frac{\If}{\If^{\rm (N)}}=-\frac{1}{2} \frac{\If}{\If^{\rm (N)}},
\end{align}
where we have used $C_f^{\rm (N)}=-1/l$ and for quadrupolar f-modes, $l=2$. 
This gives good agreement with $C_f$ estimated from \cite{Kruger:20} or \cite{Steinhoff:21} using their quasi-universal relations.  

The leading order $\Omega_A^2$ effect is the forcing of the $m_a=0$ f-mode by the centrifugal potential at $\mathcal{\epsilon_A}$ (the first term in $H_{\rm cen}$ in Eq. \ref{eq:H_rot}). This induces a mass quadrupole of the NS which can couple with the tidal tensor (via the first term in $H_{\rm tide}$ in Eq. \ref{eq:H_tide}), thereby affecting the orbital evolution. This is the quadrupole-monopole interaction studied in, e.g., \cite{Poisson:98}. Under Newtonian gravity, we have the identity (Eq. \ref{eq:Q_A_vs_Lam_A_I_A})
\begin{equation}
    \bar{\mathcal{Q}}_A^{(\rm N)}=\frac{\bar{\lambda}_A^{(\rm N)}}{\left[\bar{I}_A^{(\rm N)}\right]^2}, 
    \label{eq:bar_Q_A_vs_Lam_A_I_A}
\end{equation}
where $\bar{\mathcal{Q}}_A=\mathcal{Q}_A(R_A/M_A)$ and $\bar{I}_A=I_A/M_A^3$, as the quadrupole's coupling to the tidal and centrifugal potentials are identical up to the difference in an overall constant. In GR, the situation is more complicated, causing small deviations from Eq. (\ref{eq:bar_Q_A_vs_Lam_A_I_A}), though $\bar{\mathcal{Q}}_A$ is still fully determined by ${\lambda}_A$ based on the argument of Sec. \ref{sec:UR_in_GR}. To ensure our spin-induced quadrupole matches the known value in GR \cite{Poisson:98}, we rescale the centrifugal potential via a replacement
\begin{equation}
    \Omega_A^2\to\tilde{\Omega}_A^2 \equiv \frac{\bar{\mathcal{Q}}_A \bar{I}_A^2}{\bar{\lambda}_A}\Omega_A^2.
    \label{eq:cen_scale}
\end{equation}
Under the Newtonian limit, $\tilde{\Omega}_A=\Omega_A$. In GR,
we use the value of $k_{2A}$ and $M_A/R_A$ to first determine $\bar{\lambda}_A$, and then use empirical fits from \cite{Yagi:13} to compute $\bar{I}_A$ and $\bar{Q}_A$. The replacement applies to only $\Omega_A^2$ terms in the centrifugal potential but not those in $H_{\rm mode}$.

The values of other coupling coefficients are not directly constrained in relativity at the moment. Furthermore, they depend on a new NS parameter, $\Gamma_{\rm ad}$ (or $\omega_r^2$), that is not constrained by $(k_{2A}, \ddot{k}_{2A}, p_{2A})$. 
Therefore, we simply assume without further justification that
\begin{align}
    &\frac{\II_r}{\II_r^{(N)}} \simeq \frac{\II_f}{\II_f^{(N)}} \frac{\bar{\lambda}_A}{\bar{\mathcal{Q}}_A\bar{I}_A^2}, \\
    &\frac{\JJ_{2r}}{\JJ_{2r}^{\rm (N)}} \simeq \frac{\JJ_2}{\JJ_2^{\rm (N)}}, \\
    &\frac{\JJ_{r2}}{\JJ_{r2}^{\rm (N)}}\simeq \frac{\JJ_{rr}}{\JJ_{rr}^{\rm (N)}} \simeq \frac{\JJ_2}{\JJ_2^{\rm (N)}}\frac{\bar{\lambda}_A}{\bar{\mathcal{Q}}_A\bar{I}_A^2}, \\
    &\frac{\kappa_{2r}}{\kappa_{2r}^{\rm (N)}} 
    \simeq \frac{\kappa_{rr}}{\kappa_{rr}^{\rm (N)}} \simeq \frac{\kappa_{2}}{\kappa_{2}^{\rm (N)}}. 
\end{align}
The $\bar{\lambda}_A/(\bar{\mathcal{Q}}_A\bar{I}_A^2)$ factor in $\II_r$, $\JJ_{r2}$, and $\JJ_{rr}$ is to keep the rescaling of Eq. (\ref{eq:cen_scale}) only to the $l=2$ part of the centrifugal potential.
For the four-mode coupling coefficients $\zeta_{abcd}$, we simply assume 
\begin{equation}
    \zeta_{abcd}\simeq \zeta_{abcd}^{(\rm N)} \exp[-2(M_A/R_A)],
\end{equation}
as an educated guess based on Eq. (\ref{eq:kap2_vs_kap2_N}). 
While the GR values of nonlinear coupling coefficients related to spin effects at quadratic or higher powers in $\Omega_A$ and hydrodynamics at the four-wave order have fractional uncertainties of $\sim (M_A/R_A)\lesssim 30\%$, they are nonetheless subdominant in the tidal problem, as we will show in Sec. \ref{sec:nl_mode_amp}.  
The only exception is in an NS with rapid, anti-aligned spin, where resonant excitation of the f-mode becomes possible, which we will discuss in Sec. \ref{sec:res_locking} and \ref{sec:Gamma_ad}. 

For the rest of the paper, we will refer to an NS whose coupling coefficients are calibrated following the outlined procedure as a relativistic NS, while one with values derived in Sec. \ref{sec:UR_nl_tide_N} as a Newtonian one. 


\section{Equilibrium solutions of the hydrodynamics}
\label{sec:eq_sol_hydro}

In this section, we describe the solutions to the internal hydrodynamics in terms of mode amplitudes. They will then be used in the next section to derive the tidal back-reactions to the orbit, and consequently, the final tidal corrections to the GW phase.
Our analytical solutions (Sec. \ref{sec:nl_mode_amp}) will focus on the ``equilibrium tide'' following the convention of \cite{Yu:24a}, defined as the component in $c_a$ that has a phase dependence of $\exp(-im_a\phi)$ (i.e., the particular solution of a driven harmonic oscillator). It contains finite-frequency corrections in the form of a Lorentzian. In contrast, the dynamical tide is the component that varies at the mode's natural frequency, or approximately as $\exp[-i\omega_{a}t -im_a\Omega_A t]$ (the homogeneous solution excited at mode resonance), and its impact will only be numerically investigated in Secs. \ref{sec:res_locking} and \ref{sec:Gamma_ad}. 
All numerical integrations start at $r=8R_A$, using the analytical equilibrium tide solutions as the initial conditions, and terminate at $r=2R_A$. 
For the mode amplitude evolutions presented in this section, a $\Gamma=1+1/n=2$ polytrope will be assumed (either Newtonian or relativistic), whereas our discussion on tidal dephasing in Sec. \ref{sec:sol_orb} will cover both a polytropic $\Gamma=2$ (very soft) and the SLy \cite{Douchin:01} equation of state (medium-soft). 
In most cases, we take $\Gamma_{\rm ad}=2$, which has little impact on the result as it does not enter at the three-wave (NLO) dynamics for a slowly-spinning NS as discussed in Sec. \ref{sec:UR_nl_tide_N}, except for in Sec. \ref{sec:Gamma_ad} where we will discuss how $\Gamma_{\rm ad}$ may be probed in a rapidly spinning NS.

\subsection{Nonlinear mode amplitudes}
\label{sec:nl_mode_amp}

The explicit equations of motion derived from $H_{\rm npp}$ are provided in this subsection (see Eq. \ref{eq:EOM_phase} for the general form). Here, we will focus on the positive-frequency branch and drop the ``+'' in the subscripts for notational simplicity. The negative-frequency branch is obtained through a complex conjugation \cite{Schenk:02}. 
Since the radial mode does not directly couple to the tide at the $\epsilon_A$ order, and the $m_a=0$ mode is only adiabatically forced by the tide, we will compute their solutions to $\epsilon_A^2$, corresponding to three-wave interactions (NLO) in the Hamiltonian. The $|m_a|=2$ modes are affected by the finite-frequency corrections and dominate the tidal response in the late inspiral (also when the f-modes' tidal impact becomes significant in the GW waveform).  Therefore, we will solve the amplitudes of $|m_a|=2$ modes to the $\epsilon_A^3$ order (NNLO). This is also needed to assess whether the resummation leading to eq. (1.10) of \cite{Pitre:25} is accurate (Sec. \ref{sec:effective_Love}), and
whether f-mode can experience resonance locking in rapidly spinning NSs as claimed in \cite{Kuan:24} (Sec. \ref{sec:res_locking}). 

We will start with the radial mode, whose equation of motion is 
\begin{align}
    \dot{c}_r = - i\omega_rc_r + i\omega_r&\left[\frac{4\sqrt{\pi}}{3}\mathcal{I}_r\frac{\tilde{\Omega}_A^2}{\omega_A^2}
    +\sqrt{\frac{3\pi}{10}}\mathcal{J}_{2r}(b_2+b_{-2}^\ast + b_{-2} + b_2^\ast)\epsilon_A  - \sqrt{\frac{\pi}{5}}\mathcal{J}_{2r}(c_0+c_0^\ast)\left(\epsilon_A +\frac{2}{3}\frac{\tilde{\Omega}_A^2}{\omega_A^2}\right)
    \right. \nonumber\\
    &\left.+\frac{4\sqrt{\pi}}{3}\mathcal{J}_{rr}(c_r+c_r^\ast)\frac{\tilde{\Omega}_A^2}{\omega_A^2} +2\kappa_{2r}(b_2+b_{-2}^\ast)(b_{-2}+b_{2}^\ast) + \kappa_{2r}(c_0+c_{0}^\ast)^2 + \kappa_{rr}(c_r+c_r^\ast)^2  \right],
    \label{eq:EOM_cr}
\end{align}
where we have defined $b_{\pm2}=c_{\pm 2} e^{\pm 2i\phi }$ so that $b_{\pm 2}$ are slow varying prior to the $|m_a|=2$ modes' resonance. This also means $|\dot{c}_r|\ll \omega_r c_r$ before resonance. We can therefore solve for the adiabatic solution of $c_r$ to NLO,
\begin{subequations}
\begin{align}
    c_r\simeq& \frac{4\sqrt{\pi}}{3}\mathcal{I}_r\frac{\tilde{\Omega}_A^2}{\omega_A^2} \\
    +& 2 \kappa_{2r}(b_2+b_{-2}^\ast)(b_{-2} + b_2^\ast) 
    + \sqrt{\frac{3\pi}{10}}\JJ_{2r}(b_2+b_{-2}^\ast+b_{-2} + b_2^\ast)\epsilon_A \nonumber \\
    + &\frac{2\pi}{45}\If\left(\JJ_{2r} + 2\kappa_{2r}\If\right)\left(3\epsilon_A + 2\frac{\tilde{\Omega}_A^2}{\omega_A^2}\right)^2
    + \frac{32\pi}{9} \II_r (\JJ_{rr} +  2\kappa_{rr}\II_r) \frac{\tilde{\Omega}_A^4}{\omega_A^4}. 
    \label{eq:cr_sol_3m}
\end{align}
\end{subequations}
where the first line shows the linear solution, and the subsequent lines are the nonlinear corrections. 
We have substituted the linear tide solution into the $c_0$ and $c_r$ terms in the bracket in Eq. (\ref{eq:EOM_cr}).
To $\epsilon_A$ order, the radial mode corresponds to an expansion of the NS caused by the centrifugal force. 
At $\epsilon_A^2$ order (from three-wave interactions in the Hamiltonian), $c_r$ will be non-zero even when $\Omega_A=0$, as quadratic couplings between $l=2$ modes and tides can lead to an expansion of the NS. When $c_r$ back-reacts on the $l=2$ modes, it corresponds to the modes propagating in an expanded background, which tends to lower their oscillation frequencies \cite{Yu:23a}. This effect is a crucial component of the nonlinear anharmonic frequency shift of $l=2$ modes at the four-wave level, and we keep the $b_{\pm 2}$ terms in the $c_r$ solution (instead of reducing them with their linear tide solutions) to capture this frequency shift (to be discussed later when the solutions of $b_{\pm2}$ are derived).  

Now turn to the $c_0$ mode, whose equation of motion is
\begin{align}
    \dot{c}_0=-i\omega_{f0}c_0 + i\omega_{f0} &\left[-\sqrt{\frac{\pi}{5}} \If(\epsilon_A + \frac{2}{3}\frac{\tilde{\Omega}_A^2}{\omega_A^2})
    \right.\nonumber \\
    & + 2\kappa_2 (b_2+b_{-2}^\ast)(b_{-2}+b_{2}^\ast) 
    -\kappa_2(c_0+c_0^\ast)^2 + 2\kappa_{2r} (c_0+c_0^\ast)(c_r+c_r^\ast) 
    \nonumber \\
    &+\sqrt{\frac{3\pi}{10}}\JJ_2(b_2+b_{-2}^\ast + b_{-2} + b_{2}^\ast)\epsilon_A
    +\sqrt{\frac{\pi}{5}}\JJ_2(c_0+c_0^\ast)\left(\epsilon_A + \frac{2}{3}\frac{\tilde{\Omega}_A^2}{\omega_A^2}\right)
    \nonumber \\
    &-\sqrt{\frac{\pi}{5}}\JJ_{2r}(c_r+c_r^\ast)(\epsilon_A + \frac{2}{3}\frac{\tilde{\Omega}_A^2}{\omega_A^2}) 
    +\frac{4\sqrt{\pi} }{3}\JJ_{r2}(c_0 + c_0^\ast)\frac{\tilde{\Omega}_A^2}{\omega_A^2}
\end{align}
The three-wave equilibrium (i.e., with $\dot{c}_0=0$) solution reads
\begin{subequations}
\begin{align}
    c_0\simeq& -\sqrt{\frac{\pi}{5}}\If(\epsilon_A+\frac{2}{3}\frac{\tilde{\Omega}_A^2}{\omega_A^2}) \label{eq:c0_sol_lin}\\
    &+ 2\kappa_2 (b_2+b_{-2}^\ast)(b_{-2}+b_{2}^\ast)  
    + \sqrt{\frac{3\pi}{10}}\JJ_2(b_2+b_{-2}^\ast + b_{-2}+b_2^\ast)\epsilon_A \nonumber \\
    &-\frac{2\pi}{5}\If(\JJ_2 + 2\If\kappa_2)\epsilon_A^2
    -\frac{8\pi}{15}[ \If(\JJ_2+2\kappa_2\If + \sqrt{5}\JJ_{2r}) + \sqrt{5}\II_r(\JJ_{2r}+4\kappa_{2r}\If)]\epsilon_A\frac{\tilde{\Omega}_A^2}{\omega_A^2} \nonumber \\
    &-\frac{8\pi}{45}\left[\If\left(\JJ_2 + 2\kappa_2\If
    +2\sqrt{5}\JJ_{r2}
    \right) + 2\sqrt{5} \II_r(J_{2r}+4\kappa_{2r}\If)\right]\frac{\tilde{\Omega}_A^4}{\omega_A^4}.
    \label{eq:c0_sol_3m}
\end{align}
\end{subequations}
Again, we write the linear tide solution in line (\ref{eq:c0_sol_lin}) and the nonlinear corrections in line (\ref{eq:c0_sol_3m}).
We have reduced $c_0$ and $c_r$ in the nonlinear couplings by their linear solutions while keeping the $b_{\pm 2}$ terms. This way, when the three-wave $c_0$ solution is plugged into the equation of motion of $b_{\pm 2}$, these terms can lead to anharmonic frequency shifts. 

\begin{figure}
    \centering
    \includegraphics[width=0.7\linewidth]{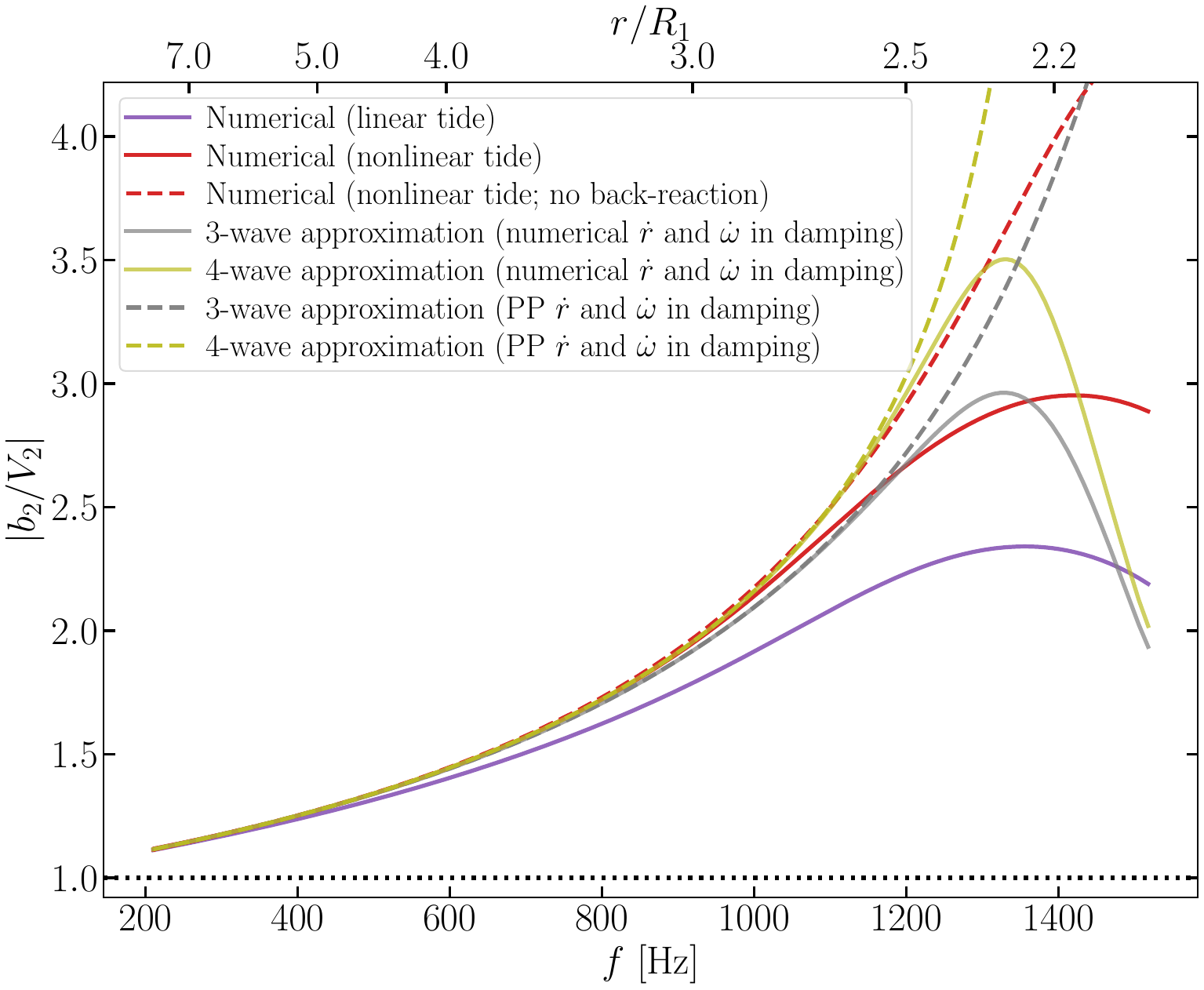}
    \caption{Amplitude of the $l=m=2$ f-mode that dominates the tidal interaction. A non-spinning, Newtonian $\Gamma=\Gamma_{\rm ad}=2$ polytrope is assumed. 
    Below 1000\,Hz, the four-wave approximation agrees well with numerical simulations. At higher frequencies, whether tidal back-reaction is included can have significant impacts on the mode dynamics (red-solid vs. red-dashed). This is mainly through the effective damping of the mode that depends on $\dot{\omega}$, which can be enhanced by an order of magnitude by the tide compared to the PP value when the f-mode approaches resonance. Near resonance, the equilibrium tide solution also tends to overestimate the mode amplitude.}
    \label{fig:b2_vs_f_N}
\end{figure}

We now consider the $|m_a|=2$ f-modes. Since they dominate the tidal response, 
we solve them to $\epsilon_A^3$ order (NNLO). Note that two channels contribute to the equation of motion at $\epsilon_A^3$. The first originates from the direct four-wave interaction Hamiltonian (last term in Eq. \ref{eq:H_mode}). The other comes from substituting the three-wave solutions of $c_0$ (line \ref{eq:c0_sol_3m}) and $c_r$ (line \ref{eq:cr_sol_3m}) into the three-wave forcing terms of $b_{\pm 2}$.
The equation of motion reads
\begin{align}
    &\dot{b}_{2} + i[\omega_{2} +(\omega-\Omega_A)]b_2+ i\sigma_2 b_{-2}^\ast = i \omega_{f0}\left[\sqrt{\frac{3\pi}{10}}\If \epsilon_A - \frac{\Delta \omega_2}{\omega_{f0}}(b_2+b_{-2}^\ast)+\Delta V_{2}\right],
    \label{eq:db2}\\
    &\dot{b}_{-2}^\ast - i[\omega_{-2} -(\omega-\Omega_A)]b_{-2}^\ast - i\sigma_2 b_2 = -i \omega_{f0}\left[\sqrt{\frac{3\pi}{10}}\If \epsilon_A - \frac{\Delta \omega_2}{\omega_{f0}}(b_2+b_{-2}^\ast)+\Delta V_{2}\right],
    \label{eq:dbn}
\end{align}
where $\omega_{\pm2}$ and $\sigma_2$ are given by Eqs. (\ref{eq:freq_shift_coriolis}) and (\ref{eq:sigma_a_coriolis}), respectively accounting for the frequency shift and coupling caused by the Coriolis effect.
For the nonlinear corrections, we split them into two parts. 
We have grouped all the nonlinear terms that are proportional to $(b_{2} + b_{-2}^\ast)$ into the $[-(\Delta \omega_{2}/\omega_{f0})(b_{2} + b_{-2}^\ast)]$ term. It corresponds to the nonlinear shift of the $m_a=2$ f-mode's natural frequency. When writing the mode solution into a Lorentzian form, the frequency shift introduces corrections to the denominator. In contrast, the rest of the nonlinear terms grouped in $\Delta V_{2}$ correct numerators of the Lorentzian. Specifically, we have
\begin{subequations}
\begin{align}
    \frac{\Delta \omega_2}{\omega_{f0}}&\equiv\frac{1}{2}\frac{\Delta \omega_2^2}{\omega_{f0}^2} \nonumber \\
    &= \sqrt{\frac{\pi}{5}}(\JJ_2 + 4\kappa_2\If)(\epsilon_A+\frac{2}{3}\frac{\tilde{\Omega}_A^2}{\omega_A^2}) 
    - \frac{4\sqrt{\pi}}{3}(\JJ_{r2}+4\kappa_{2r}\II_r)\frac{\tilde{\Omega}_A^2}{\omega_A^2} 
    \label{eq:dw2_wf_3m}\\
    & - [8(\kappa_{2}^2 + \kappa_{2r}^2)+3\zeta_{22}](b_2+b_{-2}^\ast)(b_{-2}+b_2^\ast) \label{eq:dw2_wf_4m_b2sq} \\
    &-2\sqrt{\frac{6\pi}{5}}(\kappa_2\JJ_2 + \kappa_{2r}\JJ_{2r})(b_2+b_{-2}^\ast + 2b_{-2} + 2b_2^\ast)\epsilon_A \nonumber \\
    &-\frac{\pi}{5}\left[ 3\JJ_2^2 + 3\JJ_{2r}^2 
    - 8\kappa_2\If\JJ_2 + 8 \kappa_{2r}\If \JJ_{2r} - 4\If^2(4\kappa_2^2 - 4\kappa_{2r}^2-3\zeta_{20}) 
    \right]\epsilon_A^2 \nonumber \\
    &+\frac{16\pi}{15}\left[
    \If^2(4\kappa_2^2 - 4\kappa_{2r}^2 -3 \zeta_{20}) + 
    2\If(\kappa_2\JJ_2+\sqrt{5}\kappa_2\JJ_{r2}-\kappa_{2r}\JJ_{2r})
    \right.\nonumber \\
    &\left.
    \quad\quad 
    + 2\sqrt{5} \II_r(\kappa_2 \JJ_{2r} + 4\kappa_2\kappa_{2r} \If + 3 \zeta_{20r}\If)
    \right]\epsilon_A \frac{\tilde{\Omega}_A^2}{\omega_A^2} \nonumber \\
    &+\frac{16\pi}{45}\left[
    \If^2(4\kappa_2^2-4\kappa_{2r}^2 - 3\zeta_{20})+
    2\If(\kappa_2\JJ_2 + 2\sqrt{5}\kappa_2 \JJ_{r2} - \kappa_{2r}\JJ_{2r})
    + 4\sqrt{5}\If\II_r(4\kappa_2\kappa_{2r} + 3\zeta_{20r})
    \right. \nonumber \\
    &\left. 
    \quad\quad
    \II_r(4\sqrt{5} \kappa_2 \JJ_{2r} -40 \kappa_{2r}\JJ_{rr})
    -20\II_r^2(4\kappa_{2r}\kappa_{rr} + 3 \zeta_{2r})
    \right]\frac{\tilde{\Omega}_A^4}{\omega_A^4},
    \label{eq:dw2_wf}
    \end{align}
\end{subequations}
and
\begin{subequations}
    \begin{align}
    \Delta V_2 &= -\frac{\sqrt{6}\pi}{5} \If \JJ_2\epsilon_A(\epsilon_A + \frac{2}{3}\frac{\tilde{\Omega}_A^2}{\omega_A^2}) + 4\pi\sqrt{\frac{2}{15}} \II_r\JJ_{2r} \epsilon_A \frac{\tilde{\Omega}_A^2}{\omega_A^2} \label{eq:dV2_3m} \\
    &
    +\frac{\pi}{5}\sqrt{\frac{6\pi}{5}} \If \left[
    \JJ_2^2 - 4\kappa_2\If\JJ_2 + 5 \JJ_{2r}^2 + 4\kappa_{2r}\If \JJ_{2r}
    \right]\epsilon_A^3 \nonumber \\
    &-\frac{8\pi}{25}\sqrt{\frac{10\pi}{3}}\left[
    2\If^2(\kappa_2 \JJ_2 -\kappa_{2r}\JJ_{2r}) 
    +\If(\JJ_2^2 - \JJ_{2r}^2 + \sqrt{5}\JJ_2 \JJ_{2r} + 4\sqrt{5}\kappa_{2r}\II_r\JJ_2) + \sqrt{5}\II_r \JJ_2 \JJ_{2r} 
    \right]\epsilon_A^2\frac{\tilde{\Omega}_A^2}{\omega_A^2}\nonumber \\
    &-\frac{8\pi}{75}\sqrt{\frac{10\pi}{3}} \left[
    2\If^2(\kappa_2\JJ_2 - \kappa_{2r}\JJ_{2r}) 
    +\If (\JJ_2^2 - \JJ_{2r}^2 + 2\sqrt{5}\JJ_2\JJ_{r2} +8\sqrt{5}\kappa_{2r}\II_{r}\JJ_2) \right.
    \nonumber \\
    & \left.  \quad\quad\quad\quad\quad
    +2\sqrt{5}\II_r \JJ_{2r}(\JJ_2 - 2\sqrt{5}\JJ_{rr} - 4\sqrt{5}\kappa_{rr}\II_r)
    \right]\epsilon_A \frac{\tilde{\Omega}_A^4}{\omega_A^4}.
    \label{eq:dV2}
\end{align}
\end{subequations}
The corrections at the three-wave level (NLO) are presented in lines (\ref{eq:dw2_wf_3m}) and (\ref{eq:dV2_3m}), while the subsequent lines are corrections due to the four-wave interactions (NNLO). 
When one seeks a perturbative solution, the $b_{\pm 2}$ terms appearing in $\Delta \omega_2/\omega_{f0}$ can be replaced by their linear solution. We retain them because they are needed for the discussion on resonance locking in Sec. \ref{sec:res_locking}.
One can show $\Delta \omega_{-2}=\Delta \omega_2^\ast$ and $\Delta V_{-2}=\Delta V_2^\ast$. 
Note that when computing $\Delta\omega_2$ and $\Delta V_2$ entering at NLO and NNLO, one can use the PP Keplerian law to replace $\epsilon_A$ by $(M_B/M_t) (\omega/\omega_A)^2$. This introduces a fractional error in $b_{2}$ of the order $\epsilon_A (R_A/r)^5\propto (R_A/r)^8 \propto \omega^{16/3}$. However, for the linear tide solution
\begin{equation}
    b_{\pm 2}^{(\rm lin)} = \sqrt{\frac{3\pi}{10}}\frac{\omega_{f0}}{\omega_{\pm2} \mp 2(\omega+\Omega_{A})} \If \epsilon_A,
\end{equation}
replacing $r$ in $\epsilon_A$ by $\omega$ using the Keplerian relation will introduce an error of lower order than the four-wave correction. 

For the analytical expressions, we will focus on the case where resonance does not occur, in which case it is sufficient to consider only the equilibrium tide solution following \cite{Yu:24a}. 
Since $b_{\pm 2}$ are slow varying before resonances, we can first drop $\dot{b}_{\pm 2}$ in Eqs. (\ref{eq:db2}) and (\ref{eq:dbn}). Then, solving $b_2$ and $b_{-2}^\ast$ together leads to 
\begin{align}
    &b_2\simeq\frac{\omega_{f0}[\omega_{f0} + 2(\omega-\Omega_A-C_f\Omega_A)  ](V_2+\Delta V_2)}{\varpi^2},\\
    &b_{-2}^\ast \simeq\frac{\omega_{f0}[\omega_{f0} - 2(\omega-\Omega_A-C_f\Omega_A)  ](V_2+\Delta V_2)}{\varpi^2},
\end{align}
where 
\begin{equation}
    \varpi^2 = \omega_{f0}^2\left(1+\frac{\Delta \omega_2^2}{\omega_{f0}^2}\right) - 4 (\omega-\Omega_A)[\omega-(1+2 C_f)\Omega_A], 
    \label{eq:detuning_sq}
\end{equation}
describing the nonlinearly corrected denominator of a Lorentzian. This becomes transparent when we take the sum of $(b_2 + b_{-2}^\ast)$, which becomes
\begin{equation}
    q_2e^{2i(\phi-\Omega_At)}=b_2+b_{-2}^\ast\simeq \frac{
    2\omega_{f0}^2 (V_2 + \Delta V_2)
    }{
    \varpi^2},
    \label{eq:b2_p_bnc_no_damping}
\end{equation}
Compared to what was derived in \cite{Yu:23a}, the result is a better solution to the sum $(b_2 + b_{-2}^\ast)$ entering the mass quadrupole of the NS (Eq. \ref{eq:Q_lm}). It makes the connection between $\Delta \omega_2^2/\omega_{f0}^2<-1$ and the onset of hydrodynamical instability (i.e., when the NS exceeds its Roche limit \cite{Yu:26}) transparent.
It also incorporates the spin corrections valid for all values of $\Omega_A$. 

Ignoring spin, focusing on the three-wave solution, and replacing $\epsilon_A$ by $(M_B/M_t)(\omega^2/\omega_A^2)$, we have
\begin{align}
    q_2 e^{2i\phi} &\simeq \frac{2\omega_{f0}^2 (V_2 + \Delta V_2)}{\omega_{f0}^2 + 2\sqrt{\frac{\pi}{5}}(\JJ_2 + 4\kappa_2\If) \frac{\omega_{f0}^2}{\omega_A^2}\frac{M_B}{M_t}\omega^2 - 4\omega^2}\nonumber \\
    &=\frac{2\omega_\ast^2(V_2+\Delta V_2)}{\omega_{\ast}^2 - 4\omega^2},
    \label{eq:b2_p_bnc_3w_recast}
\end{align}
where
\begin{equation}
    \omega^2_\ast = \frac{\omega_{f0}^2}{1-\frac{1}{2}\sqrt{\frac{\pi}{5}} (\JJ_2 + 4\kappa_2\If)\frac{\omega_{f0}^2}{\omega_A^2}\frac{M_B}{M_t}}.
    \label{eq:eff_freq_Lorentzian_3m}
\end{equation}
In other words, when one further ignores the $\Delta V_2$ term, the three-wave solution can be recast into a Lorentzian form similar to what the linear theory would predict, with a constant shift of the mode frequency given by Eq. (\ref{eq:eff_freq_Lorentzian_3m}). This observation is important to illustrate why the three-wave frequency shift cannot lead to resonance locking, a point we will discuss further in Sec. \ref{sec:res_locking}. Note our $\omega_\ast^2$ differs from eq. (1.10) of \cite{Pitre:25}, because part of the nonlinear corrections goes to the numerator (the $\Delta V_2$ term); see also Sec. \ref{sec:vs_PP25}. 

One can improve the solution by taking $\dot{b}_{\pm 2}$ into account. 
Following \cite{Lai:94c}, we define the approximate solution of $q_2$ in Eq. (\ref{eq:b2_p_bnc_no_damping}) as $q_2^{(0)}$. Taking the time derivative leads to
\begin{align}
    \frac{d}{dt}\left[q_2^{(0)}e^{2i(\phi-\Omega_A t)}\right] 
    = \gamma_{2n} q_2^{(0)}e^{2i(\phi-\Omega_A t)},
\end{align}
where
\begin{align}
    \gamma_{2n} &= \frac{ d(V_2 + \Delta V_2)/dt}{V_2+\Delta V_2} - \frac{d(\varpi^2)/dt}{\varpi^2}, \nonumber \\
    &\simeq -(l_2 +1)\left(1+ \frac{\Delta V_2^{(\rm 3w,\,ns)}}{V_2}\right)\frac{\dot{r}}{r} 
    + 2\left\{
    m_2^2\left[1-(1+C_f)\frac{\Omega_A}{\omega}\right]
    -\frac{\Delta (\omega_2^2)^{(\rm 3w,\,ns)}}{\omega_{f0}^2}\frac{\omega_{f0}^2}{\omega^2}
    \right\}\frac{\omega\dot{\omega}}{\varpi^2},\nonumber \\
    &\simeq -3\left(1-2\sqrt{\frac{\pi}{5}}\JJ_2\epsilon_A\right)\frac{\dot{r}}{r}
    +2\left\{4\left[1-(1+C_f)\frac{\Omega_A}{\omega}\right]
    -2\sqrt{\frac{\pi}{5}}(\JJ_2 + 4\kappa_2\If)\frac{M_B}{M_t}\frac{\omega_{f0}^2}{\omega_A^2}
    \right\}\frac{\omega\dot{\omega}}{\varpi^2}.
    \label{eq:gam_2n}
\end{align}
The derivation includes NLO corrections to the hydrodynamics. Terms with a superscript ``(3w, ns)'' are the non-spinning ($\Omega_A=0$), three-wave corrections from Eqs. (\ref{eq:dw2_wf_3m}) and (\ref{eq:dV2_3m}), and the explicit form is given in the final expression.
One may further assume that 
\begin{equation}
    \frac{d}{dt}\left[q_2e^{2i(\phi-\Omega_A t)}\right] 
    \simeq \gamma_{2n}q_2e^{2i(\phi-\Omega_A t)}.
    \label{eq:dq2_orb_eq_gam2n_q2_orb}
\end{equation}
Removing the superscript ``(0)'' has the equivalent effect as the resummation of the equilibrium solution proposed in \cite{Yu:24a}. Plug the above $d\left[q_2e^{2i(\phi-\Omega_A t)}\right]/dt$ into the equation of motion for $q_2 e^{2i(\phi-\Omega_A t)}$ and keep terms to $\mathcal{O}(\dot{r}/r)\sim \mathcal{O}(\dot{\omega}/\omega)$ leads to an improved equilibrium solution 
\begin{align}
    q_2e^{2i(\phi-\Omega_A t)} = b_2+b_{-2}^\ast \simeq\frac{
    2\omega_{f0}^2 (V_2 + \Delta V_2)
    }{
    \varpi^2 +i\omega_{f0}\gamma_{2d}}.
    \label{eq:b2_p_bnc}
\end{align}
This means the evolution of the orbit introduces a dynamical tidal lag \cite{Lai:94c} that behaves like an effective damping \cite{Yu:24a} (likely dominating over fluid dissipation) 
denoted by $\gamma_{2d}$, whose explicit form is given by
\begin{align}
    \omega_{f0}\gamma_{2d} =m_2\left[\omega - (1+C_f) \Omega_A\right]&
    \left[
    -\frac{\dot{\omega}}{\omega-(1+C_f)\Omega_A}-2\gamma_{2n}
    \right] \nonumber \\
    \simeq 2\left[\omega - (1+C_f) \Omega_A\right]&
    \left[
    -\frac{\dot{\omega}}{\omega-(1+C_f)\Omega_A}
    +6\left(1-2\sqrt{\frac{\pi}{5}}\JJ_2 \epsilon_A\right)\frac{\dot{r}}{r} 
    \right. \nonumber \\
    &
    \left.
    -4\left\{4\left[1-(1+C_f)\frac{\Omega_A}{\omega}\right]
    -2\sqrt{\frac{\pi}{5}}(\JJ_2 + 4\kappa_2\If)\frac{M_B}{M_t}\frac{\omega_{f0}^2}{\omega_A^2}
    \right\}
    \frac{\omega\dot{\omega}}{\varpi^2}
    \right]
    \label{eq:gam_2d}
\end{align}
Our result improves over eq. (3.6) of \cite{Lai:94c} by including rotation and the NLO hydrodynamics. 
While the derivation is conducted at NLO, all nonlinear corrections should be included in $\varpi^2$ in practice to ensure the damping term diverges at mode resonance, thereby zeroing the equilibrium tide solution \cite{Yu:24a}. 

To evaluate the $\dot{r}$ and $\dot{\omega}$ terms appearing in $\gamma_{2n}$ and $\gamma_{2d}$, a perturbative calculation would use the PP prediction
\begin{align}
    &\frac{\dot{\omega}_{\rm pp}}{\omega^2} = \frac{96}{5} \frac{M_AM_B}{M_t^2} \left(M_t \omega\right)^{5/3}, 
    \label{eq:dwdt_pp}
    \\
    &\frac{\dot{r}_{\rm pp}}{r} = -\frac{2}{3}\frac{\dot{\omega}_{\rm pp}}{\omega}. 
    \label{eq:drdt_pp}
\end{align}
However, as we will see shortly, the tidal corrections to $\dot{r}$ and $\dot{\omega}$ can be significant, even dominant, when the effective damping term becomes important (as the $m_a=2$ f-mode is close to resonance). We will thus present also results that utilize numerically extracted $\dot{r}$ and $\dot{\omega}$ in the effective damping. These values are not known until the differential equations are solved, representing a technical challenge that needs to be addressed in future studies when efficient and accurate analytical waveforms are to be developed. 

For the phase space mode amplitudes, we use the definition that $q_2 e^{2i(\phi-\Omega_At)}=b_2 + b_{-2}^\ast$ together with (cf. Eq. \ref{eq:dq_vs_c})
\begin{equation}
    \frac{d}{dt}\left[q_2e^{2i(\phi-\Omega_At)}\right]
    =-i\left\{\omega_{f0} - 2[\omega-(1+C_f)\Omega_A] \right\} b_{2} 
    +i\left\{\omega_{f0} + 2[\omega-(1+C_f)\Omega_A] \right\} b_{2}^\ast,
    \label{eq:dq_vs_b_orb_frame}
\end{equation}
to solve for $b_{\pm 2}$. Eq. (\ref{eq:dq2_orb_eq_gam2n_q2_orb}) is further utilized to replace the left-hand side, leading to
\begin{align}
    &b_{\pm 2} = 
    \frac{\omega_{f0}\left\{\omega_{f0} \pm 2[\omega-(1+C_f)\Omega_A] + i\gamma_{2n}\right\} (V_2 + \Delta V_{\pm 2})}{\varpi^2 \pm i\omega_{f0}\gamma_{2d}}.
    \label{eq:b_pm2}
\end{align}

To assess the significance of the effective damping, we note that the damping enters the denominator of the mode amplitude Lorentzian as $\dot{\omega}$ (see the first two terms of Eq. \ref{eq:gam_2d}), whereas a typical four-wave interaction produces a correction $\omega_{f0}^2\epsilon_A^2$ in the mode amplitude. The ratio of the two effects is
\begin{equation}
    \frac{\dot{\omega}}{\omega_{f0}^2 \epsilon_A^2} \propto 
    \frac{M_t^2}{M_A M_B}\left(\frac{M_A}{R_A}\right)^3 (M_t\omega)^{-1/3}.
\end{equation}
Consequently, the effective damping due to orbital evolution is typically more important than the four-wave interaction for small $(M_t\omega)$. 
Both effects enter the denominator of the Lorentzian, and thus are significant in the late inspiral when the f-mode is close to resonance. 
There, $\gamma_{2d}$ will instead be dominated by the last term of Eq. (\ref{eq:gam_2d}), which diverges when $\varpi^2=0$, making the equilibrium tide zero. The physical mode amplitude is finite, corresponding to the excitation of the dynamical tide. 
Note further that the effective damping will lead to an imaginary component in the mode amplitude and consequently, the effective Love number (see later in Eq. \ref{eq:k22_resum_3m}), which is responsible for the tidal torque that dominates energy flow from the orbit to the NS near resonance (eq. 64 of \cite{Yu:24a} and Fig. \ref{fig:radial_tangential_Enpp}). The four-wave effect determines whether the f-mode can evolve into a state of resonance locking, and the correct form of the effective Love number beyond the low-frequency limit, topics we will address respectively in Secs. \ref{sec:res_locking} and \ref{sec:effective_Love}. 

To assist comparison with the results of \cite{Pitre:25} in Sec. \ref{sec:effective_Love}, we also present a low-frequency expanded version of Eq. (\ref{eq:b_pm2})
\begin{align}
    b_{\pm2} &\simeq W_2 \If\epsilon_A\left\{1 \pm \frac{2[\omega - (1+C_f)\Omega_A]}{\omega_{f0}} + 4\frac{(\omega-\Omega_A)[\omega-(1+2C_f)\Omega_A]}{\omega_{f0}^2} \right.
    \nonumber \\
    &\left.
    -4\sqrt{\frac{\pi}{5}}(\JJ_2 + 2\kappa_2\If)(\epsilon_A+\frac{2}{3}\frac{\Omega_A^2}{\omega_A^2})+\frac{8\sqrt{\pi}}{3}(\JJ_{r2}+\frac{\II_r}{\If}\JJ_{2r} + 4\kappa_{2r}\II_r) \frac{\Omega_A^2}{\omega_A^2}
    \right\}. \label{eq:b_2_low_f}
\end{align}
For this part, we only keep terms at the three-wave level (NLO). 

\begin{figure}
    \centering
    \includegraphics[width=0.7\linewidth]{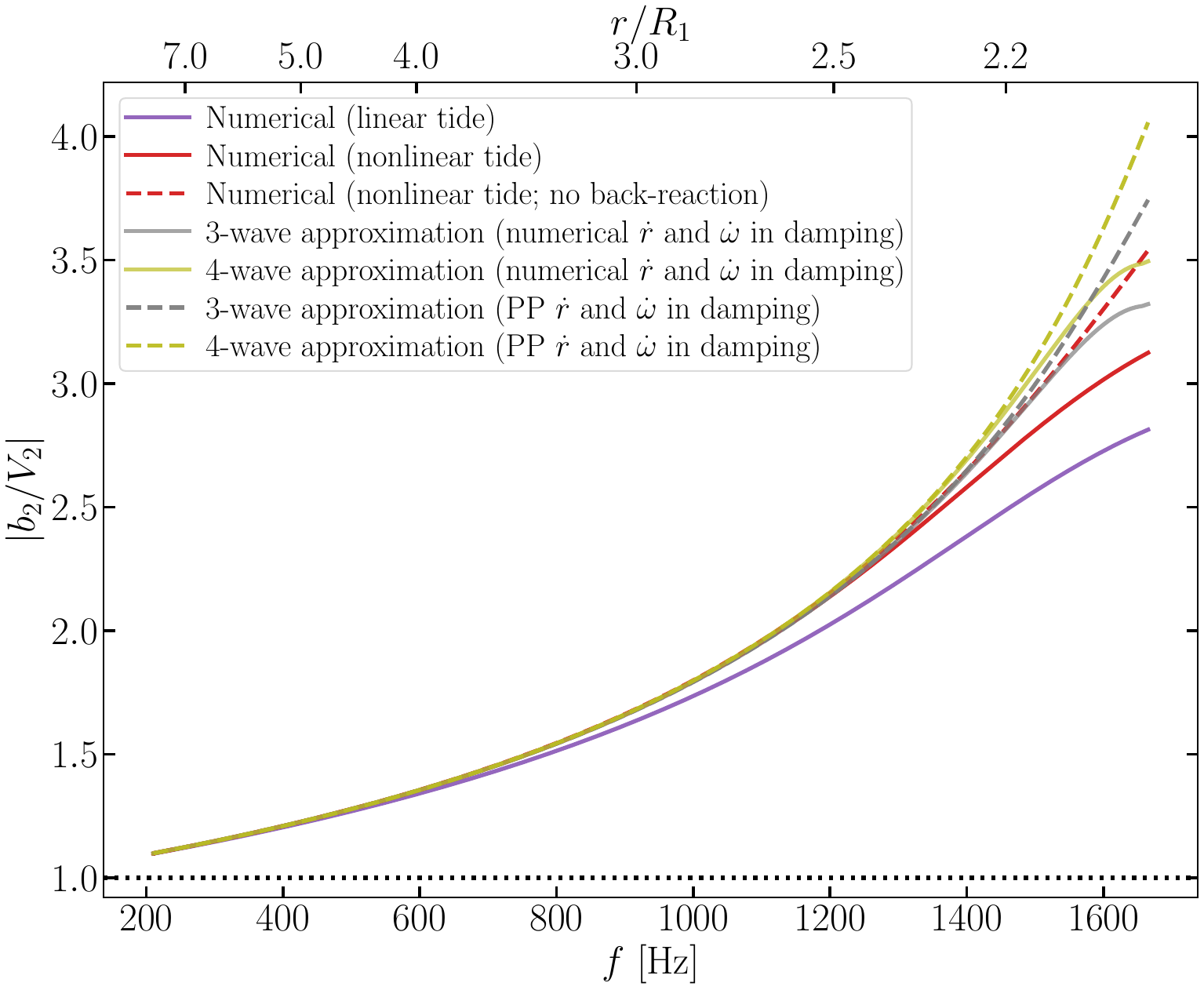}
    \caption{Similar to Fig. \ref{fig:b2_vs_f_N} but for a relativistic $\Gamma=\Gamma_{\rm ad}=2$ polytrope whose parameters are set following Sec. \ref{sec:calibrations}. 
    As tidal coupling is weaker in a relativistic NS than in a Newtonian one, the three-wave (NLO) solution shown in gray lines is sufficient to match the numerical result in red lines. 
    }
    \label{fig:b2_vs_f_G}
\end{figure}

We now compare our analytical solution with that obtained numerically in 
Figs. \ref{fig:b2_vs_f_N} and \ref{fig:b2_vs_f_G}. 
We show the mode amplitude normalized by its adiabatic tide response, $V_2$, as a function of the GW frequency $f=\omega/\pi$. Fig. \ref{fig:b2_vs_f_N} is for a Newtonian $\Gamma=2$ polytrope with coupling coefficients from Sec. \ref{sec:UR_nl_tide_N}, while Fig. \ref{fig:b2_vs_f_G} is a relativistic NS with coefficients scaled to match \cite{Pitre:24, Pitre:25} as described in Sec. \ref{sec:calibrations}.   
The purple line is obtained from numerical integration, assuming a linear tidal response. The red-solid line, on the other hand, has nonlinear interactions included to four-wave order. In the red-dashed line, the tidal back-reaction is turned off in the numerical integration. The gray and yellow lines are analytical approximations to the solution under the equilibrium tide assumption, with dashed lines corresponding to using the PP $\dot{r}$ and $\dot{\omega}$ in computing the effective damping, and solid lines using numerical $\dot{r}$ and $\dot{\omega}$ that also include tidal back-reactions. 

In the Newtonian case, the coupling coefficients are large, and four-wave hydrodynamics is needed to accurately describe the tidal response from $\sim 600-1000\,{\rm Hz}$. Above $1000\,{\rm Hz}$, the mode is close to resonance, and the Lorentzian form assumed in the equilibrium solution loses accuracy and tends to overestimate the amplitude before it goes to zero at resonance (see Eq. \ref{eq:gam_2d} and \cite{Yu:24a}) 
The situation can be mitigated by adding the dynamical component derived in \cite{Yu:24a}, which first cancels the overshooting of the equilibrium tide before it rises at resonance (see the lower panel of fig. 2 in \cite{Yu:24a}). However, as shown in \cite{Yu:24a}, the strong tidal back-reaction near mode resonance due to the tidal torque can dominate over the PP estimation of $\dot{r}$ and $\dot{\omega}$ by an order of magnitude. This can be seen by comparing the solid and dashed lines in Figs. \ref{fig:b2_vs_f_N} and \ref{fig:b2_vs_f_G}. The dynamical tide, therefore, cannot be faithfully built with only the PP estimation of $\dot{r}$ and $\dot{\omega}$. 

When the coefficients are calibrated to the GR calculations of \cite{Pitre:24}, the linear tidal overlap $\If$ is smaller by $\sqrt{k_{2A}/k_{2A}^{\rm (N)}}\simeq 0.46$ for an $n=1$ polytrope, and the nonlinear coupling coefficients satisfy $|\kappa_2| \sim |\JJ_2/\If|\propto\exp(-M_A/R_A)$. Consequently, the tide is much weaker than in the Newtonian limit, significantly reducing the size of nonlinear corrections. 
The four-wave (NNLO) solution does not improve over the three-wave (NLO) one, given that the equilibrium solution characterized by the Lorentzian already overestimates the amplitude near resonance. 
Nonetheless, the three-wave contribution is still a significant, $\mathcal{O}(10\%)$ correction to the linear solution. Even with the reduced coupling coefficients, the tidal effect still dominates the evolution of $\dot{r}$ for a relativistic NS, making it greater than the PP estimation by a factor of $\sim2$.

\subsection{Can f-modes experience resonance locking?}
\label{sec:res_locking}

While we focus on non-resonant systems for our analytical solutions, we nonetheless study, in this and the subsequent subsections, the resonance of f-modes, including nonlinear hydrodynamical interactions. Here, our goal is to investigate, using a combination of theoretical arguments and numerical experiments, whether resonance locking can occur in f-modes.

First, the orbital frequency where the $m_a=2$ f-mode hits resonance, $\omega_{\rm res}$, can be estimated by solving $\varpi^2(\omega)=0$ with $\varpi$ given by Eq. (\ref{eq:detuning_sq}). We use the three-wave correction of $\Delta \omega_2^2/\omega_{f0}^2$ in Line (\ref{eq:dw2_wf_3m}) and replace $\epsilon_A$ by $(M_B/M_t)(\omega^2/\omega_A^2)$. Now  $\varpi^2=0$ is a quadratic equation about $\omega$ whose positive root can be easily obtained as 
\begin{align}
    \omega_{\rm res} &\simeq \frac{(1+C_f)\Omega_A + \frac{1}{2}\sqrt{(1-\mathcal{A}/4)\omega_{f0}^2 + (\mathcal{A} + 2\mathcal{A}C_f+4C_f^2)\Omega_A^2 +(1-\mathcal{A}/4)\mathcal{B}\}}}{1-\mathcal{A} /4},
    \label{eq:f_res}
\end{align}
where 
\begin{align*}
    &\mathcal{A} = 2\sqrt{\frac{\pi}{5}}(\JJ_2 + 4 \kappa_2 \mathcal{I}_f)\frac{M_B}{M_t}\frac{\omega_{f0}^2}{\omega_A^2},\\
    &\mathcal{B} = \left[\frac{4}{3}\sqrt{\frac{\pi}{5}}(\JJ_2 + 4 \kappa_2 \mathcal{I}_f)\frac{M_B}{M_t}\frac{\omega_{f0}^2}{\omega_A^2} - \frac{8\sqrt{\pi}}{3}(\JJ_{r2} + 4\kappa_{2r}\II_r) \frac{\omega_{f0}^2}{\omega_A^2}\right]\tilde{\Omega}_A^2. 
\end{align*}
From this analysis, we estimate that f-mode resonance during the inspiral would require $\Omega_A/2\pi\lesssim-400\,{\rm Hz}$ for an $\Gamma=2$ relativistic polytrope. A stiffer equation of state typically requires a less negative $\Omega_A$ to reach resonance. 

In order for resonance locking to occur, the detuning $\varpi^2$ should be able to first approach zero so that resonance could occur, and then stay approximately constant so that the mode is locked to a resonant state \cite{Burkart:14, Kwon:24, Kwon:25}. As discussed around Eq. (\ref{eq:eff_freq_Lorentzian_3m}), the three-wave (NLO) frequency shift can be viewed as a constant modification of the linear mode frequency, which cannot produce $\varpi^2$ evolution required by resonance locking. Consequently, the four-wave, NNLO frequency shift would be needed. More specifically, only NNLO terms involving $b_2$ could potentially lead to locking, as they are initially small so $\varpi^2$ can approach zero to enable resonance. Once the mode is resonantly excited, the detuning might adjust in a fine manner to maintain the locking. 

Can the process described above occur for the f-mode in a rapidly rotating NS? Given the discussion, we can approximate the NNLO frequency shift as\footnote{
Note that under a canonical description, the background spin frequency $\Omega_A$ is a constant, and all the spin evolution of the NS is captured by the canonical spin $\mathcal{S}_{a,s_a}$. Physically, about half of the angular momentum is carried by the mode, and the rest causes a background spin evolution, as discussed in appendix A of \cite{Yu:25a}. 
We verify in Appendix \ref{appx:Coriolis} of this work that the background spin evolution, corresponding to an axial mode, has a subdominant effect in the evolution of $\varpi^2$. 
}
\begin{align}
    \frac{\Delta \omega_{2}}{\omega_{f0}} 
    &\equiv\frac{1}{2}\frac{\Delta\omega^2_2}{\omega_{f0}^2} \nonumber \\
    &\simeq - [8(\kappa_{2}^2 + \kappa_{2r}^2)+3\zeta_{22}] b_2b_2^\ast
    \equiv \zeta_{\rm eff} b_2b_2^\ast
    \simeq
    -\frac{15}{784\pi\If^2 }\frac{\omega_A^4}{\omega_{f0}^4}
    \left(80 + 147\frac{\omega_{f0}^2}{\omega_r^2}\right)b_2b_2^\ast.
    \label{eq:dw2_4m_E2}
\end{align}
We have also used the fact that $|b_2| \gg \epsilon_A \sim  |b_{-2}|$ near resonance to select the dominant term (see line \ref{eq:dw2_wf_4m_b2sq}). 
Eq. (\ref{eq:dw2_4m_E2}) corresponds to the anharmonic frequency shift of an oscillator that is proportional to the oscillator's energy. 
The last equality uses nonlinear universal relations (Sec. \ref{sec:UR_nl_tide_N}) to explicitly show that $\zeta_{\rm eff}<0$.
On the other hand, to maintain resonance locking after the mode hits resonance, we would need
\begin{align}
    &\frac{d(\varpi^2
    )}{dt}\simeq 0 \nonumber \\
    \implies & \frac{d(\Delta \omega^2_2)}{dt} =2\zeta_{\rm eff} \frac{d}{dt}(b_2b_2^\ast) \simeq 8[\omega-(1+C_f)\Omega_A]\dot{\omega}.
\end{align}
Since both $\omega$ and $b_2b_2^\ast$ increases as the orbit shrinks, $\zeta_{\rm eff}$ would need to be positive in order to lock the mode near resonance. Clearly, this contradicts Eq. (\ref{eq:dw2_4m_E2}) which showed that $\zeta_{\rm eff}<0$. 

As discussed around Eq. (\ref{eq:eff_freq_Lorentzian_3m}), the three-wave (NLO) frequency shift can be viewed as a constant modification of the linear mode frequency, which cannot produce resonance locking. Consequently, the four-wave, NNLO frequency shift is required.  
Not all the NNLO terms are equally important when the f-mode reaches resonance. Only the term in line (\ref{eq:dw2_wf_4m_b2sq}) sees two factors of resonance amplification, making it more significant than other terms. Therefore, we can approximate (note $|b_2|\gg |b_{-2}|$)

More specifically, we note that the direct four-wave interaction (the $\zeta_{22}$ term in Eq. \ref{eq:dw2_4m_E2}) in the Hamiltonian takes a form $-\zeta_{22}c_2c_2^\ast c_2c_2^\ast \propto (\mathcal{S}_{2+}/I_A) \mathcal{S}_{2+}$ (cf. Eq. \ref{eq:tidal_spin}), which is formally similar to the $\Omega_A \mathcal{S}_{2+}$ term in Eq. (\ref{eq:H_mode}) originating from transforming from the corotating frame to the inertial one \cite{Yu:25a}. Because the direct four-wave interaction has $\zeta_{22}<0$, this term increases the f-mode frequency in the inertial frame, as one might expect when the NS spins up due to the tidal torque. The physical background spin also increases, though its impact on $\varpi^2$ is of higher order as discussed in Appendix \ref{appx:Coriolis}. This could be what caused \cite{Kuan:24} to postulate that the f-mode was resonantly locked when $\omega>\omega_{\rm res}$. However, as the tide acquires energy together with angular momentum, it also causes the background to expand, as shown in the three-wave solutions of $c_r$ and $c_0$, lines (\ref{eq:c0_sol_3m}) and (\ref{eq:cr_sol_3m}). When the f-mode propagates in this expanded background, its eigenfrequency is always lowered, as it corresponds to the $-8(\kappa_{2}^2 + \kappa_{2r}^2)$ term in Eq. (\ref{eq:dw2_4m_E2}) and dominates over the direct four-wave coupling. The combined effect, characterized by $\zeta_{\rm eff}<0$, is a lowering of the f-mode's eigenfrequency as the mode's energy increases.

Even if the sign of $\zeta_{\rm eff}$ were flipped to be positive, resonance locking is still prohibited due to the effective damping of the f-mode, Eq. (\ref{eq:gam_2d}). If damping were sufficiently small, the overall tidal field would cause the mode energy $\propto b_2b_2^\ast$ to increase as $\epsilon_A^2\propto \omega^4$, which could then be reduced by slightly increasing the detuning so that the frequency shift would match the evolution of the tidal drive to enable locking \cite{Burkart:14, Kwon:24, Kwon:25}. However, effective damping induced by the orbital decay is always present and can be significant. Suppose a mode were initially at $\varpi^2\simeq \omega_{f0}\gamma_{2d}$, its amplitude would be given by (Eq. \ref{eq:b2_p_bnc} with $|b_{-2}|\ll|b_2|$)
\begin{equation}
    b_2\simeq q_2e^{2i(\phi-\Omega_A t)} \simeq 2W_{2} \If \frac{\omega_{f0}}{\gamma_{2d} + i\gamma_{2d}} \epsilon_A,
\end{equation}
and its energy normalized by $E_A$
\begin{equation}
    b_2b_{2}^\ast \simeq 2 W_{2}^2 \If^2 \frac{\omega_{f0}^2}{\gamma_{2d}^2}\epsilon_A^2.
\end{equation}
In this case, the mode energy is insensitive to $\varpi^2$ but instead set by $\gamma_{\rm 2d}$. 
Even if we adopt a conservative estimate of $\gamma_{2d} \sim \dot{\omega}/\omega_{f0} \propto \omega^{11/3}$, which would significantly underestimate the damping, as Figs. \ref{fig:b2_vs_f_N} and \ref{fig:b2_vs_f_G} have shown, we would still have $b_2b_{2}^\ast \propto \omega^{-10/3}$ at best, insufficient to keep up with the evolution of the orbit. Should resonance locking be developed, $\dot{\omega}$ would increase more due to the rapid energy transfer into the mode, causing even further discrepancy in the predicted and required energy evolution. The situation is different from the resonant locking of g-modes studied in \cite{Kwon:24, Kwon:25}, where there is also a significant ``fggg'' four-wave interaction (a g-mode self-interacting three times and then coupling to an f-mode) that acts as an anti-damping term for the g-mode, corresponding to it extracting an increasing amount of energy from the f-mode to counteract the effective damping of the orbital evolution. For the f-mode, including its coupling with g-modes would enhance the effective damping even further. 

\begin{figure}
    \centering
    \includegraphics[width=0.7\linewidth]{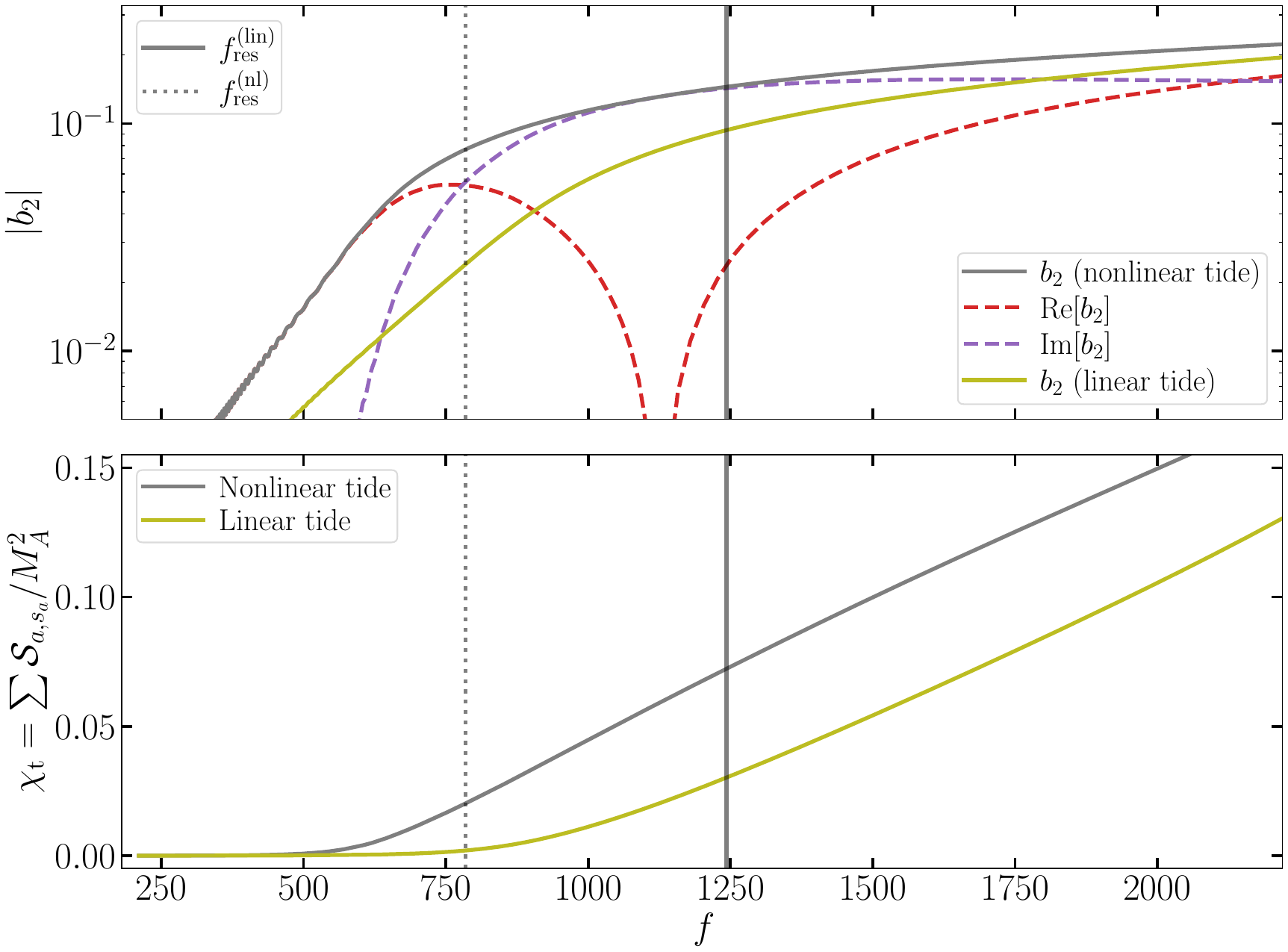}
    \caption{Top: numerical solutions of the $l_a=m_a=2$ f-mode in a rapidly spinning NS with $\Omega_A/2\pi=-800\,{\rm Hz}$. A relativistic $\Gamma=2$ polytrope is assumed for the coupling coefficients. The gray line shows the mode solution including all nonlinear interactions. Its real and imaginary parts are shown respectively in the red-dashed and purple-dashed lines, indicating how the mode oscillates beyond resonance. The linear tide solution is represented by the yellow curve, which is qualitatively similar to the nonlinear case. The two vertical lines indicate the estimated f-mode resonance frequency from linear and three-wave approximations using Eq. (\ref{eq:f_res}). 
    Bottom: Evolution of the tidal spin $\chi_t$. In both the linear and nonlinear cases, the tidal spin evolves roughly linearly with frequency, which is due to the transient response of a driven harmonic oscillator, instead of nonlinear resonance locking.}
    \label{fig:no_res_lock}
\end{figure}

In Fig. \ref{fig:no_res_lock}, we show numerical examples of the $l_a= m_a=2$ f-mode's evolution in a rapidly spinning NS with anti-aligned spin, $\Omega_A/2\pi=-800\,{\rm Hz}$. The top panel shows the evolution of the mode amplitude, where the gray line includes all nonlinear interactions, while the yellow line uses only the linear tide. 
The frequency of resonance are indicated by the two vertical lines. The solid one sets $\mathcal{A}=\mathcal{B}=0$ in Eq. (\ref{eq:f_res}) and corresponds to the resonance of the linear system. The dotted one instead includes nonlinear corrections from the $\mathcal{A}$ and $\mathcal{B}$ terms. 
We let both systems evolve slightly beyond $r=2R_A$ to explore more of the post-resonance regions. 
Qualitatively similar behaviors are observed, indicating that the nonlinear tide introduces only perturbative corrections to the solutions. 
We further show the real and imaginary parts of the nonlinear tide solution of $b_2$ in the dashed curves, which develop oscillatory features (the development of an imaginary part and the sign change of the real part) beyond resonance. 
This is another piece of evidence that the mode evolves through resonance (as the dynamical tide whose phase evolution does not follow the tidal field is excited \cite{Yu:24a}) instead of being locked before resonance (cf. fig. 6 of \cite{Kwon:25}). Note, however, that the mode amplitude reaches only half of its asymptotic value at resonance, and its peak amplitude further overshoots the asymptotic value \cite{Yu:24a} (also \cite{Pnigouras:25}). Therefore, the mode amplitude continues to rise beyond resonance. In terms of the tidal spin (Eq. \ref{eq:tidal_spin} or its dimensionless version normalized by $M_A^2$) plotted in the bottom panel, we observe an approximate linear rise with respect to frequency, similar to what was reported in \cite{Kuan:24}. Such a behavior can already be qualitatively explained with the linear tide model (also fig. 2 of \cite{Yu:25a}). With the nonlinear interaction, the tidal spin is greater by about a factor of 2 at $f=1500\,{\rm Hz}$ (the precise result depends on the equation of state).   

We therefore conclude that resonance locking of the f-mode is highly unlikely. Theoretically, the anharmonicity shifts the f-mode frequency to a lower value as the mode energy increases, in the opposite direction of what is required by resonance locking. Even if the sign were flipped, the presence of effective damping caused by the orbital decay would also prohibit the resonance from being maintained. Numerically, we find qualitatively similar results between the linear and nonlinear solutions, indicating that the nonlinear tide introduces no qualitatively new phenomenon. Importantly, the linear tide can already reproduce a roughly linear growth of the tidal spin post-resonance, as observed in \cite{Kuan:24} without invoking resonance locking (see also \cite{Yu:25a}).

\subsection{Probing $\Gamma_{\rm ad}$ with nonlinear f-mode resonances}
\label{sec:Gamma_ad}

\begin{figure}
    \centering
    \includegraphics[width=0.95\linewidth]{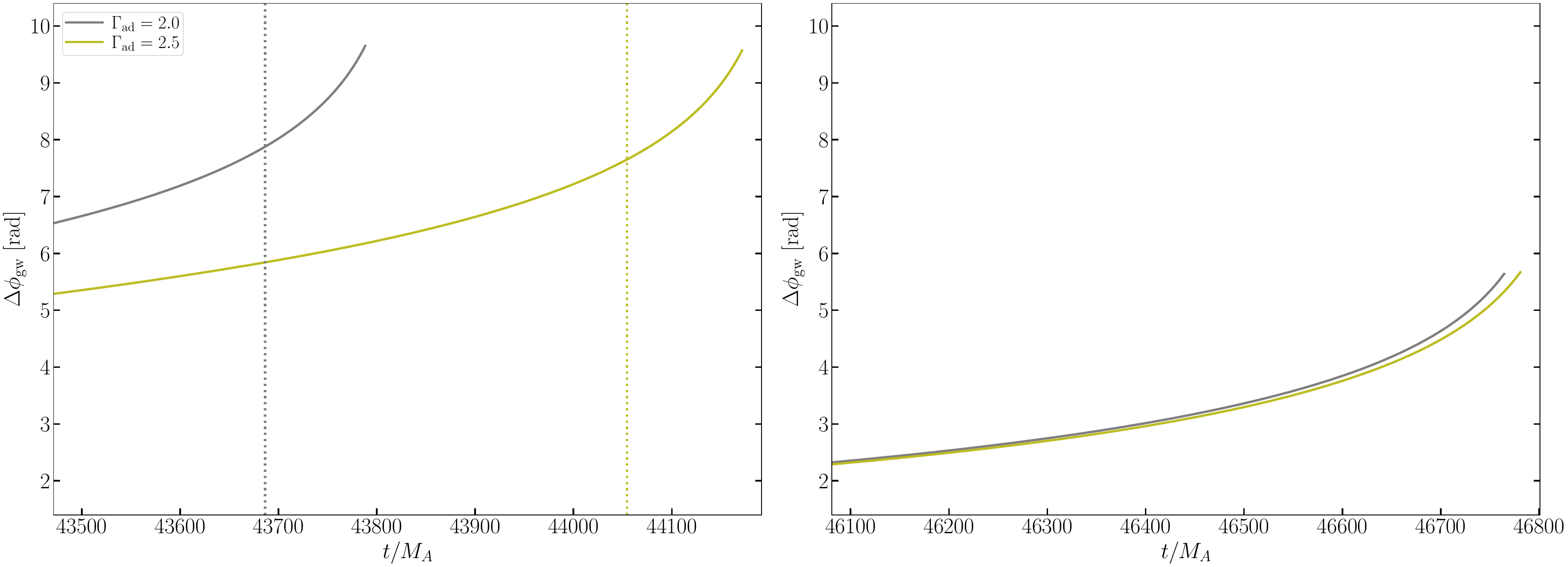}
    \caption{Tidal phase shifts for NS models with different $\Gamma_{\rm ad}$ but otherwise identical parameters. 
    In the left panel, we evolve a rapidly spinning NS with $\Omega_A/2\pi=-800\,{\rm Hz}$. The system shown in the gray line is identical to the one shown in Fig. \ref{fig:no_res_lock} with the same color. The yellow line instead has $\Gamma_{\rm ad}=2.5$.
    The estimated times of resonance, including four-wave corrections, are shown in the vertical dotted lines. 
    Right: for an NS with $\Omega_A/2\pi=-400\,{\rm Hz}$. Resonance of the f-mode does not occur until $r\simeq 2R_A$. The impact of $\Gamma_{\rm ad}$ is much smaller in such a system. 
    }
    \label{fig:dphi_vs_Gam_ad}
\end{figure}

Although nonlinear interactions do not enable resonance locking of an f-mode, they nonetheless introduce important new physics in a rapidly rotating NS with anti-aligned spin, where the f-mode can be resonantly excited. 
Crucially, three-wave interactions in a slowly spinning NS do not introduce new NS physics to be probed beyond what the linear tide already probes. That is, the coupling coefficients for f-modes, $\kappa_2$ and $\JJ_2$, are determined by the linear tidal parameters $\If = \sqrt{5k_{2A}/4\pi}$ and $\omega_{f0}^2$, see Eqs. (\ref{eq:kap2_vs_I_w}) and (\ref{eq:J2_vs_I_w}) and Sec. \ref{sec:UR_nl_tide_N}. 
Nonetheless, the nonlinear forcing of the centrifugal force (the second term in line \ref{eq:dw2_wf_3m}) and the four-wave interaction among f-modes (Eq. \ref{eq:dw2_4m_E2}) do introduce new physics, the adiabatic index $\Gamma_{\rm ad}$, which is related to the buoyancy, or the Brunt–V\"ais\"al\"a frequency squared set by NS thermal and compositional gradients,\footnote{The Brunt–V\"ais\"al\"a frequency can be directly probed by g-modes \cite{Reisenegger:92, Lai:94c, Passamonti:16, Yu:17a, Yu:17b, Andersson:19, Andersson:20} in the linear theory, which are absent in the affine model. Previous studies have suggested that they have small impacts on the GW dephasing \cite{Yu:17b}, unless resonance locking could occur \cite{Kwon:24, Kwon:25}.
In this work, we focus on probing it through the nonlinear couplings of the f-modes. }
\begin{equation}
    \mathcal{N}^2 = g^2\frac{\rho}{P}\left(\frac{1}{\Gamma} - \frac{1}{\Gamma_{\rm ad}}\right) \sim \omega_A^2 \left(\frac{1}{\Gamma} - \frac{1}{\Gamma_{\rm ad}}\right),
\end{equation}
where $g$ is the gravitational acceleration inside the NS. 
In the affine model, $\Gamma_{\rm ad}$ determines $\omega_r^2$ via Eq. (\ref{eq:omega_a}) and enters through (see later in Eq. \ref{eq:k22_w0_W2})
\begin{equation}
    J_{r2} + 2 \frac{\II_r}{\II_f} \JJ_{2r} + 4\kappa_{2r}\II_r \simeq \frac{1}{4\sqrt{\pi}}\left(\frac{2\omega_A^2}{\omega_{f0}^2} + \frac{5\omega_A^2}{\omega_r^2}\right),
    \label{eq:Omegasq_coeff_UR}
\end{equation}
at the lowest order.  

The possibility of probing $\Gamma_{\rm ad}$ is illustrated in \ref{fig:dphi_vs_Gam_ad}, where we compute the tidal shift of the GW phase, $\phi_{\rm gw}(t) = 2\phi(t)$. The y-axis is the dephasing relative to a PP system of the same masses and spins. In the left panel, the NS has $\Omega_A/2\pi=-800\,{\rm Hz}$. The gray line assumes it is described by a $\Gamma=2$, buoyancy neutral ($\Gamma_{\rm ad} = \Gamma$), relativistic polytrope (identical to the system shown in gray in Fig. \ref{fig:no_res_lock}). The system shown in yellow instead has $\Gamma_{\rm ad}=2.5$ but is otherwise identical. This leads to $\mathcal{N}/2\pi\sim 500\,{\rm Hz}$, which is consistent with the peak value inside a normal-fluid NS \cite{Kantor:14}. 
The two systems are aligned at $r=8 R_A$, corresponding to $t=0$. Greater $\Gamma_{\rm ad}$ causes a smaller phase shift at a fixed time and a longer inspiral.
This is because an increasing $\Gamma_{\rm ad}$ makes $\omega_r^2$ greater (Eq. \ref{eq:omega_a}). This delays the f-mode resonance, as shown in the vertical dotted lines. The $>1\,{\rm rad}$ difference in the GW phase indicates the possibility of constraining $\Gamma_{\rm ad}$, a parameter that cannot be probed with the linear f-modes as well as their three-wave (NLO) interactions in slowly spinning systems. 

On the other hand, if the f-mode resonance does not occur over a significant portion of the inspiral, the impact of $\Gamma_{\rm ad}$ will be challenging to measure. This is illustrated in the right panel of Fig. \ref{fig:dphi_vs_Gam_ad}, where the NS's spin is reduced to $\Omega_A/2\pi = -400\,{\rm Hz}$ so that resonance occurs only marginally when $r\simeq2R_A$. In such a system, the phase difference is at a sub-radiant level, and is likely buried by other waveform systematic uncertainties. In particular, values of the coupling coefficients in Eq. (\ref{eq:Omegasq_coeff_UR}) are computed only under Newtonian theory. Finding their values in GR would be necessary to facilitate the extraction of $\Gamma_{\rm ad}$.

\section{Tidal back-reaction on the orbit}
\label{sec:sol_orb}
After describing solutions to the fluid in terms of mode amplitudes, we now describe the tidal back reaction to the orbit. We resume our analytical investigation and restrict our analyses to slow-spinning (no f-mode resonance) NS under the equilibrium solution at the three-wave order (as justified in Fig. \ref{fig:b2_vs_f_G}). 

\subsection{Mass quadrupole and the effective Love number}
\label{sec:effective_Love}

In the GW community, describing the tide in terms of an effective Love number is an especially popular (though its use in general  requires some caution; see Sec. \ref{sec:issue_w_effective_Love}). For a particular $(l, m)$ harmonic, we can define an effective Love number through
\begin{equation}
    Q_{lm}=-k_{lm,{\rm eff}} \lambda_A E_{lm} \text{, or }
    k_{lm,{\rm eff}} = -\frac{Q_{lm} E_{lm}^\ast}{\lambda_AE_{lm}E_{lm}^\ast}.
    \label{eq:klm_eff_def}
\end{equation}
 
Using Eq. (\ref{eq:Q_lm}), we can write the $l=m=2$ component of the mass moment in the inertial frame in terms of mode amplitudes as
\begin{equation}
    \frac{Q_{22}}{M_AR_A^2} = [\If + \mathcal{J}_2(c_0+c_0^\ast) + \mathcal{J}_{2r}(c_r+c_r^\ast) ](c_2+c_{-2}^\ast)
    \label{eq:Q22}
\end{equation}
One can further replace $c_{\pm 2}$ by $b_{\pm 2}$ to get the moment in a frame corotating with the orbit.
Plugging in the equilibrium solutions of the modes from Sec. \ref{sec:nl_mode_amp}, we can write, in a low-frequency expansion,
\begin{equation}
    k_{22,{\rm eff}} = 1 + \sum_{i,j}k_{22,{\rm eff}}^{(\omega^i\Omega_A^j)} \frac{\omega^i\Omega_A^j}{\omega_A^{i+j}}.
    \label{eq:k22_eff}
\end{equation}
Note that in the equation above, we use $i$ and $j$ to indicate the power in $\omega$ and $\Omega_A$, instead of indices of Cartesian coordinates. 
If the background star is not rotating, we have at the $\omega^2$ order
\begin{subequations}
\begin{align}
    k_{22,{\rm eff}}^{(\omega^2\Omega_A^0)} &= 4\frac{\omega_A^2}{\omega_f^2}
    -2\sqrt{\frac{\pi}{5}}(3 \mathcal{J}_2 + 4\kappa_2\If)\frac{M_B}{M_t}, \label{eq:k22_w2_W0}\\
    &\simeq 4\left(1 + \frac{5}{7}\frac{M_B}{M_t}\right) \frac{\omega_A^2}{\omega_f^2},
\end{align}
\end{subequations}
where in the first line, the expression is written in the nonlinear coupling coefficients to assist comparisons with \cite{Yu:23a}, while in the second line, Newtonian nonlinear universal relations have been applied (Sec. \ref{sec:UR_nl_tide_N}). We have replaced $\epsilon_A$ by $(M_B/M_t) (\omega^2/\omega_A^2)$. The error of this replacement will enter at $(\omega/\omega_A)^{16/3}$ in the expansion of $\kappa_{22, {\rm eff}}$.

Including rotation, we have
\begin{align}
    k_{22,{\rm eff}}^{(\omega^1\Omega_A^1)} &= -8(1+C_f)\frac{\omega_A^2}{\omega_{f0}^2},\\
    k_{22,{\rm eff}}^{(\omega^0\Omega_A^2)} &=
    4(1+2C_f)\frac{\omega_A^2}{\omega_{f0}^2} 
    +
    \left[- \frac{4}{3}\sqrt{\frac{\pi}{5}}(3\JJ_2+4\kappa_2\If) + \frac{8\sqrt{\pi}}{3} \left(\JJ_{r2} + 2 \frac{\II_r}{\If}\JJ_{2r} + 4\kappa_{2r}\II_r \right)\right]\frac{\mathcal{\bar{Q}}_A \bar{I}_A^2}{\bar{\lambda}_A}. 
    \label{eq:k22_w0_W2}
\end{align}

The full effective Love number also requires the $m=0$ f-mode's contribution, whose mass quadrupole is given by
\begin{equation}
    \frac{Q_{20}}{M_A R_A^2}=\If(c_0+c_0^\ast) + \frac{1}{2}[-\JJ_2(c_0+c_0^\ast)^2 + 2\JJ_2(c_2+c_{-2}^\ast)(c_{-2}+c_{2}^\ast) + 2\JJ_{2r}(c_0+c_0^\ast)(c_{r}+c_r^\ast)]. 
\end{equation}
Solving the corresponding effective Love number perturbatively to $\epsilon_A^2$, we have
\begin{align}
    k_{20,{\rm eff}}&=  1+ \frac{2}{3} \frac{M_t}{M_B}\frac{\tilde{\Omega}_A^2}{\omega^2} - 2\sqrt{\frac{\pi}{5}}\left(3\JJ_2 + 4\kappa_2\If\right)\frac{M_B}{M_t} \frac{\omega^2}{\omega_A^2} \nonumber \\
    &\quad+\frac{4\sqrt{\pi} }{15}\left(
    3\sqrt{5}\JJ_2 + 4\sqrt{5}\kappa_2\If + 10 \JJ_{r2}  + 40 \kappa_{2r}\II_r+ 20 \frac{\II_r}{\If} \JJ_{2r}
    \right)\frac{\tilde{\Omega}_A^2}{\omega_A^2}\nonumber \\
    &\quad +\frac{4\sqrt{\pi}}{45}\left(
    3\sqrt{5}\JJ_2 + 4\sqrt{5} \kappa_2\If + 20 \JJ_{r2}  + 80 \kappa_{2r} \II_r + 20\frac{\II_r}{\If} \JJ_{2r}
    \right)\frac{M_t}{M_B} \frac{\tilde{\Omega}_A^2}{\omega^2}\frac{\tilde{\Omega}_A^2}{\omega_A^2} 
    \label{eq:k20_eff}
\end{align}
Note that terms involving $(\Omega_A^2/\omega^2)$ would count as $\mathcal{O}(1)$ based on Eq. (\ref{eq:order_counting}). It corresponds to the leading-order spin-induced quadrupole contribution studied in \cite{Poisson:98}. 

The total effective Love number is 
\begin{align}
    k_{r,{\rm eff}}&= \frac{1}{4} k_{20,{\rm eff}} + \frac{3}{8}k_{22,{\rm eff}} + \frac{3}{8}k_{2-2,{\rm eff}} = \frac{1}{4} k_{20,{\rm eff}} + \frac{3}{4}{\rm Re}[k_{22,{\rm eff}}] \nonumber \\
    &\simeq1 + \frac{M_t}{6M_B} \frac{\tilde{\Omega}_A^2}{\omega^2} + \frac{3\omega^2}{\omega_{f0}^2} - \frac{6(1+C_f)\omega\Omega_A}{\omega_{f0}^2} 
    -2\sqrt{\frac{\pi}{5}}(3\JJ_2+4\kappa_2\If)\frac{M_B}{M_t}\frac{\omega^2}{\omega_A^2}.
    \label{eq:k_r_eff}
\end{align}
The last two lines of Eq. (\ref{eq:k20_eff}) were omitted for conciseness. 
The effective Love number introduced this way preserves the value of the interaction potential, satisfying
\begin{equation}
    \frac{1}{2}Q_{ij}E_{ij} = - \frac{1}{2} k_{r, {\rm eff}}\lambda_AE_{ij}E_{ij} = -3 M_B^2 k_{r, {\rm eff}} \frac{\lambda_A}{r^6}. 
\end{equation}
The subscript $r$ in the effective Love number indicates that it also preserves the radial tidal back-reaction; see later in Eq. (\ref{eq:g_r_tide}). Note that $k_{r,{\rm eff}}$ is real but $k_{2\pm2, {\rm eff}}$ are complex due to Eq. (\ref{eq:gam_2d}) being nonzero, meaning the tidal interaction is not purely radial. The tidal torque is $ \propto \sum_m m{\rm Im}[k_{lm,{\rm eff}}]$, which we will discuss further in Sec. \ref{sec:issue_w_effective_Love}.


\subsection{Comparisons with \citeauthor{Pitre:25} \cite{Pitre:25}}
\label{sec:vs_PP25}

We can compare our expression of the effective Love number under the low-frequency, non-spinning limit (Eq. \ref{eq:k22_w2_W0}) to that obtained in a relativistic calculation by \cite{Pitre:25}, especially their eq. (8.8), or 
\begin{equation}
    k_{22, {\rm eff}}^{(\omega^2\Omega_A^0)} = 4 \frac{\ddot{k}_{2A}}{k_{2A}} + \frac{p_{2A}}{k_{2A} } \frac{M_B}{M_t}. 
\end{equation}
We see that the two take the same form, and matching the coefficients directly leads to Eqs. (\ref{eq:ddk2_vs_k2_wf}) and (\ref{eq:p2_vs_J2_kap2}).

\begin{figure}
    \centering
    \includegraphics[width=0.7\linewidth]{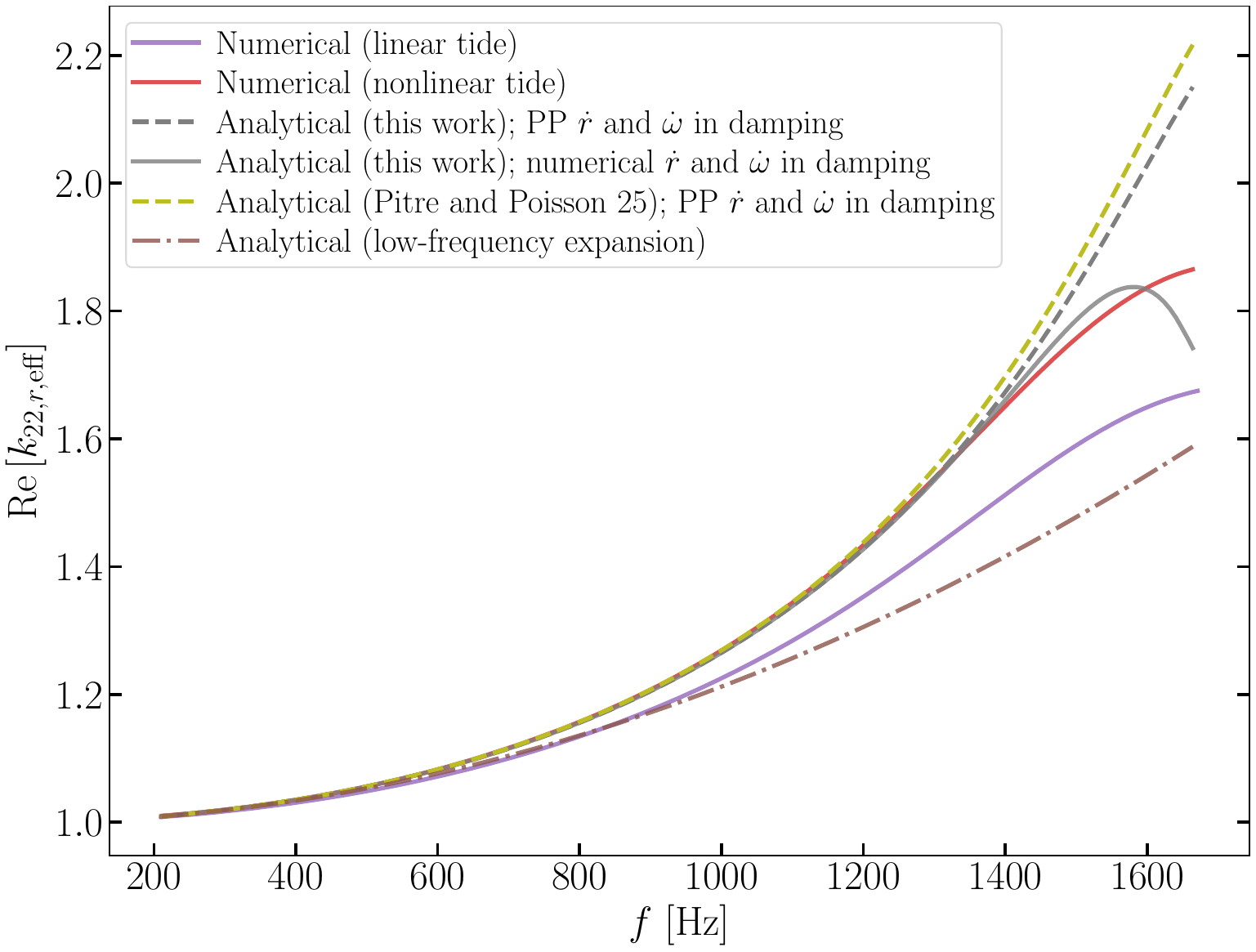}
    \caption{Comparison between analytical and numerical calculations of the effective Love number. The background NS is the same as the one considered in Fig. \ref{fig:b2_vs_f_G}. }
    \label{fig:eff_love}
\end{figure}

When we go beyond the low-frequency limit, Eq. (\ref{eq:Q22}) and the equilibrium solution of the modes, Eq. (\ref{eq:b2_p_bnc}), suggest that
\begin{align}
    \kappa_{22,  {\rm eff}} \simeq \frac{(1 -4\sqrt{\frac{\pi}{5}}\JJ_2\frac{M_B}{M_t} \frac{\omega^2}{\omega_A^2} )}{1- 4\frac{\omega^2}{\omega_{f0}^2}+2\sqrt{\frac{\pi}{5}}(\JJ_2+4\kappa_2\If)\frac{M_B}{M_t}\frac{\omega^2}{\omega_A^2}   + i\frac{\gamma_{2d}}{\omega_{f0}}},
    \label{eq:k22_resum_3m}
\end{align}
where we have plugged in three-wave nonlinear corrections from lines (\ref{eq:dw2_wf_3m}) and (\ref{eq:dV2_3m}) with $\Omega_A=0$, as well as an effective damping from Eq. (\ref{eq:gam_2d}). 
Therefore, we think the resummation done in eq. (8.9) of \cite{Pitre:25}, or
\begin{equation}
    \kappa_{22,  {\rm eff}}^{\rm (PP25)} \simeq
    \frac{1}{1-\frac{4\ddot{k}_{2A} + (M_B/M_t)p_{2A}}{k_{2A}} \frac{\omega^2}{\omega_A^2}}
    =
    \frac{1}{1- 4\frac{\omega^2}{\omega_{f0}^2}+2\sqrt{\frac{\pi}{5}}(3\JJ_2+4\kappa_2\If)\frac{M_B}{M_t}\frac{\omega^2}{\omega_A^2}  },
    \label{eq:k22_resum_3m_PP}
\end{equation}
oversimplifies the problem. The main difference is that, in \cite{Pitre:25}, all the nonlinear corrections are treated as a frequency shift. Our hydrodynamical analysis (following \cite{Yu:23a}), on the other hand, shows that the nonlinear corrections enter through a combination of corrections to the tidal forcing of each mode (the $\Delta V_2$ term), the mapping from mode amplitude to mass quadrupole (Eq. \ref{eq:Q22}), and the shift of the mode's natural frequency (the $\Delta \omega_2$ term). The first two lead to a modification in the numerator of $\kappa_{22,{\rm eff}}$, while only the last one is a correction to the denominator. This is similar to the 1-PN orbital corrections to the tidal interaction. There are corrections to both the mode frequency (the redshift) and the overall coupling strength; see \cite{Steinhoff:16}. Additionally, we also include an effective damping term that is of higher power in $\omega$ but significant in the late inspiral. 

Fig. \ref{fig:eff_love} compares analytical predictions, Eq. (\ref{eq:k22_resum_3m}) in gray (from this work) and Eq. (\ref{eq:k22_resum_3m_PP}) in yellow (from \cite{Pitre:25}), as well as a low-frequency expansion using Eqs. (\ref{eq:k22_eff}) and (\ref{eq:k22_w2_W0}) in the brown-solid-dotted line. 
The numerically extracted effective Love numbers are shown in red (fully nonlinear) and purple (linear tide only), obtained by plugging the numerical solutions of the mode amplitudes in Eq. (\ref{eq:Q_lm}) and then using Eq. (\ref{eq:klm_eff_def}) to extract $k_{22}$. 
The background NS is a non-spinning, relativistic polytrope (same as the one shown in Fig. \ref{fig:b2_vs_f_G}). For a fair comparison, we also include the same effective damping term, $\gamma_{2d}/\omega_{f0}$, in the denominator of Eq. (\ref{eq:k22_resum_3m_PP}). 
As in Fig. \ref{fig:b2_vs_f_G}, we use solid (dashed) lines when numerical values (PP estimations) of $\dot{r}$ and $\dot{\omega}$ are used in $\gamma_{2d}$. 
We see that the low-frequency expansion from combining Eqs. (\ref{eq:k22_eff}) and (\ref{eq:k22_w2_W0}) is insufficient to capture the dynamics in the late inspiral with $f>600\,{\rm Hz}$.  On the other hand, the resummation proposed by \cite{Pitre:25}, or Eq. (\ref{eq:k22_resum_3m_PP}), overestimates the value of $\kappa_{22, {\rm eff}}$, though its deviation from Eq. (\ref{eq:k22_resum_3m}) is small because it enters at NNLO. As discussed around Eq. (\ref{eq:gam_2d}), the effective damping caused by the orbital evolution is in fact a more dominant effect there. Using the numerical $\dot{r}$ and $\dot{\omega}$ is critical as energy transfer through tidal torque is very efficient near the late inspiral and can dominate over the values caused by PP GW evolutions. 

For completeness, we also show a modification to Eq. (\ref{eq:k22_resum_3m}) that accounts for spin to $\Omega_A^2$,  
\begin{equation}
    \kappa_{22, {\rm eff}} \simeq \frac{1 -4\sqrt{\frac{\pi}{5}}\JJ_2(\frac{M_B}{M_t} \frac{\omega^2}{\omega_A^2} + \frac{2}{3}\frac{\tilde{\Omega}_A^2}{\omega_A^2}) + \frac{16\sqrt{\pi}}{3}\frac{\II_r}{\If} \JJ_{2r} \frac{\tilde{\Omega}_A^2}{\omega_A^2}}
    {1 - 4\frac{[\omega-(1+C_f)\Omega_A]^2}{\omega_{f0}^2}+2\sqrt{\frac{\pi}{5}}(\JJ_2+4\kappa_2\If)(\frac{M_B}{M_t}\frac{\omega^2}{\omega_A^2} + \frac{2}{3} \frac{\tilde{\Omega}_A^2}{\omega_A^2}) - \frac{8\sqrt{\pi}}{3}(\JJ_{r2}+4\kappa_{2r}\II_r)\frac{\tilde{\Omega}_A^2}{\omega_A^2}  + i\frac{\gamma_{2d}}{\omega_{f0}}}
    \label{eq:k22_resum_3m_w_spin}.
\end{equation}

\subsection{Inaccurate usage of the effective Love number}
\label{sec:issue_w_effective_Love}
The tidal interaction Hamiltonian can be written as 
\begin{align}
    H_{\rm tide} &= \frac{1}{2}Q_{ij}E_{ij} = \frac{4\pi}{15}\sum_m^{\{2,0,-2\}} Q_{2m}E_{2m}^\ast \nonumber \\
    &=-\sum_m^{\{2,0,-2\}} W_m \left[\If(c_m+c_{-m}^\ast) + \frac{1}{2}\sum_{bc}\JJ_{m,bc}c_b^\ast c_c^\ast \right] \epsilon_A e^{im\phi} \nonumber \\
    &=-\frac{M_B  }{r^3}M_AR_A^2\left\{\If[W_2(c_2+c_{-2}^\ast) e^{2i\phi} +W_2(c_{-2} + c_2^\ast)e^{-2i\phi} + W_0(c_0+c_0^\ast)] \nonumber \right. \\
    &
    \quad +W_2[\JJ_2(c_0+c_0^\ast) + \JJ_{2r}(c_r+c_r^\ast)][(c_{2}+c_{-2}^\ast)e^{2i\phi} 
    + (c_{-2} + c_{2}^\ast) e^{-2i\phi}
    ]
    \nonumber \\
    &\left.
    \quad+W_0[-\frac{1}{2}\JJ_2(c_0+c_0^\ast)^2 + \JJ_2 (c_2+c_{-2}^\ast)(c_{-2}+c_2^\ast)
    +\JJ_{2r}(c_0+c_0^\ast)(c_r+c_r^\ast)
    ]
    \right\},
    \label{eq:H_tide_explicit}
\end{align}
where in the last equality we have explicitly written down the tidal interaction Hamiltonian in terms of canonical variables $r, \phi, c_a, i(E_A/\omega_{a0})c_a^\ast$. 
We can also introduce 
\begin{equation}
    E_{\rm tide} =- \frac{1}{2}k_{r, {\rm eff}}\lambda_AE_{ij}E_{ij} = -3 M_B^2 \kappa_{r, {\rm eff}} \frac{\lambda_A}{r^6}. 
\end{equation}
whose numerical value is the same as $H_{\rm tide}$. However, $H_{\rm tide}$ (a function of canonical variables) and $E_{\rm tide}$ (reduced to be a function of $r$ or $\omega$ only using equilibrium solutions of the modes) have different meanings.

The Hamiltonian generates the equations of motion via its partial derivatives with respect to the canonical variables. 
For example, the radial tidal back-reaction is defined through
\begin{align}
    -\frac{\partial H_{\rm tide}}{\partial r}=\mu g_r^{\rm (tide)} &= -3\frac{M_B}{r^4} M_A R_A^2 
    \left\{2W_2\left(\If + 2\JJ_2{\rm Re}[c_0] + 2\JJ_{2r}{\rm Re}[c_r]\right)({\rm Re}[b_2]+{\rm Re}[b_{-2}]) \nonumber  \right.\\
    &\left.
    +2W_0(\II_f - \JJ_2{\rm Re}[c_0] + 2\JJ_{2r}{\rm Re}[c_r]){\rm Re}[c_0]
    +W_0\JJ_2(b_2+b_{-2}^\ast)(b_{-2} + b_2^\ast)
    \right\} \nonumber \\
    &= -6 \frac{M_B^2}{r^2}\left(\frac{R_A}{r}\right)^5 k_{2A}k_{r,{\rm eff}}. 
    \label{eq:g_r_tide}
\end{align}
Note that $k_{r,{\rm eff}}$ can be used only after taking the partial derivative.
If, however, one directly use $-\partial E_{\rm tide}(r)/\partial r$ computed in terms of the effective Love number $k_{r, {\rm eff}}$, with $k_{r,{\rm eff}}$ treated as a function of $r$ using $\omega^2\simeq M_t/r^3$, the result will be off by a factor of $\simeq2$. 
Moreover, Eq. (\ref{eq:H_tide}) enables a tidal torque from 
\begin{align}
    -\frac{\partial H_{\rm tide}}{\partial \phi} =\mu r g_\phi^{\rm (tide)}
    =\frac{4M_B}{r^3}M_A R_A^2 W_2\left(\II_f + 2\JJ_2{\rm Re}[c_0] + 2\JJ_{2r}{\rm Re}[c_r]\right)
    \left({\rm Im}[b_2] - {\rm Im}[b_{-2}] \right),
    \label{eq:pH_pphi}
\end{align}
which is non-zero because 
\begin{align}
    {\rm Im}[b_2] -{\rm Im}[b_{-2}] \simeq -2i\frac{\omega_{f0}^3 \gamma_{2d}}{\varpi^4} (V_2+\Delta V_2). 
\end{align}
If $E_{\rm tide}(r)$ is used as the tidal interaction Hamiltonian, then it would support no torque because $\partial E_{\rm tide}(r)/\partial\phi =0$. This is because when $Q_{ij}$ is replaced by $-\kappa_{r, {\rm eff}} \lambda_A E_{ij}$ and then $E_{ij}$ contracts with itself, $E_{\rm tide}$ is now a function of $r$ only.
This subtlety of replacing a canonical variable by its equilibrium solution was recognized in the original derivation of \cite{Steinhoff:16}. In subsequent applications, the effective Love number has nonetheless often been used to reduce the tidal interaction Hamiltonian to a function of $r$ alone. This is a convenient simplification whose quantitative consequences have, to our knowledge, not been carefully assessed. As we show here, this reduction misestimates the radial tidal back-reaction by a factor $\simeq 2$ and supports no tidal torque, the latter becoming the dominant channel of energy transfer in the late inspiral (Fig. 8; also see the theoretical study of \cite{Yu:25a} and the numerical simulations of \cite{Kuan:24}).

\begin{figure}
    \centering
    \includegraphics[width=0.7\linewidth]{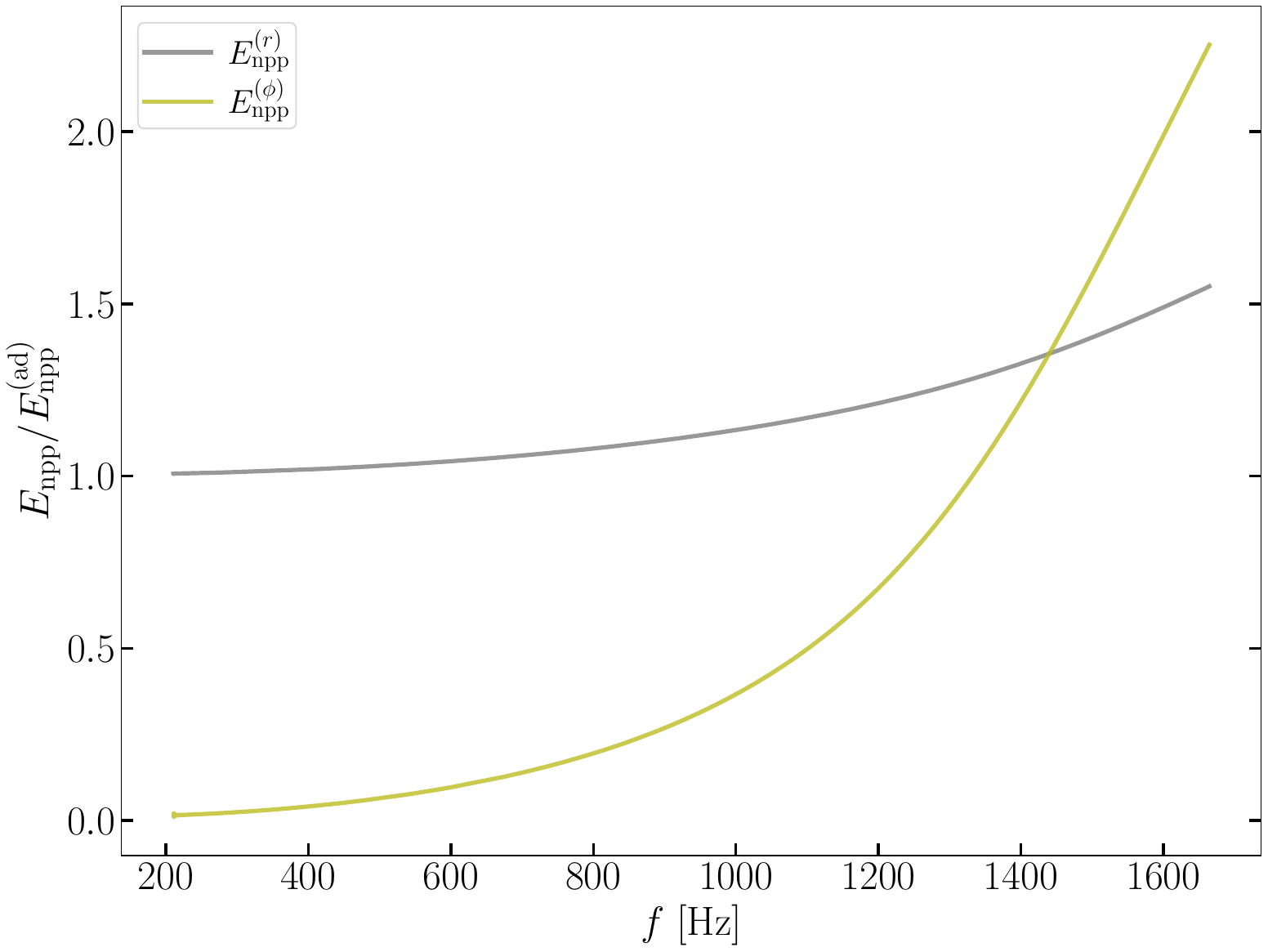}
    \caption{Comparison of non-PP energies driven by radial (gray) and tangential (i.e., tidal torque; yellow) interactions. The values are numerically extracted for a non-spinning $\Gamma=2$ relativistic polytrope. Tidal torque's contribution is initially negligible but becomes significant, even dominant, in the late inspiral. }
    \label{fig:radial_tangential_Enpp}
\end{figure}

More specifically, we note that the orbital energy, $E_{\rm orb}$, changes due to the tide as \cite{Yu:24a} 
\begin{equation}
    \dot{E}_{\rm orb}^{\rm (tide)} = \mu \dot{r} g_{r}^{\rm (tide)} + \mu\omega r g_{\phi}^{\rm (tide)}. 
\end{equation}
In the absence of GW radiation, one can verify that $-\dot{E}_{\rm orb}^{\rm (tide)}=\dot{E}_{\rm npp}$ as expected by energy conservation, where
\begin{equation}
    E_{\rm npp} = (H_{\rm tide} + H_{\rm mode} + H_{\rm cen})|_{c_a=c_a(\omega)} = E_{\rm tide} + E_{\rm ns},
\end{equation}
is the sum of all non-PP energies, including tidal interaction, internal modes, and centrifugal (cf. eq. 62 of \cite{Yu:24a}).  
We further group the latter two as $E_{\rm ns}$. 
The $c_a=c_a(\omega)$ part in the subscript means the mode amplitudes are replaced by their equilibrium solutions at a particular $\omega$ (or $r$; Sec. \ref{sec:nl_mode_amp}). 
While each energy term has the same value as the corresponding Hamiltonian component, we intentionally denote them differently to highlight their different roles as discussed above.  
We further split $E_{\rm npp}$ according to
\begin{align}
    E_{\rm npp}^{(r)} (\omega) = -\int \frac{\mu \dot{r} g_{r}^{\rm (tide)}}{\dot{\omega}} d\omega, \text{\ \ \ and\ \ \ }E_{\rm npp}^{(\phi)} (\omega) = -\int \frac{\mu \omega r g_{\phi}^{\rm (tide)}}{\dot{\omega}} d\omega,
\end{align}
to examine the significance of the radial and tangential interactions. 
In Fig. \ref{fig:radial_tangential_Enpp}, we compare each term's evolution as a function of the GW frequency for a nonspinning $\Gamma=2$ relativistic polytrope. When $\omega \ll \omega_{\rm res}$, the mode acquires nearly all of its energy through the radial interaction. However, the contribution from the torque rises sharply after 1000 Hz and eventually becomes dominant before the merger.

\subsection{GW phase shift from energy balance}
\label{sec:GW_phase}

\begin{figure}
    \centering
    \includegraphics[width=0.95\linewidth]{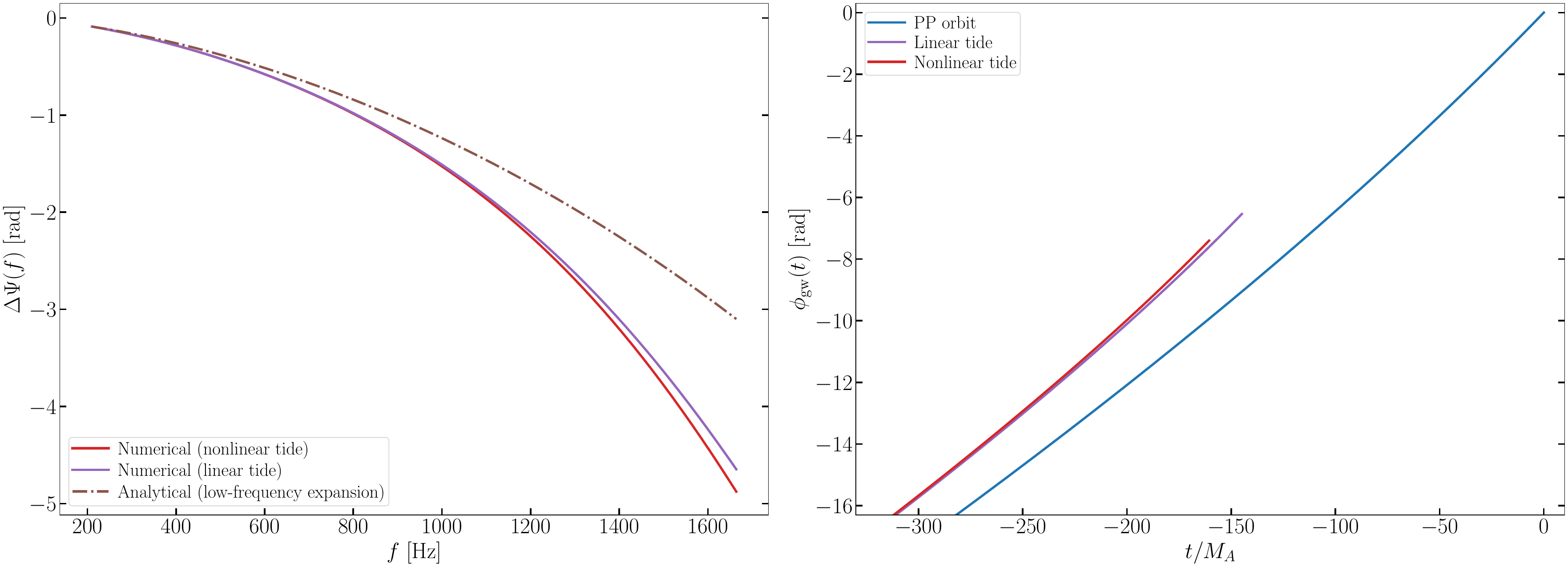}
    \caption{GW phase shift for a relativistic $\Gamma=2$, non-spinning polytrope. 
    The left panel shows the frequency-domain GW phase shift relative to a PP orbit.
    The red and purple lines show the phase shift with and without the nonlinear tide. The brown-dashed line shows the low-frequency limit including NLO nonlinear tide (Eqs. \ref{eq:phase_shift_LF} and \ref{eq:k_psi_eff}), which significantly underestimates the result. 
    The right panel shows the time-domain GW phases (blue for PP, purple for linear tide, and red for nonlinear tide), which demonstrates that the actual effect of the nonlinear tide (red) is more significant than what the left panel shows.
    At the last time point where the system with the nonlinear tide merges ($t/M_A\simeq-160$), the nonlinear tide causes a phase shift of 0.22 rad relative to the linear tide system, same as the phase shift at the last frequency data point in the left up to a minus sign).
    However, the total GW phase accumulated to the merger ($r=2R_A$) is less by 0.81 rad due to the nonlinear tide. 
    }
    \label{fig:phase_shift}
\end{figure}

\begin{figure}
    \centering
    \includegraphics[width=0.95\linewidth]{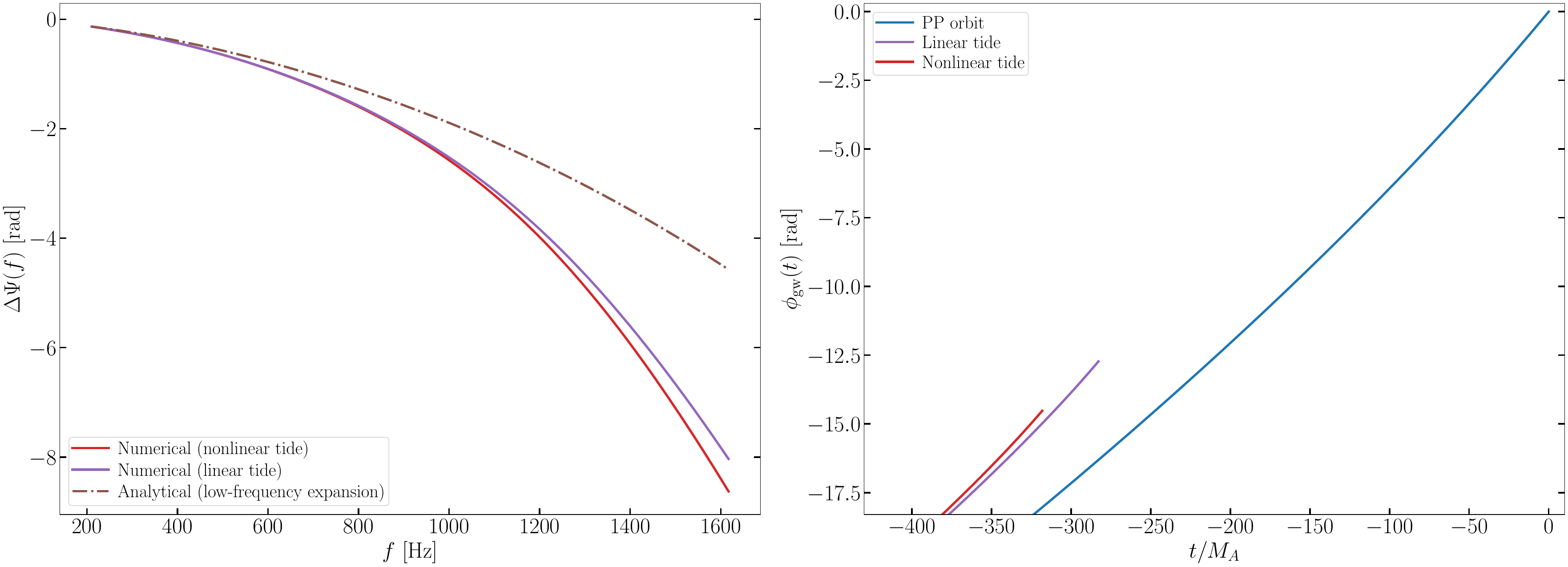}
    \caption{Similar to Fig. \ref{fig:phase_shift} but $M_A$ is now assumed to be described by the SLy equation of state with $(k_{2A}, \ddot{k}_{2A}, p_{2A})=(0.084, 0.060, 0.086)$. 
    Compared to the linear tide prediction, the nonlinear tidal response of a single NS causes a phase shift of 0.59 rad measured at the time when the nonlinear system merges, and 1.8 rad when considering the total accumulated phase to $r=2R_A$.  
    }
    \label{fig:phase_shift_SLy}
\end{figure}

\begin{figure}
    \centering
    \includegraphics[width=0.95\linewidth]{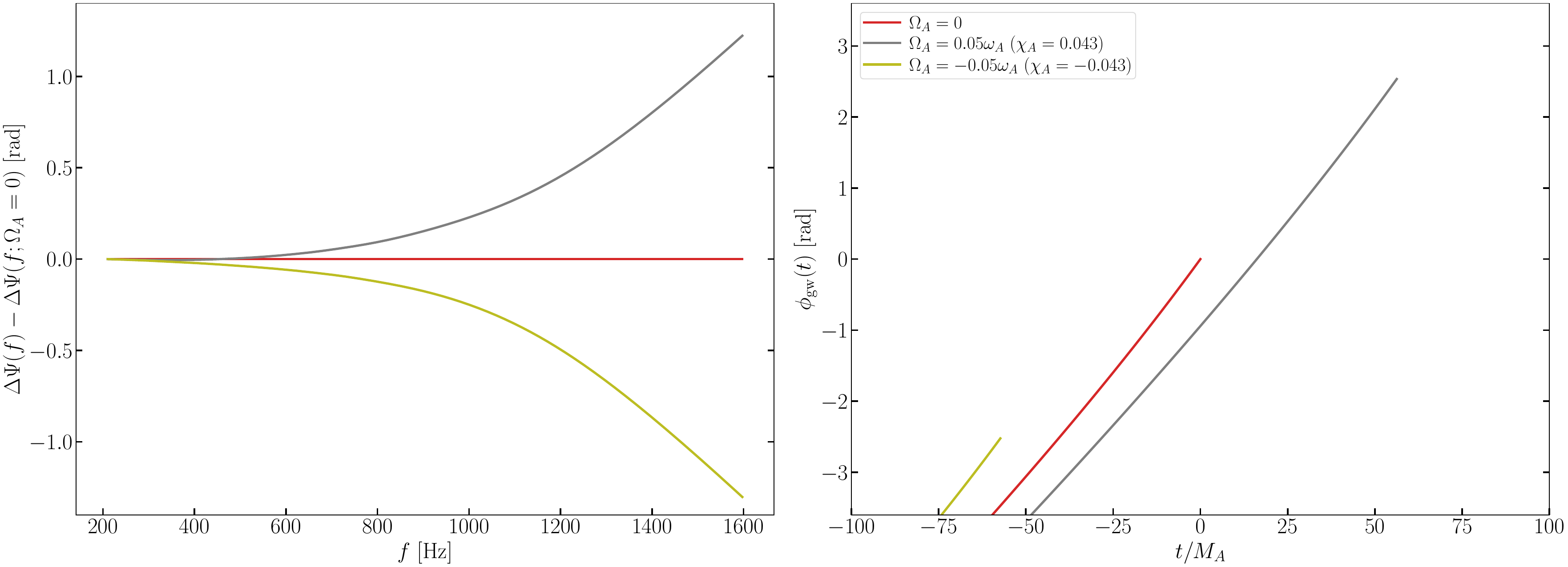}
    \caption{GW phase shift caused by relativistic SLy NSs (parameters same as Fig. \ref{fig:phase_shift_SLy}) but with different spins (with colors gray, red, and yellow for positive, zero, and negative spins). Nonlinear tides are included in all models. 
    The left panel shows the frequency-domain phase shift relative to the non-spinning system.
    The right panel shows the time-domain phase shift near the merger. 
    For slowly rotating NSs, a small change in the spin rate of $|\Omega_A|/\omega_A\simeq0.05$ ($|\chi_A|\simeq 0.04$) leads to a total phase shift similar to that caused by the nonlinear tide. 
    }
    \label{fig:phase_shift_spin}
\end{figure}

One may bypass the subtleties in computing the tidal back reaction by instead computing the phase shift from an energy-balance argument \cite{Lai:93, Lai:94a, Lai:94b, Lai:94c, Flanagan:08, Yu:23a}. In particular, the duration of inspiral can be computed by
\begin{equation}
    t(f) = \int \frac{dE/df}{\dot{E}} df, 
\end{equation}
where $E$ is the total energy of the system and $\dot{E}$ the total energy loss rate.
At the three-wave level (NLO), and before f-mode resonance, one may compute the time shift caused by the tide as linear perturbations to $E$ and $\dot{E}$, with
\begin{equation}
    \Delta t\simeq \int \left(\frac{d\Delta E/df}{\dot{E}_{\rm pp}} - \frac{dE_{\rm pp}/df}{\dot{E}_{\rm pp}^2}\Delta \dot{E} \right)df,
    \label{eq:delta_t_vs_f}
\end{equation}
where $\Delta E=E-E_{\rm pp}$ is the energy change relative to a PP binary of the same mass and spin, and similarly for $\Delta \dot{E}$. Consistent with Eqs. (\ref{eq:dwdt_pp})  and (\ref{eq:drdt_pp}), in this work we estimate $\dot{E}_{\rm pp}/E_{\rm pp}\simeq (2/3)\dot{\omega}_{\rm pp}/\omega$. 
The accumulated GW phase can be computed from $d\Delta \phi_{\rm gw}/df =2\pi fd\Delta t/df$, which, upon integration over $df$, leads to $\Delta \phi_{\rm gw}(f)$, or the time-domain phase shift computed at a given frequency. More naturally entering the data analysis is the time-domain phase shift computed at a given time, $\Delta \phi_{\rm gw}(t)$, which, up to a minus sign, is the same as the frequency-domain phase shift computed at a given frequency, $\Delta \Psi(f) \simeq -\Delta \phi_{\rm gw}(t)$ (see eq. 94 of \cite{Yu:24a}). Under a stationary phase approximation, we have
\begin{equation}
    \Delta \Psi(f) \simeq 2\pi f \Delta t(f) - \Delta \phi_{\rm gw}(f). 
    \label{eq:delta_Psi}
\end{equation}
Note, however, $\Delta \phi_{\rm gw}(t)$ is a meaningful measurement only over times when both waveforms (e.g., tidal vs. PP, or linear vs. nonlinear) are defined. There is an additional effect that two waveforms being compared typically have different durations,\footnote{Note that this corresponds to a time (or phase) shift measured at the same orbital separation (e.g., the location of merger with $r\simeq R_A+R_B\simeq 2R_A$ for a BNS), which is different from the $\Delta t$ from Eq. (\ref{eq:delta_t_vs_f}) measured at the same frequency.} a point we will illustrate further in Figs. \ref{fig:phase_shift} to \ref{fig:phase_shift_spin} later in this section. 

The rest of the calculation becomes finding $\Delta E$ and $\Delta \dot{E}$, whose dependence on mode amplitudes has been worked out in \cite{Yu:23a} to three-wave order. 
One can then substitute the improved mode amplitude solutions derived in this work (Sec. \ref{sec:eq_sol_hydro}) to obtain closed-form expressions for $\Delta \Psi$. 
In this section, we further provide low-frequency expansions of the intermediate steps to assist the development of analytical intuitions. 

First, we compute the quasi-circular separation at a given $\omega$ by solving $\partial H/\partial r=0$, or 
\begin{align}
    \frac{\Delta r}{r}&\simeq \frac{1}{3}\frac{M_t}{\mu} \frac{1}{(M_t\omega)^{4/3}}\frac{\partial H_{\rm tide}}{\partial r} \nonumber \\
    &=3 \frac{M_A^4M_B}{M_t^5} (M_t\omega)^{10/3} \frac{\lambda_A}{M_A^5} k_{r,{\rm eff}}(\omega), 
    \label{eq:delta_r_r}
\end{align}
where one can use Eqs. (\ref{eq:k22_resum_3m_w_spin}) and (\ref{eq:k20_eff}) to compute $k_{r, {\rm eff}}$ with high accuracy (Fig. \ref{fig:eff_love}). 
We also provide the explicit form in the low-frequency limit, as
\begin{equation}
    k_{r, {\rm eff}}\simeq 1 + \frac{M_t^3}{6M_A^2M_B} \frac{\bar{\mathcal{Q}}_A}{\bar{\lambda}_A}\frac{ \chi_A^2}{(M_t \omega)^2} + \frac{3\omega^2}{\omega_{f0}^2} - \frac{6(1+C_f)\omega\Omega_A}{\omega_{f0}^2} 
    +\frac{M_B}{M_t}\frac{p_{2A}}{k_{2A}}\frac{\omega^2}{\omega_A^2}.
\end{equation}
In this section, we replace $\tilde{\Omega}_A^2$ by $(\bar{\mathcal{Q}}_A/\bar{\lambda}_A)(\chi_A^2/M_A^2)$ so that our result matches the spin-induced quadrupole computed in GR \cite{Poisson:98, Harry:18}, where $\chi_A\simeq I_A \Omega_A/M_A^2$ is the dimensionless spin. 
We drop $\Omega_A^2$ terms entering at NLO ($\sim \epsilon_A$ in $k_{r, {\rm eff}}$) for conciseness. Those terms are already derived in Secs. \ref{sec:nl_mode_amp} and \ref{sec:effective_Love}. 

Once the equilibrium separation is obtained, the orbital energy change is \cite{Yu:23a}
\begin{equation}
    \frac{\Delta E_{\rm orb}}{E_{\rm pp}} = -4\frac{\Delta r}{r}.
\end{equation}
This can be seen by noticing that the kinetic part of the orbital energy scales as $\propto r^2$ and the potential part as $\propto 1/r$. 
The energy in the tidal interaction is 
\begin{equation}
    \frac{E_{\rm tide}}{E_{\rm pp}} = 2\frac{\Delta r}{r}, 
\end{equation}
which follows directly from Eq. (\ref{eq:delta_r_r}) as $H_{\rm tide}\propto 1/r^3$.  
We only need to find the energy internal to the NS (mode plus centrifugal), given by
\begin{align}
    E_{\rm ns} &= (H_{\rm mode} + H_{\rm cen})|_{c_a = c_a(\omega)}\nonumber \\
    &\simeq\frac{3}{2}\left(\frac{M_B}{M_t}\right)^2\lambda_A\omega^4 
    \left[1+\frac{3\omega(3\omega-4\Omega_A-4C_f\Omega_A)}{\omega_{f0}^2} 
    +\frac{4}{3}\frac{M_B}{M_t}\frac{p_{2A}}{k_{2A}}\frac{\omega^2}{\omega_A^2}
    \right].
\end{align}
For an analytical solution with high accuracy, mode solutions from Sec. \ref{sec:nl_mode_amp} should be plugged into $E_{\rm ns}$. The second equality is provided for an intuitive interpretation of the results. 
Total energy change is thus 
\begin{align}
    \frac{\Delta E}{E_{\rm pp}}= \frac{\Delta E_{\rm orb} + E_{\rm tide} + E_{\rm ns}}{E_{\rm pp}} = -9\frac{M_A^4 M_B}{M_t^5}(M_t\omega)^{10/3}\frac{\lambda_A}{M_A^5} k_{E,{\rm eff}}(\omega),
    \label{eq:Delta_E}
\end{align}
where, in the low-frequency limit, 
\begin{equation}
    k_{E,{\rm eff}}(\omega)\simeq 1+ \frac{M_t^3}{9M_A^2M_B}\frac{\bar{\mathcal{Q}}_A}{\bar{\lambda}_A}\frac{\chi_A^2}{(M_t\omega)^2} 
    + \frac{\omega[5\omega-8(1+C_f)\Omega_A]}{\omega_{f0}^2} + \frac{10}{9}\frac{M_B}{M_t}\frac{p_{2A}}{k_{2A}}\frac{\omega^2}{\omega_A^2}.
\end{equation}

We use the quadrupole formula to estimate the energy loss from the system $\dot{E}$, as 
\begin{equation}
    \dot{E} = -\frac{1}{5}\langle \dddot{Q}_{ij}^{(\rm sys)}\dddot{Q}_{ij}^{(\rm sys)} \rangle 
    \simeq -\frac{1024\pi}{75} \omega^6 Q_{2-2}^{(\rm sys)}Q_{22}^{(\rm sys)},
    \label{eq:dot_E_sys}
\end{equation}
where $Q_{ij}^{\rm (sys)}=Q_{ij}^{\rm (orb)}+Q_{ij}^{\rm (ns)}$ is the total system mass quadrupole tensor, including both orbital and NS contributions, and $Q_{22}^{\rm (sys)}=\mathcal{Y}_{22}^{ij} Q_{ij}^{\rm (sys)}=Q_{2-2}^{\ast \rm (sys)}$ its $l=m=2$ harmonic. In the last equality, we have assumed $Q_{lm}^{\rm (orb)}\sim Q_{lm}^{\rm (ns)}\sim \exp[-im\phi]$, which is a reasonable assumption when focusing on the equilibrium tide, as it is defined to be the component that is phase-coherent with the orbit. When the dynamical tide is present, $Q_{lm}^{\rm (ns)}$ would vary at the mode's eigenfrequency and does not contribute to the energy loss at $\mathcal{O}(\epsilon_A)$, which is the main reason behind the separation adopted by \cite{Yu:24a}.

When tide is present, the system quadrupole is perturbed in two ways. First, the perturbed NS itself carries a mass quadrupole $Q_{22}^{\rm (ns)}$, which introduces
\begin{equation}
    \frac{\Delta \dot{E}_{\rm ns}}{\dot{E}_{\rm pp}} = 2 \frac{{\rm Re}[Q_{22}^{\rm (ns)}Q_{\rm 2-2}^{\rm (orb)}]}{Q_{\rm 22}^{\rm (orb)}Q_{\rm 2-2}^{\rm (orb)}}=6 \frac{M_A^4}{M_t^4} (M_t\omega)^{10/3} \frac{\lambda_A}{M_A^5}k_{22,{\rm eff}}(\omega).  
\end{equation}
Meanwhile, the orbital quadrupole is perturbed because the orbital separation is changed by $\Delta r$, leading to 
\begin{equation}
    \frac{\Delta \dot{E}_{\rm orb}}{\dot{E}_{\rm pp}}=2 \frac{\Delta Q_{22}^{\rm (orb)}}{Q^{\rm (orb,pp)}_{22}} = 4 \frac{\Delta r}{r}. 
\end{equation}
The total change in the energy loss rate is the sum of the two, 
\begin{align}
    \frac{\Delta \dot{E}}{\dot{E}_{\rm pp}} = 
    6\frac{M_A^4}{M_t^4}\frac{\lambda_A}{M_A^5}(M_t\omega)^{10/3}\left[k_{22,{\rm eff}}(\omega) + 2\frac{M_B}{M_t}k_{r, {\rm eff}}(\omega)\right]
    = 6\frac{M_A^4}{M_t^4}\left(1 + 2\frac{M_B}{M_t}\right)\frac{\lambda_A}{M_A^5}(M_t\omega)^{10/3} k_{h, {\rm eff}},
    \label{eq:Delta_dot_E}
\end{align}
where 
\begin{equation}
    k_{h, {\rm eff}} \simeq 1 + \frac{M_t^3}{3M_A^2(M_t + 2M_B)}\frac{\bar{\mathcal{Q}}_A}{\bar{\lambda}_A}\frac{\chi_A^2}{(M_t\omega)^2}
    +\frac{2(2+3M_B/M_t)\omega[\omega-2(1+C_f)\Omega_A]}{(1+2M_B/M_t)\omega_{f0}^2}
    +\frac{M_B}{M_t}\frac{p_{2A}}{k_{2A}}\frac{\omega^2}{\omega_A^2}
\end{equation}
is an effective Love number that preserves the form of energy loss from its adiabatic limit (see eq. 145 of \cite{Yu:25a}). 

With these ingredients, we are ready to obtain
\begin{align}
    \Delta\Psi(\omega) = -\frac{9}{16}\frac{M_A^3}{M_BM_t^3}\left(11M_B + M_t\right)\frac{\lambda_A}{M_A^5}(M_t\omega)^{5/3} k_{\Psi,{\rm eff}}(\omega),
    \label{eq:phase_shift_LF}
\end{align}
where 
\begin{align}
    k_{\Psi, {\rm eff}} \simeq& 1 
    + \frac{25 M_t^3}{12 M_A^2(11M_B+M_t)}\frac{\bar{\mathcal{Q}}_A}{\bar{\lambda}_A} \frac{\chi_A^2}{(M_t\omega)^2}
    -\frac{(1+C_f)(51 M_B + 4 M_t)}{2(11 M_B + M_t)}\frac{\omega\Omega_A}{\omega_{f0}^2} \nonumber \\    
    &+\frac{5(147 M_B + 8 M_t)}{88(11M_B + M_t)} \frac{\omega^2}{\omega_{f0}^2}
    +\frac{5M_B(17M_B+M_t)}{44 M_t(11M_B+M_t)}\frac{p_{2A}}{k_{2A}} \frac{\omega^2}{\omega_A^2}, 
    \label{eq:k_psi_eff}
\end{align}
is the low-frequency limit of an effective Love number that preserves the form of adiabatic tidal dephasing. 
Note that the three-wave (NLO) nonlinear tide's contribution (the $p_{2A}$ term) to the phase shift can also be written as 
\begin{align}
    \Delta \Psi_{\rm NLO, LF}= - \frac{15}{352}\frac{M_A^5 (17 M_B+M_t)}{M_t^6} \bar{p}_{2A} (M_t\omega)^{11/3}. 
    \label{eq:dPsi_LF_barp2A}
\end{align}
This is why $\bar{p}_{2A} = p_{2A}/(M_A/R_A)^8$ is introduced in Sec. \ref{sec:UR_in_GR} and a practical quasi-universal relation is provided in Eq. (\ref{eq:p2bar_vs_lambar}). 
The NLO nonlinear tide is degenerate with the linear dynamical tide in the low-frequency expansion. One may define an ``effective'' f-mode frequency, $\omega_{f,\ {\rm LF\ eff}}^2$, that absorbs the nonlinear correction (the $p_{2A}$ term) and preserves the linear tide form in the low-frequency expansion, such that (ignoring spin terms)
\begin{equation}
    k_{\Psi, {\rm eff}} \simeq 1 +\frac{5(147 M_B + 8 M_t)}{88(11M_B + M_t)} \frac{\omega^2}{\omega_{f, {\rm LF, eff}}^2},
\end{equation}
with
\begin{equation}
    \frac{\omega_{f,\, {\rm LF\ eff}}^2}{\omega_{f0}^2} = \left[1+\frac{2M_B(17M_B+M_t)}{M_t(147 M_B+8 M_t)} \frac{p_{2A}}{k_{2A}}\frac{\omega_{f0}^2}{\omega_A^2}\right]^{-1} \simeq \left[1+\frac{40M_B(17M_B+M_t)}{7M_t(147 M_B+8 M_t)}\exp(-4M_A/R_A) \right]^{-1},
    \label{eq:omega_f_LF_eff}
\end{equation}
In the last equality, we have used the universal relation, Eq. (\ref{eq:p2_vs_ddk2_GR}). For a typical BNS systems with $M_B=M_A$, $M_A/R_A\simeq 0.17$, and $\Omega_A=0$, $\omega_{f,\, {\rm LF\ eff}}^2/\omega_{f0}^2\simeq 0.85$. 

Importantly, it is a common practice in the existing waveforms to ignore the nonlinear hydrodynamics while using the linear $\omega_{f0}-\lambda_A$ quasi-universal relation \cite{Chan:14} to eliminate $\omega_{f0}$. This process introduces a systematic bias, because what is measured, in the low-frequency limit, is not $\omega_{f0}$ but instead $\omega_{f,\,{\rm LF\,eff}}$ and this effective frequency does not follow the same relation with $\lambda_A$. See \cite{Bretz:26} for a preliminary investigation of the biases induced. 
 
Note, however, that $\omega_{f,\, {\rm LF\ eff}}^2$ is not the actual nonlinearly corrected frequency of the $l_a=m_a=2$ f-mode, as it mixes the numerator $\Delta V_2$ and denominator $\Delta \omega_2$ corrections to the Lorentzian in the low-frequency expansion (see also Sec. \ref{sec:vs_PP25}). The appropriate frequency shift beyond the low-frequency expansion is given by line (\ref{eq:dw2_wf}). Related to the separation between $\Delta V_2$ and $\Delta \omega_{2}$, it is also important to note that while in the low-frequency limit, the nonlinear tide enters only with a single coefficient $\bar{p}_{2A}$ (and masses), a more appropriate treatment should separate the $\JJ_2$ and $\kappa_2\If$ contributions. Nonetheless, Eqs. (\ref{eq:kap2_vs_I_w}) and (\ref{eq:J2_vs_I_w}) show that they probe the same NS physics. 

We now present resultant phase shifts. In this section, we consider $M_A$ described by both the relativistic $\Gamma=2$ polytrope (very soft with very weak tidal perturbations), and also the SLy equation of state (medium soft with stronger, potentially more realistic, tidal deformations). We also remind the readers that $R_A=11.7\,{\rm km}$ in both cases, and $M_B$ is treated as a PP throughout the analysis, so the phase shift for a realistic BNS with $M_B\simeq M_A$ will be about twice as large as shown in this work. 

Our most conservative estimate of the tide is presented in Fig. \ref{fig:phase_shift} for a nonspinning, relativistic $\Gamma=2$ polytrope. 
In the left panel, we consider $\Delta \Psi(f)$, with numerical phase shifts presented in the red (including four-wave, NNLO, nonlinear tides) and purple (linear tide only) lines, and the low-frequency expansion (Eqs. \ref{eq:phase_shift_LF} and \ref{eq:k_psi_eff}; see also \cite{Pani:25}) shown in the brown-solid-dotted line. As in Fig. \ref{fig:eff_love}, the low-frequency limit significantly underestimates the tidal impact. We present it in the analytical derivation only to show the limitation of using the linear $\lambda_A-\omega_{f0}$ relation explicitly in Eq. (\ref{eq:omega_f_LF_eff}). Even in this nonspinning case where the f-mode never reaches resonance, an appropriate Lorentzian form is still needed to capture the dynamics. 
At $f=1664\,{\rm Hz}$ (the merger frequency of the nonlinear tide system), the nonlinear tide introduces a correction of $\simeq0.23\,{\rm rad}$ relative to the linear tide model from a single NS. 
Nonetheless, the complete impact of the nonlinear tide may be more significant. Note $\Delta \Psi(f)\simeq -\Delta \phi_{\rm gw}(t)$ \cite{Yu:24a} is well-defined only when both waveforms being compared are defined. Meanwhile, different waveforms also have different durations, as the right panel demonstrates. Instead of showing shifts relative to the PP orbit, we present here the full time-domain GW phase $\phi_{\rm gw}$ as a function of time. The GW phase of a PP orbit is shown in the blue line. The same initial time and phase shifts are applied to all systems such that the PP system merges at $t=0$ with $\phi_{\rm gw}=0$. Note that the phase shift measured at the last frequency point in the left panel corresponds to the shift measured at $t/M_A\simeq-160$ (up to a minus sign) in the right when the nonlinear tide system merges. After that, the linear tide system evolves another $\sim 10M_A$ until it merges, accumulating nearly another radian of GW phase. The difference in duration is another measurable, and it should make the impact of the nonlinear tide more prominent than what the left panel shows. When measured at the same separation of $r=2R_A$ (which occurs at different times), the nonlinear tide system has $0.81\,{\rm rad}$ less GW phase accumulated compared to the linear case. 

Fig. \ref{fig:phase_shift_SLy} is similar to Fig. \ref{fig:phase_shift}, but the tidally deformed NS $M_A$ is now described by the SLy equation of state with $\lambda_A/M_A^5=390$ or $k_{2A}=0.084$. We then use the universal relations presented in Sec. \ref{sec:UR_in_GR} to estimate $\ddot{k}_{2A}\simeq 0.060$ and $p_{2A}\simeq 0.086$. These values are then converted to our $\If, \omega_{f0}, \JJ_2, \kappa_2, ...$ according to Sec. \ref{sec:calibrations}. For this less soft (and potentially more realistic) model, the nonlinear tide leaves a greater impact. Relative to the linear tide system, including nonlinear tides leads to a phase shift of 0.59\,rad measured at the merger time of the nonlinear system ($t/M_A\simeq -316$), and 1.8\,rad at $r=2R_A$. Accounting for the companion's contribution, the values will be approximately doubled.  

Lastly, we consider the effect of slow spins in Fig. \ref{fig:phase_shift_spin}. The NS's equation of state is SLy, and we vary its spin by $\pm 0.05\omega_A\simeq\pm2\pi\times 84\,{\rm Hz}.$ The dimensionless spin is $\chi_A=\pm 0.04$, within the prior considered for GW170817 \cite{GW170817, GW170817eos}. We use the color gray (yellow) to represent the positive (negative-)spin case. This time, the phase shifts are measured relative to the non-spinning system shown in red. Nonlinear effects are included in all cases. The linear in $\Omega_A$ term shifting the inertial frame f-mode frequency is the dominant effect when $\Omega_A$ is small, making the gray and yellow lines approximately symmetric about the non-spinning line. The chosen spins can create phase shifts similar to the nonlinear tide, highlighting the need to include all effects in the data analysis to avoid misinterpretation of the signal.

\section{Discussion}
\label{sec:discussion}

The main results of the paper were summarized in Sec. \ref{sec:exe_sum}. We highlight again that the nonlinear tide from a single SLy NS can cause a GW phase shift of $\simeq 1.8\,{\rm rad}$ at merger than predicted by the linear theory. 
In a low-frequency expansion, the nonlinear tide is degenerate with the finite-frequency correction of the linear tide (Eq. \ref{eq:omega_f_LF_eff}). If a BNS waveform reduces the f-mode frequency as a function of the tidal deformability using their linear universal relation without accounting for the impact of the nonlinear tide, the inferred tidal deformability will be biased higher by $\sim 10\%$ systematically \cite{Bretz:26}. 
The effect exists in all BNSs and NSBHs to be detected, and therefore, can have a significant impact on the NS equation of state inference through stacking a population of detections \cite{DelPozzo:13, Pang:20}. 
The presence of slow spins, however, can potentially complicate the problem. 
In the rare case where a rapidly spinning NS is detected with f-mode resonantly excited, we also showed that such a system offers a good opportunity to probe new NS physics, $\Gamma_{\rm ad}$, which can hardly be probed through f-modes (but potentially through g-modes \cite{Yu:17b, Ho:23, Kwon:24}) in slowly spinning systems.

Our study also suggests multiple directions for future studies. 
As the affine model is a powerful tool to study the nonlinear tide analytically, it is interesting to upgrade our Newtonian model to a relativistic one, or at least include PN corrections as done in \cite{Ferrari:09, Ferrari:12}. This may theoretically explain why the scalings introduced in Sec. \ref{sec:UR_in_GR} work well in GR. 
Such a calculation would also provide the relativistic values of $\kappa_2\If$ and $\JJ_2$ separately, as they enter $\Delta \omega_2$ (line \ref{eq:dw2_wf_3m}) and $\Delta V_2$ (line \ref{eq:dV2_3m}) in different combinations than $p_{2A}$ (Eq. \ref{eq:p2_vs_J2_kap2}). This separation is needed for NLO beyond the low-frequency limit. Currently, we assume that they scale the same way in GR based on their Newtonian universal relations with $\omega_{f0}^2$ (Sec. \ref{sec:calibrations}), although the validity of this assumption should be assessed more carefully.

Since our study focused primarily on the nonlinear tidal response, we adopted a Newtonian description of the orbit with quadrupolar GW radiation. Formally, the resulting error in the GW waveform is subleading in both tidal and PN effects. In practice, however, tidal effects can become sizable during the late inspiral, requiring a more sophisticated treatment of the orbital dynamics to produce faithful GW waveforms. For example, the Hamiltonian developed in this work can be readily resummed into an EOB form \cite{Buonanno:99} to incorporate relativistic orbital dynamics, similar to the approach of \cite{Yu:25a}.

We remind the reader that some theoretical uncertainties arise from an unknown 3PN boundary term, which causes the GR values of $\ddot{k}_2 \propto 1/\omega_{f0}^2$ to differ between \cite{Pitre:24} and \cite{AbhishekHegade:24} (see Fig. \ref{fig:ddk2_vs_k2}). 
We adopted the results of \cite{Pitre:24} in this work, as they provide a more conservative estimate of the tidal effects. The same uncertainty may also affect the relativistic value of $p_{2A}$ and, consequently, Eqs. (\ref{eq:p2_vs_ddk2_GR}) and (\ref{eq:p2bar_vs_lambar}).
The issue may be resolved by comparing the to-be-developed EOB waveform with nonlinear tides to numerical relativity simulations to fit for the values of $\ddot{k}_{2A}$ and $p_{2A}$.

For data-analysis purposes, computationally efficient phenomenological models are essential.  
As in \cite{Pani:25}, our Eq. (\ref{eq:phase_shift_LF}) can be treated as an additive correction to any baseline binary-black-hole waveform. 
When doing so, one should first compute the full analytical solution for the mode amplitudes (Sec. \ref{sec:nl_mode_amp}) and then evaluate $\Delta E$ and $\Delta \dot{E}$ in terms of these amplitudes. The low-frequency expansion, Eq. (\ref{eq:k_psi_eff}), significantly underestimates the tidal corrections.  
In developing such waveform models, two issues must be addressed. 
First, a more sophisticated treatment is required to compute $\gamma_{2d}$, accounting for the tidal contributions to $\dot{r}$ and $\dot{\omega}$, which dominate over the point-particle predictions in regimes where tidal effects become important.
Second, one should account for the fact that different waveforms generally have different durations in the time domain, an effect not sufficiently captured by $\Delta \Psi$ in Eq. (\ref{eq:delta_Psi}) or the left panels of Figs. \ref{fig:phase_shift} and \ref{fig:phase_shift_SLy}. Since $\Delta \Psi(f)\simeq -\Delta \phi_{\rm gw}(t)$ (eq. 94 of \cite{Yu:24a}), the issue is better illustrated in the time-domain phase shown in the right panels of Fig. \ref{fig:phase_shift} and \ref{fig:phase_shift_SLy}. In addition to measuring the phase shift at a given time, the phase shift at a given orbital separation (reached at different times) also has physical significance, as the location of the merger is more naturally defined by an orbital separation. At times when only one waveform is defined, phase shift may not be a sufficient description.

\begin{acknowledgments}
We thank Abhishek Hegade for the clarification of the theoretical uncertainties on the relativistic value of $\ddot{k}_{2A}$. 
This work is supported by NSF through Awards No. PHY-2308415, No. PHY-2541579 (CAREER Award), and No. AST-2606125. H.Y. and S.Y.L. are supported in part by Montana NASA EPSCoR Research Infrastructure Development under award No. 80NSSC22M0042.
T.V. and K.J.K. acknowledge support from NSF
grant no. 2309360, the Alfred P. Sloan Foundation through grant number FG-2023-20470, and the BSF through award number 2022136.
NA gratefully acknowledges support from the STFC via Grant No.~ST/Y00082X/1.
PP acknowledges support from the grant PID2021-127495NB-I00, funded by MCIN/AEI/10.13039/501100011033 and by the European Union, from the Prometeo 2023 excellence programme grant CIPROM/2022/13, funded by the Ministry of Education, Culture, Universities, and Occupation of the Generalitat Valenciana, and from the grant PID2025-171322NB-C21, funded by the Spanish Ministry of Science and Innovation (MCIN).
FG acknowledges funding from the European Union’s Horizon Europe research and innovation programme under the Marie Sk{\l}odowska-Curie Grant Agreement No.~101151301.
\end{acknowledgments}

\appendix
\section{Coriolis effect and axial modes}
\label{appx:Coriolis}
In general, to handle hydrodynamics in a rotating star, a phase space expansion, as done in the main text, is preferred over a configuration space expansion to have an orthonormal basis and uncoupled (at least in the leading order) equations of motion for the modes \cite{Schenk:02}. Under the affine model with only a finite number of modes and fixed mode eigenfunctions, however, a configuration space expansion can yield insightful results. In this appendix, we will use the affine model to 1). derive the Coriolis effect's impact on each mode's equation of motion beyond linear in $\Omega_A$; 2). assess the impact of the ignored axial modes. 

We note that the equations of motion of the Lagrangian fluid displacement $\vxi$ in a frame corotating with the NS can be derived from a Lagrangian  \cite{Schenk:02}
\begin{equation}
    \mathcal{L} = \int\frac{1}{2}\dvect{\xi} \cdot \dvect{\xi} dM_A + \int \dvect{\xi} \cdot(\vect{\Omega}_A \times \vxi) dM_A - \mathcal{C}(\vxi),
    \label{eq:Lagrangian}
\end{equation}
where the $\mathcal{C}(\vxi)$ term includes all the potential energies. 
We still use Eqs. (\ref{eq:mode_amp_def}) and (\ref{eq:qij_def}) to express $\vxi$ in terms of amplitudes of configuration space eigenmode, $q_a$, under the affine model. 

To address the Coriolis effect, we note that its contribution to the Lagrangian can be evaluated through
\begin{equation}
    \frac{1}{E_A}\int \dvect{\xi} \cdot(\vect{\Omega}_A \times \vxi) dM_A
    =-\Omega_A\Ma \varepsilon_{i3j}(q_{ik} - \delta_{ik})\dot{q}_{jk}, 
\end{equation}
where $\varepsilon_{ijk}$ is the Levi-Civita symbol.
Plugging the values of $q_{ij}$ in terms of mode amplitudes, the Lagrangian equation yields the following equation for the $m_a=2$ mode
\begin{equation}
    \ddot{q}_2 +2 i m_2 C_f\Omega_A \dot{q}_2 + \omega_{f0}^2q_2 = 2\omega_{f0}^2 K_2,
    \label{eq:ddq2}
\end{equation}
where $m_2=2$ and $C_f=-1/l_2=-1/2$. We keep them to respect the general definition of $C_a$ given by Eq. (\ref{eq:Ca}).

The second-order equation of $q_2$ above is equivalent to two first-order equations respectively for $c_{2+}$ and $c_{2-}=c_{-2+}^\ast$, satisfying $q_2=c_{2+} + c_{-2+}^\ast$ with
\begin{align}
    &\dot{c}_{2+}+i\omega_{2} c_{2+}
    + i\sigma c_{-2+}^\ast
    =i\omega_{f0}K_2,\\
    &\dot{c}_{-2+}^\ast - i \omega_{-2}c_{-2+}^\ast 
    - i\sigma c_2=-i\omega_{f0}K_{-2}^\ast =- i\omega_{f0}K_2. 
\end{align}
Note in this appendix, $c_{m_a s_a}$ is the phase space mode amplitude evaluated in the \emph{frame corotating with the NS}, whereas in the main text, $c_a$ is in the inertial frame. The frequencies in the first-order equations are
\begin{align}
    &\omega_{m} = \omega_{f0} +mC_f\Omega_A+\frac{m^2C_f^2\Omega_A^2}{2\omega_{f0}}, \\
    &\sigma=\frac{m^2C_f^2\Omega_A^2}{2\omega_{f0}}.
\end{align}
To complete the mapping, we also need
\begin{align}
    \dot{q}_2=\dot{c}_{2+} + \dot{c}_{-2+}^\ast 
    =- i(\omega_{f0} + m_2 C_f \Omega_A)c_{2+} 
    + i(\omega_{f0} - m_2 C_f\Omega_{A})c_{-2+}^\ast,
\end{align}
where in the second equality, we have used the equations of motion to replace the time derivatives. Similar relations in the inertial frame and in the frame corotating with the orbit are respectively given by Eq. (\ref{eq:dq_vs_c}) and Eq. (\ref{eq:dq_vs_b_orb_frame}). 
The conversion between $(q_{-2}, \dot{q}_{-2})$ and $(c_{-2+}, c_{+2+}^\ast)$ follows similarly by taking complex conjugations.  

The corresponding Hamiltonian in the corotating frame is 
\begin{align}
    \frac{H^{\rm(corot)}}{E_A} &= \frac{1}{E_A}\left(\dvect{\vxi} \frac{\partial \mathcal{L}}{\partial \dvect{\xi}} - \mathcal{L}\right) 
\end{align}
The $|m|=2$ modes' contribution, truncated to terms quadratic in the mode amplitude (which covers all the terms related to the Coriolis effect; note that here we do not treat $\Omega_A$ as a small parameter), are explicitly give below
\begin{align}
    \frac{H_{|m|=2,\, {\rm quadratic}}^{\rm(corot)}}{E_A}&=\frac{1}{2}\left(q_2q_{-2} + \frac{\dot{q}_2 \dot{q}_{-2}}{\omega_{f0}^2}\right)  \nonumber \\
    &=c_2 c_2^\ast + \frac{m_2 C_f \Omega_A}{\omega_{f0}}(c_2 c_2^\ast-c_{-2}c_{-2}^\ast) + \frac{m_2^2 C_f^2\Omega_A^2}{2\omega_{f0}^2}(c_2 + c_{-2}^\ast)(c_{-2} + c_{2}^\ast).
\end{align}
We omit the $s_a$ part in the subscript, and all phase-space mode amplitudes presented above have positive eigenfrequencies. 
To get the Hamiltonian in the inertial frame, we can apply the canonical transformation described in \cite{Yu:25a}, which corresponds to adding the $(\Omega_A/2) \sum_{a, s_a}S_{a,s_a}$ term to the Hamiltonian.  

To incorporate the axial mode ignored in the main text, we can modify Eq. (\ref{eq:q_ij_vs_q_a}) to include an anti-symmetric component of $q_{ij}$, whose amplitude is labeled $\Xi$, as
\begin{subequations}
    \begin{align}
        q_{12} &= \frac{\omega_A}{\omega_{f0}} \frac{i(q_2-q_{-2})}{2\sqrt{2\Ma}} - \Xi,\\
        q_{21} &= \frac{\omega_A}{\omega_{f0}} \frac{i(q_2-q_{-2})}{2\sqrt{2\Ma}} + \Xi, 
    \end{align}
\end{subequations}
All other components in $q_{ij}$ remain the same as in Eq. (\ref{eq:qij_def}). 
Evidently, when $\Xi$ is small, the axial mode corresponds to an overall rotation of the background star about the z-axis by an angle $\Xi$.

We will study the impact of $\Xi$ directly in the configuration space. 
One needs to add the following terms to the Lagrangian shown in Eq. (\ref{eq:Lagrangian}) due to $\Xi$ and its time derivative, 
\begin{align}
    \frac{\mathcal{L}_{\Xi}}{E_A} &= 
    \underbrace{\frac{4\pi}{15}\If^2\frac{\omega_{f0}^2}{\omega_A^4}\dot{\Xi}^2}_{\text{kinetic}}
    \underbrace{+\frac{2}{3} \sqrt{\frac{\pi}{5}}\If\frac{\Omega_A}{\omega_A^2}\left[\left(\dot{q}_0\Xi - q_0\dot{\Xi}\right)-\sqrt{2}\frac{\omega_{f0}}{\omega_{r}}\left(\dot{q}_r \Xi - q_r \dot{\Xi}\right)\right]}_{\text{Coriolis}} \nonumber \\
    &\underbrace{+\frac{1}{3}\sqrt{\frac{\pi}{5}}\If\left(\frac{\omega_{f0}^2}{\omega_A^2}q_0 - \sqrt{2}\frac{\omega_{f0}\omega_r}{\omega_A^2} q_{r}\right) \Xi^2}_{\text{three-wave interaction}}
    \underbrace{+\frac{2\pi}{15}\If^2\frac{\omega_{f0}^2}{\omega_A^2}\epsilon_A \Xi^2
    + \frac{4\pi}{15}\If^2 \frac{\omega_{f0}^2}{\omega_A^2}\frac{\Omega_A^2}{\omega_A^2}\Xi^2}_{\text{nonlinear tidal/centrifugal coupling with $m=0$}}
    \nonumber \\
    &\underbrace{-i \sqrt{\frac{3\pi}{10}} \If \left[q_2e^{2i(\phi-\Omega_At)} - q_{-2}^\ast e^{-2i(\phi-\Omega_A)}\right]\epsilon_A \Xi}_{\text{nonlinear tidal coupling with $|m|=2$}}.
\end{align}
Four-wave and higher-order interactions are ignored. From the Lagrangian, we first study the evolution of $\Xi$ itself. 
\begin{align}
    2\Ma \dot{\Omega}_\Xi+\frac{4}{3}\sqrt{\frac{\pi}{5}}\If\frac{\Omega_A}{\omega_A^2}\left(\dot{q}_0 -\sqrt{2}\frac{\omega_{f0}}{\omega_r}\dot{q}_r\right)
    =&2\sqrt{\frac{3\pi}{10}}\If \epsilon_A{\rm Im}[q_2 e^{2i(\phi-\Omega_At)}]
    \nonumber \\
    &+\frac{2}{15}\If\frac{\omega_{f0}^2}{\omega_A^2}\left[2\pi \If \left(\epsilon_A + 2\frac{\Omega_A^2}{\omega_A^2}\right) 
    + \sqrt{5\pi}\left(q_0 - \sqrt{2}\frac{\omega_r}{\omega_{f0}}q_r\right)
    \right]\Xi. 
\end{align}
where we have defined $\Omega_\Xi\equiv\dot{\Xi}$, which can be viewed as a modification to the constant background spin rate $\Omega_A$. The above equation can be simplified significantly. First, we note that the second term on the left-hand side (from the Coriolis effect) should be small, as $|\dot{q}_0|\sim|\dot{q}_r|\sim |\dot{r}/r|$ and $|\Omega_A|<\omega_A$. Second, the second term on the right-hand side vanishes when the leading order solutions of $q_0=2c_0=-2\sqrt{\pi/5}\If[\epsilon_A + (2/3)(\Omega_A^2/\omega_A^2)]$ and $q_r=2c_r=(8/3) \sqrt{\pi/10}(\omega_{f0}/\omega_r)\If (\Omega_A^2/\omega_A^2)$ are substituted in.
Third, the surviving term on the right-hand side is half of the tidal torque acting on the NS (minus of Eq. \ref{eq:pH_pphi} divided by $M_A R_A^2$) when the nonlinear terms are dropped.  
Noticing further that $2\Ma$ is just the unperturbed moment of inertia measured in $M_AR_A^2$ (Eq. \ref{eq:IA_vs_Ma}), we arrive at
\begin{equation}
    I_A \dot{\Omega}_\Xi\simeq -\frac{1}{2}\frac{\partial \mathcal{L}}{\partial \phi},
    \label{eq:bg_spin_evol}
\end{equation}
which has an intuitive interpretation. That is, the background spin of an NS is actually evolving due to about half of the total tidal torque acting on the NS. This is exactly what \cite{Yu:25a} found in their appendix A. \cite{Yu:25a} arrived at the same result by finding that the physical spin, $\sim\int  \uvect{e}_z\cdot[\vxi\times (\dot{\vxi}+\vect{\Omega}_A\times \vxi)]dM_A$, is about half of the canonical spin (which is the tidal spin in Eq. (\ref{eq:tidal_spin}) driven by $\partial H/\partial \phi=-\partial \mathcal{L}/\partial\phi$), and therefore, the background must carry the other half of the angular momentum input to the NS. In this work, we verify the conclusion directly from the equation of motion for $\Xi$.

How does $\Xi$ impact the dynamics of other modes studied in the main text? Since $\Xi$ is related to an overall rotation of the background, by axial symmetry, we may argue that the impact of $\Xi$ itself on the interaction potentials should have little physical significance. 

What may be dynamically relevant is its time derivative, $\Omega_\Xi$. 
Eq. (\ref{eq:bg_spin_evol}) also tells us that $I_A\Omega_{s} \simeq (\mathcal{S}_{2+} + \mathcal{S}_{-2+})/2\propto (|b_2|^2-|b_{-2}|^2)$. Because $b_2=b_{-2}=W_2 \If\epsilon_A$ under the adiabatic limit, $\Omega_\Xi$ appears only at $\mathcal{O}[(\omega/\omega_{f0}) \epsilon_A^2]\sim \mathcal{O}(\epsilon_A^{5/2})$. It affects other modes in two ways. For the $m_a=0$ f- and radial modes, $\Omega_\Xi$ appears via the Coriolis effect, so that their equations of motion take a form
\begin{align}
    &\ddot{q}_0 + \frac{8\sqrt{5\pi}}{15} \If \frac{\omega_{f0}^2}{\omega_A^2}\Omega_A\Omega_\Xi = 2\omega_{f0}^2 K_0,\\
    &\ddot{q}_r - \frac{16\sqrt{\pi}}{3}\II_r\frac{\omega_r^2}{\omega_A^2} \Omega_A\Omega_\Xi = 2 \omega_r^2 K_r.
\end{align}
The Coriolis terms on the left-hand side are both formally $\mathcal{O}(\epsilon_A^3)$ terms at NNLO, beyond the order we consider for the quasi-adiabatic modes. 

For the $m_a=2$ f-mode, the $\Xi$ term modifies its coupling with the tidal field, as
\begin{equation}
    \ddot{q}_2 +  \omega_{f0}^2 q_2= 2\omega_{f0}^2 V_2 (1+i\Xi) e^{-im_2\phi}\simeq 2\omega_{f0}^2 V_2 e^{-i(m_2\phi-\Xi)}.
\end{equation}
We have ignored the terms involving the background spin $\Omega_A$, as well as nonlinear couplings of $q_2$ with $q_0$ and $q_r$ for simplicity. 
As $\Xi$ corresponds to a change of the background rotation, it creates a Doppler shift of the tidal drive. Now, the solution of $q_2$ may take the form of 
\begin{align}
    q_2\simeq \frac{2\omega_{f0}^2 V_2}{\omega_{f0}^2 - m_a^2\omega^2 + m_2\omega\Omega_\Xi}. 
\end{align}
The $\omega \Omega_\Xi$ term is formally $\mathcal{O}(\epsilon_A^{3})$, which is of higher order than the nonlinear frequency shift considered in Eq. (\ref{eq:dw2_wf_4m_b2sq}). Therefore, we have justified why the axial mode $\Xi$ (or the NS's background spin evolution) may be dropped, at least for a slowly spinning NS without the f-mode resonance, as done in the main text.

In the special case where the $m_a=2$ f-mode is resonantly excited, can the $m_2\omega\Omega_\Xi$ term modify the conclusions drawn on resonance locking in Sec. \ref{sec:res_locking}? 
It has the correct sign, which tends to reduce the detuning of the Lorentzian, but we would also need to compare its size with the $\Delta \omega_2$ term computed in Eq. (\ref{eq:dw2_4m_E2}). 
For this, we approximate (using eq. \ref{eq:bg_spin_evol} and the fact that $|b_2|\gg |b_{-2}|$ near mode resonance)
\begin{equation}
    \Omega_\Xi\simeq \frac{\mathcal{S}_{2+}}{2I_A}=\frac{m_2}{4\Ma} \frac{\omega_A^2}{\omega_{f0}} b_2b_2^\ast=\frac{15m_2}{16\pi\If^2} \frac{\omega_A^4}{\omega_{f0}^3} b_2b_2^\ast,
\end{equation}
where we have also used Eq. (\ref{eq:I_f}) to replace $\Ma$ in terms of $(\If, \omega_{f0})$. 
Meanwhile, Eq. (\ref{eq:dw2_4m_E2}) leads to
\begin{equation}
    \Delta \omega_2^2= \omega_{f0}^2\left(2\frac{\Delta \omega_2}{\omega_{f0}}\right) \simeq -\frac{150}{49\pi\If^2}\frac{\omega_A^4}{\omega_{f0}^2} \left(1 + \frac{147}{80}\frac{\omega_{f0}^2}{\omega_{r}^2}\right)b_2b_2^\ast. 
\end{equation}
Consequently, 
\begin{equation}
    \frac{m_2\omega\Omega_\Xi}{|\Delta \omega_2^2|}\simeq \frac{49}{40}\left(\frac{\omega}{\omega_{f0}}\right) \frac{1}{1+\frac{147}{80}\frac{\omega_{f0}^2}{\omega_r^2}}. 
\end{equation}
Even at the end of the inspiral,  $\omega \lesssim \omega_{f0}/2$ for a typical BNS with realistic equations of state with $n\sim0.5-1$. As a result, although the $m_2\omega\Omega_\Xi$ term has the correct sign, we have shown both formally and in terms of explicit numerical values that it would be subdominant compared to the $\Delta \omega_2^2$ term considered in Sec. \ref{sec:res_locking}. Furthermore, a Doppler shift due to the background spinning up cannot compensate for the increase in the effective damping. Our conclusion that the f-mode cannot experience resonance locking during the inspiral of a BNS should still hold.

\bibliography{ref}

\end{document}